\documentclass[12pt,a4paper]{article}
\pdfoutput=1
\usepackage[utf8]{inputenc}
\usepackage[T1]{fontenc}
\usepackage{mathrsfs}
\usepackage{tikz-feynman}   
\usetikzlibrary{snakes}     
\usepackage{tikz}
\usepackage{style}
\usepackage[english]{babel}
\usepackage{url}
\usepackage{footnote}

\usepackage{subcaption}
\usepackage{amsmath}
\usepackage{amssymb}
\usepackage{amsthm}
\usepackage{psfrag}
\usepackage{graphicx}
\usepackage{hyperref}
\usepackage{feynmp}

\newcommand{\be}{\begin{equation}}
\newcommand{\ee}{\end{equation}}
\def\ba{\begin{aligned}}
\def\ea{\end{aligned}}
\newcommand{\beqa}{\begin{eqnarray}}
\newcommand{\eeqa}{\end{eqnarray}}

\newcommand\m{\mu}

\renewcommand\k{{\bf k}}

\newcommand\g{\gamma}

\newcommand\n{\nu}
\renewcommand\r{\rho}

\newcommand{\SL}{\text{SL}\left(2,\mathbb{R}\right)}
\newcommand{\sla}{\mathfrak{sl}\left(2,\mathbb{R}\right)}

\newcommand{\vev}[1]{\ensuremath{\left\langle #1 \right\rangle}}

\def\d{\partial}
\newcommand{\bseq}{\begin{subequations}}
\newcommand{\eseq}{\end{subequations}}

\renewcommand{\ln}{\mathop{\rm ln}\nolimits}

\title{Love numbers of black  p-branes: fine tuning,
Love symmetries, and their geometrization}

\author[a,b]{Panagiotis Charalambous\footnote{\texttt{pcharala@sissa.it}}}
\author[c]{Sergei Dubovsky\footnote{\texttt{dubovsky@nyu.edu}}}
\author[d]{Mikhail M. Ivanov\footnote{\texttt{ivanov99@mit.edu}}}

\affiliation[a]{International School for Advanced Studies (SISSA), \\ Via Bonomea 265, 34136 Trieste, Italy}
\affiliation[b]{National Institute for Nuclear Physics (INFN), \\ Sezione di Trieste, Via Valerio 2, 34127, Italy}
\affiliation[c]{Center for Cosmology and Particle Physics, Department of Physics, New York University, \\ New York, NY 10003, USA}
\affiliation[d]{Center for Theoretical Physics, Massachusetts Institute of Technology, \\ Cambridge, MA 02139, USA}

\abstract{
We compute scalar static response coefficients (Love numbers)
of non-dilatonic black $p$-brane solutions in higher dimensional
supergravity. 
This calculation revels a fine-tuning behavior similar to that 
of higher dimensional black holes,
which we explain by 
``hidden'' near-zone Love symmetries.
In general, these 
symmetries
act on equations for perturbations but they are not background isometries.  
The Love symmetry
of charged $p=0$ branes
is described by the 
usual $SL(2,\mathbb{R})$ algebra.
For $p=1$ the Love symmetry has
an algebraic structure $SL(2,\mathbb{R})\times SL(2,\mathbb{R})$.
The $p=0,1$ Love symmetries 
reduce to isometries of the 
near-horizon Schwarzschild-AdS$_{p+2}$ 
metric in the near-extremal 
finite temperature limit. 
They further reduce to the 
AdS$_{p+2}$ isometries
in the extremal zero-temperature 
limit. 
We call this process geometrization.  
In contrast, for the $p>1$ cases, 
the Love symmetry is always an $SL(2,\mathbb{R})$, and
there is no limit 
in which it becomes geometric. 
We interpret geometrization
and its 
absence
as a consequence of the 
local equivalence between
the Schwarzschild-AdS$_{p+2}$ and pure AdS$_{p+2}$ spaces for $p=0,1$,
which does not hold for $p>1$.
We also show that 
the static Love numbers of extremal $p$-branes
are always zero regardless of  spacetime 
dimensionality, which contrasts starkly with the non-extremal case.
Overall, our results suggest that 
the Love symmetry is hidden by nature,
and it can acquire a geometric 
meaning only 
if the background has an 
AdS$_{2}$ or AdS$_{3}$
limit. 
}

\begin{document}

\vspace{-1.5cm}

\begin{flushright}
MIT-CTP/5836
\end{flushright}

\maketitle
\flushbottom

\section{Introduction}
\label{sec:int}

The detection of gravitational waves from black hole mergers with the LIGO-Virgo-KAGRA detectors~\cite{LIGOScientific:2016aoc,LIGOScientific:2017vwq} has started a new era in experimental gravitational physics. The sensitivity of these experiments to tidal deformation~\cite{Chatziioannou2020} and heating~\cite{Chia:2024bwc} of relativistic compact bodies stimulated a boom in theoretical research of these phenomena; for instance, besides probing the internal structure of the involved compact bodies, tidal response effects have been proposed for testing strong-field gravity phenomena~\cite{Cardoso:2017cfl,Franzin:2017mtq,Cardoso:2018ptl,Katagiri:2023yzm} as well as for lifting astrophysical measurement degeneracies via ``I-Love-Q'' relations~\cite{Xie:2022brn,Yagi:2013bca,Yagi:2013awa,Pani:2015tga,Yagi:2015pkc,Uchikata:2016qku,Yagi:2016qmr}. In particular, the tidal deformation of black holes parameterized by the tidal Love numbers (LNs), has become a new hot topic. The leading order (static) induced deformations of Schwarzschild~\cite{Fang:2005qq,Damour:2009vw,Binnington:2009bb,Kol:2011vg} and Kerr~\cite{LeTiec:2020spy,Chia:2020yla,LeTiec:2020bos,Charalambous:2021mea} black holes in four dimensional general relativity were found to vanish identically\footnote{Beyond the setup of an asymptotically flat and isolated general-relativistic black hole, the static Love numbers are in general non-zero, but are still expected to be small. Such deviations probe the environment of the black hole or even beyond-GR physics and their imprint on gravitational wave detectors can be magnified in extreme-mass-ration-inspirals~\cite{DeLuca:2021ite,DeLuca:2022xlz,Pani:2019cyc,Datta:2019epe}.}. This represents a formidable fine-tuning problem\footnote{The next-to-leading order, ``dynamical'' LNs are not zero for both Schwarzschild~\cite{Chakrabarti:2013lua} and Kerr~\cite{Charalambous:2021mea} black holes, see also Refs.~\cite{Saketh:2023bul,Ivanov:2024sds} for recent analyses based on scattering amplitudes. The dynamical LNs are not zero, and hence do not require any symmetry. On the other hand, a recent burst on non-perturbative computations has demonstrated the vanishing of the \textit{non-linear} static and axisymmetric LNs of both Schwarzschild and Kerr black holes~\cite{Gurlebeck:2015xpa,DeLuca:2023mio,Riva:2023rcm,Combaluzier-Szteinsznaider:2024sgb,Kehagias:2024rtz,Iteanu:2024dvx}.} in the context of effective field theory of inspiraling binaries~\cite{Goldberger:2004jt,Goldberger:2005cd,Porto:2005ac,Levi:2015msa,Porto:2016pyg,Levi:2018nxp}, which is a systematic tool for precision gravitational waveform calculations.

Recently, this naturalness problem has been resolved by the discovery of the Love symmetry~\cite{Charalambous:2021kcz,Charalambous:2022rre}. The Love symmetry is a symmetry of black hole perturbations, which provides an algebraic constraint forcing the black hole LNs to vanish. The algebraic structure of this symmetry is $\SL$. The Love symmetry only acts at the level of perturbations, not the black hole background itself. As such, it is dubbed a ``hidden'' symmetry.\footnote{See Refs.~\cite{Hui:2021vcv,Hui:2022vbh,Achour:2021dtj,BenAchour:2022uqo,BenAchour:2022fif,BenAchour:2023dgj} for different proposals.}

The discovery of the Love symmetry has left many open questions. It is unclear what is the physical origin of this symmetry, and if it can have a robust geometric interpretation.\footnote{The hidden symmetries may be interpreted in the context of ``effective geometries''~\cite{Cvetic:2011dn,Cvetic:2011hp,Cvetic:2012tr}, which, however, are not directly related to the underlying background geometries besides a defining connection with the thermodynamic properties of the horizon.} The original work~\cite{Charalambous:2022rre} showed that the Love symmetry is tighly connected with the well known near-horizon extremal (NHE) isometries of extremely charged (Reissner-Nordstr{\"o}m) and extremely rotating (Kerr) black holes. For the charged black holes, the Love symmetry exactly reduces to the NHE in the appropriate scaling near-horizon limit. For the rotating black holes, the Love symmetry is part of a larger symmetry algebra $\SL\ltimes \hat U(1)$, which also contains the Starobinsky symmetry~\cite{Charalambous:2021kcz} and the Bardeen-Horowitz NHE isometry~\cite{Bardeen:1999px,Kunduri:2007vf} as subgroups. Further connections between the Love symmetry and the near-horizon isometries of near-extremal Kerr Taub-NUT black holes have been discovered in Ref.~\cite{Guevara:2023wlr}. These results point to a deep connection between the Love symmetries and near-horizon isometries, which has not been fully uncovered yet. Our work makes a further step in this direction by considering Love numbers and Love symmetries of black $p$-branes in higher dimensions.

In the past studies, higher dimensional black holes have provided a powerful tool to understand LNs in the physical four-dimensional case. The first relevant observation to solving the Love number paradox was the absence of fine-tuning for static black holes in higher spacetime dimensions $D$ for some multipole numbers~\cite{Kol:2011vg,Hui:2020xxx}. In addition, the higher-$D$ black hole LNs provide a clear cut argument why their vanishing in $D=4$ is not directly related to the no-hair theorem~\cite{Charalambous:2022rre}. The recent study of Myers-Perry BHs~\cite{Charalambous:2023jgq, Rodriguez:2023xjd} showed that the vanishing of LNs of rotating and static black holes in general require different selection rules, which hints at extra coincidences behind the vanishing of LNs for Kerr and Schwarschild black holes in four dimensions.

These insights from higher dimensional black holes along with the connection between the Love symmetry and the extremal near-horizon isometries motivates looking into higher-dimensional $p$-brane solutions. These solutions have a richer geometric structure than black holes, and we may hope that this structure has distinctive effects on black $p$-brane perturbations and their hidden symmetries. In addition, the study of $p$-brane solutions may give new insights into the role of hidden symmetries, such as the Love symmetry, in holographic models where $p$-branes play a crucial role, see e.g.~\cite{Aharony:1999ti} for a classic review of the AdS/CFT correspondence.

The connection between the Love symmetry and NHE isometries suggests a conjecture that the Love symmetries of $p$-branes should ``inherit'' the geometric structure of the extremal near-horizon regions. This is also natural from the holographic point of view. In holographic setups, vacuum solutions enjoy conformal symmetries, which get broken spontaneously by excited states, i.e. gravity solutions at a finite Hawking temperature. This spontaneous breaking of conformal symmetry parallels the transition of the Love symmetry from being an exact isometry of the near-horizon region to becoming a hidden symmetry of perturbations only. This argument suggests that we should think of the Love symmetry as a spontaneously broken NHE isometry. This conjecture predicts that the $p$-branes should feature Love symmetries that should match underlying conformal symmetries of the vacuum solutions. Since the extremal $p$-branes posses $SO\left(p+1,2;\mathbb{R}\right)$ isometries in their near-horizon regions, one might expect the corresponding hidden Love symmetries to have the same algebraic structure. Testing this conjecture is the main goal of our work.

To that end, we consider responses of non-dilatonic black $p$-branes to external test scalar fields. The non-dilatonic $p$-branes represent special supergravity solutions that are not typically discussed in the holographic models. However, in contrast to more familiar $p$-branes that appear in the low-energy description of string theory, the non-dilatonic solutions do not have curvature singularities at the horizon in the extremal limit (when the brane mass is equal to its charge). In this sense the non-dilatonic $p$-branes appear as a more natural generalization of black holes for the purpose of tidal LN's studies. Note however that the $p=3$ non-dilatonic solution is identical to the usual self-dual $D=10$ threebrane since the latter features a constant (and consequently non-singular) dilaton profile. In addition, the non-dilatonic $p=1$ brane is identical to the well known self-dual D1-D5 brane in $D=11$. These two cases allow us to cover two familiar solutions relevant in the holographic context.

Our paper is structured as follows. We summarize our key results in Section~\ref{sec:summ}. Then in Sec.~\ref{sec:NearHorisonSymmetriesRN} we discuss in detail the case of Reissner-Nordstr{\"o}m black holes in $D$ dimensions. Their perturbations posses $\SL$ Love symmetry, which we relate to the near-extremal near-horizon isometries in the appropriate finite temperature scaling limit introduced first by Maldacena~\cite{Maldacena:1998uz}. This allows us to establish the first direct geometric interpretation of the Love symmetries. We move on to the case of black $p$-branes in Sec.~\ref{sec:WVEFTandLNs}. There, we define scalar LNs of $p$-branes using world-volume effective field theory, and match them from classical supergravity calculations. We find that scalar LNs are fine-tuned from the EFT perspective, and this fine-tuning can be explained by the hidden Love symmetry of black branes' perturbations. The Love symmetry of black strings ($p=1$) is a $\SL\times \SL$ symmetry which matches the above conjecture on the relationship to the extremal near-horizon isometries. For $p>1$, however, the only hidden symmetry one can find is an $\SL$ symmetry similar to the usual Love symmetry of higher-dimensional black holes. We also show that the scalar LNs of extremal $p$-branes are always zero regardless of $p$ and $D$. Sec.~\ref{sec:NNHEsympbrane} discusses the near-horizon near-extremal (NNHE) limit of black $p$-branes in which their geometry becomes that of the $p+2$ dimensional Schwarzschild-Anti-de-Sitter (SAdS) solution. For $p=0,1$ SAdS$_{p+2}$ is isomorphic to the vacuum AdS$_{p+2}$ which explains the existence of the AdS$_{p+2}$ Killing vectors associated with the Love symmetry. For $p>1$, however, SAdS$_{p+2}$ is not diffeomorphic to AdS$_{p+2}$, and hence the near-horizon region does not generate an enhanced symmetry. For these solutions, the Love symmetry also does not reduce to the vacuum AdS$_{p+2}$ isometry in the extremal limit. This suggests that the geometric interpretation of the Love symmetry is possible only for $p=0,1$ solutions thanks to the diffeomorphism between the relevant AdS and SAdS spaces. For general $p$, the Love symmetry exists only as a hidden symmetry without an explicit ``embedding'' into the background geometry in any limit. Finally, we draw conclusions in Sec.~\ref{sec:concl}.




\section{Summary of main results}
\label{sec:summ}




We start with a brief summary of the main results of our paper.

\textbf{1. Scalar Love numbers of black $p$-branes.}
We have computed the scalar polarizability
of black $p$-branes in classical 
(super)gravity. We define them
in analogy with scalar Love numbers
of asymptotically flat black holes, 
using the world-volume 
formalism. The Love numbers
are Wilson coefficients 
of the effective world-volume action,
which we determine by matching 
scalar response calculations
in the EFT and full supegravity. 
The scalar and homogeneous-along-the-brane Love numbers (SLNs)
appear to be identical to those
of higher-dimensional black holes,
but with the number of spacetime 
dimensions shifted by the brane codimension $p$. Specifically, the behaviour 
of $p$-brane SLNs depends on a generalized 
angular momentum $\hat \ell \equiv \ell/(D-3-p)$, where $\ell$ is the orbital number
of perturbations in the
spherically-symmetric $D-3-p$ dimensional space
transverse to the brane. SLNs 
are zero, run logarithmically, or non-zero constants provided that 
$\hat\ell$ is integer, half-integer, or a generic number, respectively. 

We define an analog of the near-zone region for $p$-branes w.r.t. to the asymptotically flat spherically symmetric 
extra dimensions. We show that within the near-zone approximation, the Love numbers
are zero even for external fields that are  non-static and 
non-homogeneous w.r.t. brane 
dimensions (i.e. having non-zero wave-vectors $\k_\parallel$ along the brane). 
This is the $p$-brane analog of the statement 
that the dynamical Love numbers of static  
black holes are zero within the near-zone 
approximation. 
In what follows, we call perturbations with $\k_\parallel=0$ brane-homogeneous.

\textbf{2. Scalar LNs of extremal $p$-branes.}
For extremal $p$-branes (when its mass and  charge satisfy the extremality condition),
the scalar Love numbers are always 
zero regardless of the number 
of spacetime dimensions $D$ or 
brane codimensionality $p$. This also holds true for the higher-dimensional Reissner-Nordstr\"{o}m black holes. In the extremal case, the $p$-branes acquire an enhanced  
conformal $SO\left(p+1,2;\mathbb{R}\right)$ symmetry in the near-horizon region.
This symmetry continues to exist 
at the level of scalar perturbations 
in the near-zone region, i.e. beyond 
the original near-horizon setup. 
We show that the static and homogeneous 
perturbations are primary vectors
in the highest weight representation
of this conformal symmetry, which 
produces solutions with zero Love numbers
for any $\hat\ell$.

\textbf{3. Love symmetries of black $p$-branes.} The intricate behaviour of static 
SLNs for non-extremal black $p$-branes
can be explained as a selection rule 
imposed by a hidden symmetry of 
$p$-brane perturbations, Love symmetry. 
This is the symmetry of the equations of 
motion for perturbations that appears in
the near-zone approximation. 
For $p=0$, this is the usual higher-dimensional $\SL$ Love symmetry 
presented previously in~\cite{Charalambous:2021kcz,Charalambous:2021mea,Charalambous:2022rre}, which is an isometry group of the AdS$_2$ space.
For $p=1$ (black strings) we present a new generalized Love symmetry whose algebraic 
structure is $\SL\times \SL$. This is 
the isometry group of AdS$_3$.
Within the near-zone expansion, the black string Love symmetry is formally 
valid even for perturbations that 
are dynamical and brane-inhomogeneous. 

For $p>1$, we were able to identify 
a hidden local symmetry for 
the near-zone $p$-brane perturbations 
only if they are 
brane-homogeneous.
(Despite the fact that they are still 
integrable in the near-zone even if they are brane-inhomogeneous.)
In that case, 
the symmetry
structure is $\SL$ similar to the 
Love symmetry of black holes. 
We call it ``homogeneous'' 
Love symmetry. 

\textbf{4. Love symmetry of black strings ($p=1$).}
The properties of the black string Love symmetry $\SL\times \SL$ 
are very similar to those of the
black hole Love symmetry $\SL$.
It outputs the vanishing of black string Love numbers as a result of the 
highest weight property of the 
relevant $\SL\times \SL$ representation
that contains the static and brane-homogeneous
solution. Moreover, in the extremal near-horizon limit
implemented by means of the standard 
scaling transformations, the Love symmetry 
reduces to the AdS$_3$ isometry 
of the black string near-horizon region. The latter is the well known
AdS$_3$ geometry of the BTZ black holes
used in many holographic 
constructions.

\textbf{5. Geometrization of Love Symmetry for $p=0,1$: 
reduction to near-horizon near-extremal isometries.}
Black $p$-branes have an interesting near-extremal limit at a formally finite 
Hawking temperature, 
introduced by Maldacena. 
In this limit,
the near-horizon region of $p$-branes,
orthogonal to the asymptotically flat dimensions, 
acquires the $p+2$ dimensional
Schwarzschild-AdS$_{p+2}$ (SAdS$_{p+2}$) geometry.
For $p=0,1$, SAdS$_{p+2}$
is diffeomorphic to AdS$_{p+2}$. This means
that in the near-extremal near-horizon (NNHE) Maldacena limit, the $p=0,1$
black branes acquire extra 
symmetries associated with 
AdS-like Killing vectors, 
at a finite 
Hawking temperature. These Killing vectors match the Love symmetry vectors. Thus, within the NNHE geometries,
one can interpret the 
appearance of the Love symmetry Killing vectors 
as a result of the diffeomorphism
between AdS and SAdS spaces. 

It is instructive to approach this problem from the opposite end, i.e. 
taking the NNHE limit of the 
Love symmetry generators. 
Then one can conclude that the hidden Love symmetries
acquire a direct geometric 
interpretation of black $p=0,1$ branes' near-horizon 
isometries in the near-extremal limit.
We call this phenomenon
``geometrization.''
We show how the Love symmetries, 
now being spacetime symmetries, 
impose geometric 
constraints on black hole and black strong perturbations. In particular, 
we show how $\SL\times \SL$ representation 
theory dictates the spectrum of 
black string quasinormal modes.


\textbf{6. Obstruction to geometrization of Love Symmetry for $p>1$.}
Unlike the black hole and black string cases, there is no general hidden symmetry 
of black $p$-branes valid in the near-zone 
for brane-inhomogeneous fluctuations. 
There is only the ``homogeneous'' black $p$-brane
Love symmetry, but it does not reduce 
to an exact geometric symmetry
in near-extremal or extremal 
near-horizon limits. In other words,
geometrization happens in no limit
for $p>1$.

From the NNHE geometric perspective, the absence of ``large'' Love symmetries can be linked to the fact that the
near-horizon
SAdS$_{p+2}$ factor of near-extremal 
black $p$-branes 
does not generate new Killing vectors, i.e. 
SAdS$_{p+2}$
is not diffeomorphic 
to the AdS$_{p+2}$ space.
Since the Love symmetry
is $\SL$, one cannot match it 
with the underlying geometry. 
We present 
an explicit proof of this statement.

On the opposite end,
if we start with the Love 
symmetry for a general $p$ case, we see that
in contrast to the $p=0,1$
branes, the $p>1$ $\SL$ Love symmetry 
does not get geometrized in the NNHE limit. 
This suggests 
that the existence of the Love symmetry 
for brane-inhomogeneous perturbations 
and its consequent geometrizations 
in the near-extremal and extremal limits 
may be a coincidental artifact of a low 
co-dimensionality
of $p$-branes.

\section{Warmup: enhanced near-horizon symmetries of near-extremal Reissner-Nordstr\"{o}m black holes}
\label{sec:NearHorisonSymmetriesRN}

Consider the geometry of an isolated, asymptotically flat and static black hole in General Relativity, described by the $D$-dimensional spherically symmetric Reissner-Nordst\"{o}m solution,
\be\label{eq:gfullRN}
	ds^2 = -f_{+}\left(r\right)f_{-}\left(r\right)dt^2 + \frac{dr^2}{f_{+}\left(r\right)f_{-}\left(r\right)} + r^2d\Omega_{\hat{d}+1}^2 \,,\quad f_{\pm}\left(r\right)=1-\left(\frac{r_{\pm}}{r}\right)^{\hat{d}} \,.
\ee
In the above expression, we have defined
\be
	\hat{d} \equiv D-3
\ee
and $d\Omega_{\hat{d}+1}^2$ is the metric on the unit radius $(\hat{d}+1)$-sphere. The outer, event, (``$+$'') and inner, Cauchy, (``$-$'') horizons are related to the ADM mass $M$ and the electric charge $Q$ of the black hole, as encoded in the Schwarzschild radius $r_{s}$ and the charge parameter $r_{Q}$,
\be
	r_{s}^{\hat{d}} = \frac{16\pi GM}{(\hat{d}+1)\,\Omega_{\hat{d}+1}}\,,\quad r_{Q}^{2\hat{d}} = \frac{32\pi^2GQ^2}{\hat{d}\,(\hat{d}+1)\,\Omega_{\hat{d}+1}^2} \,,
\ee
according to
\be
	r_{\pm}^{\hat{d}} = \frac{1}{2}\left[r_{s}^{\hat{d}}\pm\sqrt{r_{s}^{2\hat{d}}-4r_{Q}^{2\hat{d}}}\right] \,,
\ee
with $\Omega_{\hat{d}+1} = 2\pi^{(\hat{d}+2)/2}/\Gamma\big(\frac{\hat{d}+2}{2}\big)$ the surface area of $\mathbb{S}^{\hat{d}+1}$ and we are working in CGS units. The (absolute value of) the inverse surface gravity of the $D$-dimensional Reissner-Nordstr\"{o}m black hole at the outer horizon and inner horizon are given by
\be\label{eq:betaRN}
	\beta_{+} \equiv \kappa_{+}^{-1} = \frac{2r_{+}}{\hat{d}}\frac{r_{+}^{\hat{d}}}{r_{+}^{\hat{d}}-r_{-}^{\hat{d}}} \quad\text{and}\quad \beta_{-} \equiv -\kappa_{-}^{-1} = \frac{2r_{-}}{\hat{d}}\frac{r_{-}^{\hat{d}}}{r_{+}^{\hat{d}}-r_{-}^{\hat{d}}} \,,
\ee
respectively. These dictate the behavior of the tortoise coordinate, defined via $dr_{\ast}=\frac{dr}{f_{+}f_{-}}$ in terms of the areal radius, near the horizons,
\be
	r_{\ast} \sim \pm\frac{\beta_{\pm}}{2}\ln\left|\frac{r-r_{\pm}}{r_{\pm}}\right| \quad\text{as $r\rightarrow r_{\pm}$} \,.
\ee

\subsection{Near-extremal near-horizon region: equivalence to AdS$_2$}
\label{sec:nearNHEAdS2}
Let us now study the limit of approaching the inner horizon, while keeping
\be
	\frac{f_{+}}{f_{-}} = \text{fixed} \,,
\ee
as we send $r_-\to r_+$
consistent with extremality.
The Hawking temperature remains 
fixed in this limit. 
We will refer to this limit as the Maldacena limit~\cite{Maldacena:1998uz}, see also Refs.~\cite{Maldacena:1997re,Horowitz:1998pq}.
Introducing the coordinate $\rho=r^{\hat{d}}$, the full geometry gets rewritten as
\be
	\begin{gathered}
		ds^2 = -H\left(\rho\right)Z\left(\rho\right)dt^2 + \frac{r^2}{\hat{d}^2\rho^2}\frac{d\rho^2}{H\left(\rho\right)Z\left(\rho\right)} + r^2d\Omega_{\hat{d}+1}^2 \,, \\
		Z\left(\rho\right) = \frac{f_{+}}{f_{-}} \,,\quad H\left(\rho\right) = f_{-}^2 \,.
	\end{gathered}
\ee
To parameterize the near-inner-horizon region, we use the variable
\be
	z = \frac{r_{-}}{\hat{d}}\frac{\rho_{-}}{\rho-\rho_{-}}
\ee
and look at the $z\rightarrow\infty$ behavior. In the Maldacena limit, the function $H\left(\rho\right)$ behaves as
\be
	H\left(\rho\right) = \frac{r_{-}^2}{\hat{d}^2z^2}\left[1+\mathcal{O}\left(z^{-1}\right)\right] \,.
\ee
The resulting leading-order-in-$z$ near-inner-horizon (NIH) geometry,
\be\label{eq:RNNHgeometry_x}
	 ds_{\text{NIH}}^2 = \frac{r_{-}^2}{\hat{d}^2z^2}\left[-Zdt^2 + \frac{dz^2}{Z}\right] + r_{-}^2d\Omega_{\hat{d}+1}^2 \,, Z = 1 - \frac{2z}{\beta_{-}} \,,
\ee
with $\beta_{-}$ the surface gravity of the $D$-dimensional Reissner-Nordstr\"{o}m black hole at the inner horizon, given in Eq.~\eqref{eq:betaRN}, has the characteristic $\sim z^{-2}$ behavior of an $\text{AdS}_2$-like spacetime in Poincar\'{e} coordinates. In fact, fixed points on the sphere charted by the $\left(t,z\right)$ coordinates,
\be
	ds_{\text{NIH}}^2\bigg|_{\mathbb{S}^{\hat{d}+1}} = \frac{b^2}{z^2}\left[-Z dt^2 + \frac{dz^2}{Z}\right] \,,\quad Z = 1 - \frac{z}{z_{\text{h}}} \,,
\ee
precisely lie on a $2$-dimensional Schwarzschild-AdS ($\text{SAdS}_2$) spacetime in Poincar\'{e} coordinates, whose AdS radius $b$ and event horizon location $z_{\text{h}}$ are given by
\be
	z_{\text{h}} = \frac{\beta_{-}}{2} \,,\quad b = \frac{r_{-}}{\hat{d}} \,.
\ee
Interestingly, as noted in Refs.~\cite{Cadoni:1993rn,Cadoni:1994uf}, the $\text{SAdS}_2$ geometry is isomorphic to the Poincar\'{e} patch of pure $\text{AdS}_2$. Indeed, introducing coordinates $\left(\tilde{\tau},\tilde{r}\right)$ according to
\be\label{eq:SAdStoAdS}
	\tilde{\tau}\,\tilde{r} = b^2\left(\frac{2z_{\text{h}}}{z}-1\right) \,,\quad \tilde{\tau}^2 - \frac{b^4}{\tilde{r}^2} = L^2e^{t/z_{\text{h}}} \,,
\ee
with $L$ some arbitrary length scale, then
\be
	ds_{\text{NIH}}^2\bigg|_{\mathbb{S}^{\hat{d}+1}} = -\frac{\tilde{r}^2}{b^2}d\tilde{\tau}^2 + b^2\frac{d\tilde{r}^2}{\tilde{r}^2} \,.
\ee
The set of three Killing vectors $\left\{\tilde{L}_{-1},\tilde{L}_0,\tilde{L}_{+1}\right\}$ generating this pure $\text{AdS}_2$ factor are then straightforward to write down
\be\label{eq:SL2RgeneratorsNIH}
	\tilde{L}_0 = \tilde{\tau}\partial_{\tilde{\tau}} - \tilde{r}\partial_{\tilde{r}} \,,\quad \tilde{L}_{+1} = \partial_{\tilde{\tau}} \,,\quad \tilde{L}_{-1} = \left(\frac{b^4}{\tilde{r}^2}+\tilde{\tau}^2\right)\partial_{\tilde{\tau}} - 2\tilde{\tau}\tilde{r}\partial_{\tilde{r}} \,.
\ee
They satisfy an $\SL$ algebra
\be
	\left[\tilde{L}_{m},\tilde{L}_{n}\right]=\left(m-n\right)\tilde{L}_{m+n} \,,\quad m,n\in\left\{-1,0,+1\right\} \,,
\ee
and their Casimir is given by
\be
	\tilde{\mathcal{C}}_2^{\SL_{\text{NIH}}} = \tilde{L}_0^2 - \frac{1}{2}\left(\tilde{L}_{+1}\tilde{L}_{-1}+\tilde{L}_{-1}\tilde{L}_{+1}\right) = \partial_{\tilde{r}}\,\tilde{r}^2\,\partial_{\tilde{r}} - \frac{b^4}{\tilde{r}^2}\partial_{\tilde{\tau}}^2 \,.
\ee
The full isometry group of the NIH geometry is then $\SL\times SO\,(\hat{d}+2)$, with $\SL$ generated by $\tilde{L}_{m}$, and the commuting $SO\,(\hat{d}+2)$ amounting for the spherical symmetry of the geometry. In the original $\left(t,\rho\right)$ coordinates, the full NIH geometry becomes
\be\label{eq:NIHgeometryRN}
	ds_{\text{NIH}}^2 = \frac{r_{-}^2}{\hat{d}^2}\left[-\Delta\left(\frac{\hat{d}\,dt}{r_{-}\rho_{-}}\right)^2 + \frac{d\rho^2}{\Delta} + \hat{d}^2\,d\Omega_{\hat{d}+1}^2\right] \,,\quad \Delta = \left(\rho-\rho_{+}\right)\left(\rho-\rho_{-}\right) \,,
\ee
the $\text{AdS}_2$ Killing vectors read\footnote{These expressions arise from Eq.~\eqref{eq:SL2RgeneratorsNIH} after some $\SL$ automorphic redefinitions of the generators for future convenience, namely, $\tilde{L}_0 \rightarrow -\tilde{L}_0$ and $\tilde{L}_{\pm1} \rightarrow \tilde{L}_{\mp1}$. We have also performed automorphic global rescalings of the form $\tilde{L}_{m}\rightarrow \lambda^{m}\tilde{L}_{m}$ to remove the dependency on the arbitrary length scale $L$ or any signs that enter on determining various patches that cover the manifold.}
\be\label{eq:LoveGeneratorsRm}
	\tilde{L}_0 = -\beta_{-}\,\partial_{t} \,,\quad \tilde{L}_{\pm1} = e^{\pm t/\beta_{-}}\left[\mp\sqrt{\Delta}\,\partial_{\rho} + \partial_{\rho}\left(\sqrt{\Delta}\right)\,\beta_{-}\,\partial_{t}\right]
\ee
and the corresponding $\SL$ Casimir is rewritten as
\be\label{eq:LoveCasimirRm}
	\tilde{\mathcal{C}}_2^{\SL_{\text{NIH}}} = \partial_{\rho}\,\Delta\,\partial_{\rho} - \frac{\left(\rho_{+}-\rho_{-}\right)^2}{4\Delta}\beta_{-}^2\partial_{t}^2 \,.
\ee
Importantly, the $\SL$ generators $\tilde{L}_{m}$ are regular at both the future and the past inner horizons, as can be seen by employing advanced or retarded null coordinates respectively; they are, however, singular at both the future and past outer horizons.

Note that the $\text{AdS}_2$ structure we have obtained from the near-inner-horizon limits above is analogous to the $\text{AdS}_2$ structure that one obtains in the near-extremal limit, see e.g. Refs.~\cite{Bredberg2009,Hadar:2020kry,Porfyriadis:2021psx,deCesare:2024csp,Banerjee:2024zix}. However, we have never assumed that the Hawking temperature is small. Unlike the usual scaling limits, we always keep it fixed. In addition, the fact that the obtained near-inner-horizon geometry was not reached via a scaling limit also makes the resulting geometry not a solution of the vacuum field equations; rather, the near-inner-horizon geometry corresponds to a solution of the Einstein field equations supported by particular scalar and vector field profiles that screen the far-horizon region, see e.g. Refs.~\cite{Cvetic:2011dn,Cvetic:2011hp,Cvetic:2012tr} for more details.

Nevertheless, in order to make contact with near-horizon physics, we should also consider the near-extremal limit, identified in the framework of the near-inner-horizon geometry by having the inner and outer horizons approach each other at fixed  Hawking temperature. This corresponds to sending $r_{-}\rightarrow r_{+}$, while keeping the inverse surface gravity at the event horizon finite, $\beta_{+} = \text{finite}$. The resulting near-extremal NIH geometry is then a near--near-horizon-extremal (NNHE) geometry. It is described by
\be\label{eq:NNHEgeometryRN}
	ds_{\text{NNHE}}^2 = \frac{r_{+}^2}{\hat{d}^2}\left[-\Delta\left(\frac{\hat{d}\,dt}{r_{+}\rho_{+}}\right)^2 + \frac{d\rho^2}{\Delta} + \hat{d}^2\,d\Omega_{\hat{d}+1}^2\right] \,,
\ee
while the associated $\text{AdS}_2$ Killing vectors become
\be\label{eq:LoveGeneratorsRp}
	L_0 = -\beta_{+}\,\partial_{t} \,,\quad L_{\pm1} = e^{\pm t/\beta_{+}}\left[\mp\sqrt{\Delta}\,\partial_{\rho} + \partial_{\rho}\left(\sqrt{\Delta}\right)\,\beta_{+}\,\partial_{t}\right]
\ee
and the corresponding $\SL$ Casimir is given by
\be\label{eq:LoveCasimirRp}
	\mathcal{C}_2^{\SL_{\text{NNHE}}} = \partial_{\rho}\,\Delta\,\partial_{\rho} - \frac{\left(\rho_{+}-\rho_{-}\right)^2}{4\Delta}\beta_{+}^2\partial_{t}^2 \,.
\ee
As we discuss in detail shortly, these Killing vectors are identical to those of the 
Love symmetry of black holes perturbations
in the near-zone region. 
The upshot is that the new $\SL$ generators $L_{m}$ are now regular at both the future and the past event horizon, albeit singular at the inner horizon, allowing to study geometric constraints on the near-horizon modes.

\subsection{Geometric constraints on perturbations}
\label{sec:nearNHEGeometricConstraints}
Taking the NNHE geometry in Eq.~\eqref{eq:NNHEgeometryRN} as the background geometry, we can study its geometrical consequences on perturbations from external fields. For instance, consider its response to a massless scalar field $\psi$, satisfying the Klein-Gordon equation at the linear level,
\be\label{eq:KGnearNHE}
	\Box_{\text{NNHE}}\psi = \frac{\hat{d}^2}{r_{+}^2}\left[\partial_{\rho}\,\Delta\,\partial_{\rho} - \frac{\left(\rho_{+}-\rho_{-}\right)^2}{4\Delta}\beta^2\partial_{t}^2 + \frac{1}{\hat{d}^2}\Delta_{\mathbb{S}^{\hat{d}+1}}\right]\psi = 0 \,,
\ee
where we assume that sources of the scalar field reside in the far-horizon region. With respect to the full response problem of the $D$-dimensional Reissner-Nordstr\"{o}m black hole, the above equation is valid only in the NNHE region. However, the action of this NNHE wave operator has the peculiar property of becoming exact when acting on static perturbations of the non-extremal black hole.

The power of the enhanced $\SL\times SO\,(\hat{d}+2)$ isometry group of the NNHE geometry comes from the fact that the radial operator is precisely the $\SL$ Casimir in Eq.~\eqref{eq:LoveCasimirRp}, namely,
\be
	\Box_{\text{NNHE}}\psi = \frac{\hat{d}^2}{r_{+}^2}\left[\mathcal{C}_2^{\SL_{\text{NNHE}}} + \frac{1}{\hat{d}^2}\Delta_{\mathbb{S}^{\hat{d}+1}}\right]\psi \,,
\ee
deeming the problem completely integrable. To demonstrate this explicitly, we first utilize the spherical symmetry to solve the angular problem by decomposing into scalar spherical harmonic modes on $\mathbb{S}^{\hat{d}+1}$~\cite{Hui:2020xxx,Chodos:1983zi,Higuchi:1986wu},
\be
	\psi\left(t,\rho,\theta\right) = \sum_{\ell,\mathbf{m}}\psi_{\ell\mathbf{m}}\left(t,\rho\right)Y_{\ell\mathbf{m}}\left(\theta\right) \,.
\ee
The remaining radial problem reduces to
\be
	\mathcal{C}_2^{\SL_{\text{NNHE}}}\psi_{\ell\mathbf{m}} = \hat{\ell}\,(\hat{\ell}+1)\,\psi_{\ell\mathbf{m}} \,,
\ee
with the generalized orbital number defined as
\be
	\hat{\ell} \equiv \frac{\ell}{\hat{d}} \,.
\ee
Before solving these using the enhanced $\SL$ symmetry generated by the Killing vectors $L_{m}$ in Eq.~\eqref{eq:LoveGeneratorsRp}, let us first analyze the asymptotic boundary conditions. At large $\rho$, the scalar field behaves as
\be\label{eq:nearNHERCs}
	\psi_{\ell\mathbf{m}} \sim \mathcal{E}_{\ell\mathbf{m}}\left[\rho^{\hat{\ell}} + \frac{a_{\ell}}{\rho^{\hat{\ell}+1}}\right] \quad\text{as $\rho\rightarrow\infty$} \,.
\ee
These are just the boundary conditions one imposes in asymptotically $\text{AdS}_2$ spacetimes, as is the case of the NNHE geometry, for operators with scaling dimensions $\Delta_{+} = -\hat{\ell}$ and $\Delta_{-} = \hat{\ell}+1$. In this picture, the growing (non-normalizable), $\propto \rho^{-\Delta_{+}}=r^{\ell}$, mode corresponds to a source, while the decaying (normalizable), $\propto \rho^{-\Delta_{-}} = r^{-\ell-\hat{d}}$, mode corresponds to the response. The integration constants $a_{\ell}$ are then to be interpreted as ``response coefficients''. As we will see shortly,  these ``response coefficients'' will turn out to be  black hole Love numbers.

The above $\SL$ symmetry now allows us to integrate the radial equations of motion. The crucial observation is that monochromatic modes, $\psi_{\omega\ell\mathbf{m}}$ of frequency $\omega$ furnish representations of this $\SL$ factor, since
they are eigenstates of both $L_0$ and the Casimir, 
\be
	L_0\psi_{\omega\ell\mathbf{m}} = i\beta_{+}\omega\,\psi_{\omega\ell\mathbf{m}} \quad\text{and}\quad \mathcal{C}_2^{\SL_{\text{NNHE}}}\psi_{\omega\ell\mathbf{m}} = \hat{\ell}\,(\hat{\ell}+1)\,\psi_{\omega\ell\mathbf{m}} \,.
\ee
We are interested in the representations that are spanned by solutions of the NNHE wave equation regular at the black hole event horizon, i.e. they obey the ingoing boundary conditions. It turns out that the resulting differential equation for the radial wavefunction is of Fuchsian-type and is, hence, solvable in terms of Euler's hypergeometric functions. The solution that is ingoing at the future event horizon is
\be
	\psi_{\omega\ell\mathbf{m}} = e^{-i\omega t}\bar{R}_{\ell\mathbf{m}}\left(\omega\right)\left(\frac{\rho-\rho_{+}}{\rho-\rho_{-}}\right)^{-i\beta_{+}\omega/2}{}_2F_1\left(\hat{\ell}+1,-\hat{\ell};1-i\beta_{+}\omega;-\frac{\rho-\rho_{+}}{\rho_{+}-\rho_{-}}\right) \,,
\ee
where $\bar{R}_{\ell\mathbf{m}}\left(\omega\right)$
are integration constants proportional to the transmission amplitudes. Expanding around large $\rho$ by analytically continuing the hypergeometric function at large distances, we find the frequency space response coefficients,
\be
	a_{\ell}\left(\omega\right) =  \frac{\Gamma(-2\hat{\ell}-1)\Gamma(\hat{\ell}+1)\Gamma(\hat{\ell}+1-i\beta_{+}\omega)}{\Gamma(2\hat{\ell}+1)\Gamma(-\hat{\ell})\Gamma(-\hat{\ell}-i\beta_{+}\omega)}\left(\rho_{+}-\rho_{-}\right)^{2\hat{\ell}+1} \,.
\ee
The QNMs frequencies $\omega_{n\ell}$, with overtone $n\in\mathbb{N}$, can then be extracted from the roots of these response coefficients
\be\label{eq:QNMnearNHE}
	\omega_{n\ell} = -i2\pi T_{H}(n-\hat{\ell}) \,,\quad n\in\mathbb{N} \,,
\ee
where we used that $\beta_{+}^{-1} \equiv \kappa_{+} = 2\pi T_{H}$ to restore the explicit dependence on the Hawking temperature. Interestingly, these modes span a highest-weight representation of the NNHE $\SL$~\cite{Chen:2010ik}. Indeed, the primary vector $\upsilon_{-\hat{\ell},0}$ of weight $h=-\hat{\ell}$,
\be
	L_{+1}\upsilon_{-\hat{\ell},0} = 0 \,,\quad L_0\upsilon_{-\hat{\ell},0}=-\hat{\ell}\,\upsilon_{-\hat{\ell},0} \quad\Rightarrow\quad \upsilon_{-\hat{\ell},0} = \left(-e^{+t/\beta_{+}}\sqrt{\Delta}\right)^{\hat{\ell}} \,,
\ee
is a solution of the NNHE wave equation with frequency $\omega_{n=0,\ell} = i2\pi T_{H}\hat{\ell}$ that is regular at the future event horizon; this is the fundamental QNM. From the regularity of the $\SL$ generators $L_{m}$ at the horizon, we then see that this highest-weight representation is spanned by descendants,
\be
	\upsilon_{-\hat{\ell},n} = \left(L_{-1}\right)^{n}\upsilon_{-\hat{\ell},0} \,,\quad n\in\mathbb{N} \,,
\ee
that are also regular at the future event horizon, with frequencies precisely given by Eq.~\eqref{eq:QNMnearNHE}; these are the remaining overtones. In the special case where $\hat{\ell}$ is an integer, one then notices that the $\hat{\ell}$'th descendant has exactly zero frequency and is regular at the event horizon. This state is therefore a static solution of the NNHE equations of motion which, as already remarked, is an exact solution of the full Klein-Gordon equation away from the NNHE regime. We therefore obtain the phenomenologically relevant geometric constraint that the static scalar response coefficients are exactly zero for integer $\hat{\ell}$,
\be
	a_{\ell}\left(\omega=0\right)\bigg|_{\hat{\ell}\in\mathbb{N}} = 0 \,.
\ee
In $D=4$ spacetime dimensions, $\hat{\ell} = \ell\in\mathbb{N}$ and this condition is always satisfied for static scalar perturbations of the Reissner-Nordstr\"{o}m black hole. As we will shortly see, this selection rule is precisely the statement that the static scalar black hole 
Love numbers vanish.

\subsection{Extremal limit - Enhanced isometries in near-horizon throat}
The enhanced $\SL$ symmetry encountered in the NNHE region of the non-extremal Reissner-Nordstr\"{o}m black hole is reminiscent of the enhanced $\SL$ isometry group of the near-horizon throat of extremal black holes~\cite{Bardeen:1999px,Kunduri:2007vf}. Here, we briefly review this construction for completeness.

The extremal Reissner-Nordstr\"{o}m black hole is one whose electric charge acquires the critical value
\be
	Q^2 = \frac{2\hat{d}}{\hat{d}+1}GM \,,
\ee
for which the inner and outer horizons coincide at
\be
	r_{+} = r_{-} = \frac{r_{s}}{2^{1/\hat{d}}} \,.
\ee
The full extremal geometry is then given by Eq.\eqref{eq:gfullRN} with $f_{+}=f_{-}$ and the horizon is a double root of of the discriminant function $f_{+}f_{-}$. The near-horizon throat is reached by performing the change of coordinates
\be\label{eq:NHECoordsRN}
	\tilde{y} = \frac{1}{\hat{d}}\frac{r_{+}}{\lambda}\frac{r^{\hat{d}}-r_{+}^{\hat{d}}}{r_{+}^{\hat{d}}}\,,\quad \tilde{t} = \lambda t
\ee
and taking the $\lambda\rightarrow0$ scaling limit~\cite{Bardeen:1999px}. The resulting near-horizon extremal (NHE) geometry is then
\be
	ds_{\text{NHE}}^2 = -\frac{\tilde{y}^2}{b^2}d\tilde{t}^2 + b^2\frac{d\tilde{y}^2}{\tilde{y}^2} + r_{+}^2d\Omega_{\hat{d}+1}^2 \,, 
\ee
This is exactly the Poincar\'{e} patch of an $\text{AdS}_2\times\mathbb{S}^{d-2}$ manifold, with the $\text{AdS}_2$ radius given by
\be\label{eq:RvarRp}
	b = \frac{r_{+}}{\hat{d}} \,.
\ee
The $\mathbb{R}_{t}\times SO\,(\hat{d}+1)$ isometry group of the initial asymptotically flat extremal geometry now gets enhanced to the NHE isometry group $\SL\times SO\,(\hat{d}+1)$, with the $\SL$ Killing vectors being
\be
	\xi_0 = \tilde{t}\partial_{\tilde{t}} - \tilde{y}\partial_{\tilde{y}} \,,\quad \xi_{+1} = \partial_{\tilde{t}} \,,\quad \xi_{-1} = \left(\frac{b^4}{\tilde{y}^2}+\tilde{t}^2\right)\partial_{\tilde{t}} - 2\tilde{t}\tilde{y}\partial_{\tilde{y}} \,.
\ee
In fact, we see that the NNHE geometry found in the previous section is diffeomorphic to the above NHE geometry; the explicit change of coordinates that matches them is given in Eq.~\eqref{eq:SAdStoAdS}, after identifying $\tilde{\tau} = \tilde{t}$ and $\tilde{r} = \tilde{y}$. In the original $\left(t,\rho\right)$ coordinates, with $\rho=r^{\hat{d}}$, these NHE $\SL$ Killing vectors read
\be\label{eq:NHESL2R_RN}
	\begin{gathered}
		\xi_0 = t\partial_{t} - \left(\rho-\rho_{+}\right)\partial_{\rho} \,,\quad \xi_{+1} = \lambda^{-1}\partial_{t} \,, \\
		\xi_{-1} = \lambda\left[\left(\left(\frac{r_{+}}{\hat{d}}\frac{\rho_{+}}{\rho-\rho_{+}}\right)^2 + t^2\right)\partial_{t} - 2t\left(\rho-\rho_{+}\right)\partial_{\rho}\right] \,.
	\end{gathered}
\ee
Although the scaling parameter $\lambda$ enters explicitly in the above expressions, its dependence is purely automorphic from the $\SL$ perspective, since it enters in the form $\xi_{m}\left(\lambda\right) = \lambda^{m}\xi_{m}\left(\lambda=1\right)$.

It is instructive to study the highest-weight representation of this NHE $\SL$ algebra. The primary vector $v_{-\hat{\ell},0}$ of conformal weight $h=-\hat{\ell}$, defined by
\be
	\xi_{+1}v_{-\hat{\ell},0} = 0 \,,\quad \xi_0v_{-\hat{\ell},0}=-\hat{\ell}\,v_{-\hat{\ell},0} \quad\Rightarrow\quad v_{-\hat{\ell},0} = \left(\rho-\rho_{+}\right)^{\hat{\ell}} \,,
\ee
turns out to be exactly static and regular at the horizon. Its purely polynomial form also sets the static response coefficients of the extremal black hole to zero, but now without any restrictions on the range of the orbital number $\ell$. However, in contrast to the vanishings outputted from the NNHE $\SL$ symmetry, these vanishings of the static scalar response coefficients for extremal black holes are a property of the representation structure itself, rather than of the geometry, and are, hence, not to be interpreted as geometrical constraints.

\subsection{Beyond near-extremal limit - Love symmetry for black holes}
We will now show that the NNHE geometry unexpectedly arises from a completely different framework; that of the response problem of an asymptotically flat non-extremal black hole. A covariant way of studying the response problem of a compact body is within the framework of the worldline EFT, where the compact body is treated as an effective point-particle dressed with multipole moments that couple non-minimally to the worldline. The response of the compact body is then parameterized by a tower of Wilson coefficients in the effective action that are determined via  matching conditions. The details of this approach will be presented in Section~\ref{sec:WVEFTandLNs}, but we remark here how this matching is performed at the level of the microscopic computation, i.e. within black hole perturbation theory. Focusing to the linear scalar response problem of the full non-extremal Reissner-Nordstr\"{o}m black hole, the task is to solve the massless Klein-Gordon equation in the background of Eq.~\eqref{eq:gfullRN} in the presence of perturbing scalar sources $j_{\psi}$,
\be
	\Box_{\text{full}}\psi = \frac{\hat{d}^2}{r^2}\left[\partial_{\rho}\,\Delta\,\partial_{\rho} - \frac{r^2\rho^2}{\hat{d}^2\Delta}\,\partial_{t}^2 + \frac{1}{\hat{d}^2}\Delta_{\mathbb{S}^{\hat{d}+1}}\right]\psi = j_{\psi} \,,
\ee
in an expansion relevant for matching onto the worldline EFT. The relevant expansion is the near-zone expansion, defined by~\cite{Starobinsky:1973aij,Starobinskil:1974nkd,Maldacena:1997ih,Castro:2010fd,Chia:2020yla,Charalambous:2021kcz}
\be
	\omega\left(r-r_{+}\right) \ll 1 \,,\quad \omega r_{+} \ll 1 \,,
\ee
where $\omega$ is the frequency of the perturbation; this is the statement that the worldline EFT is accurate in the near-zone region, where the wavelength of the perturbation is large compared to separation between the perturbing source and the perturbed body and also large compared to the size of the perturbed body. Within the near-zone region, $j_{\psi}=0$ and the presence of the source is encoded in the asymptotic boundary conditions. Furthermore, the kinetic operator is organized in a near-zone expansion, for instance
\be\ba
	\Box_{\text{full}}\psi &= \frac{\hat{d}^2}{r^2}\left[ \mathbb{O}_{\text{NZ}}+ \frac{1}{\hat{d}^2}\Delta_{\mathbb{S}^{\hat{d}+1}} + \epsilon V_1\right]\psi \,,
\ea\ee
with
\be\label{eq:NZRN}
	\mathbb{O}_{\text{NZ}} = \partial_{\rho}\,\Delta\,\partial_{\rho} - \frac{\left(\rho_{+}-\rho_{-}\right)^2}{4\Delta}\,\beta_{+}^2\,\partial_{t}^2 \,,\quad V_1 = -\frac{r^2\rho^2-r_{+}^2\rho_{+}^2}{\hat{d}^2\Delta}\,\partial_{t}^2 \,,
\ee
and $\epsilon$ a formal expansion parameter which is equal to unity for the full kinetic operator and equal to zero for the leading-order near-zone approximation. The boundary conditions one imposes to match the response coefficients onto the worldline EFT definition are then, after expanding into monochromatic scalar spherical harmonic modes $\psi_{\omega\ell\mathbf{m}}$,
\be
	\psi_{\omega\ell\mathbf{m}}\left(t,\rho\right) \sim e^{-i\omega t}\bar{\mathcal{E}}_{\ell\mathbf{m}}\left(\omega\right)\rho^{\hat{\ell}}\left[1+k_{\ell}\left(\omega\right)\left(\frac{\rho_{s}}{\rho}\right)^{2\hat{\ell}+1}\right] \quad\text{as $\rho\rightarrow\infty$} \,,
\ee
with $k_{\ell}\left(\omega\right)$ defining\footnote{When performing the matching, one should also adapt a scheme that distinguishes between relativistic corrections to the source from actual response effects. This is typically done via analytically continuing the orbital number $\ell$ or the spacetime dimensionality $D$ to range in $\mathbb{R}$, such that $\hat{\ell}$ is analytically continued to range in $\mathbb{R}$~\cite{Kol:2011vg,LeTiec:2020spy,LeTiec:2020bos,Charalambous:2021mea,Chia:2020yla,Creci:2021rkz}, see also Ref.~\cite{Ivanov:2022hlo} for an alternative.} the dimensionless frequency space response coefficients. The conservative, even under $\omega\rightarrow-\omega$, part of these response coefficients defines the so-called Love numbers, while the part that is odd under time reversal captures dissipative effects. These two types of effects turn out to enter in the spherical harmonic modes $k_{\ell}\left(\omega\right)$ in a simple real/imaginary splitting of the response coefficients for the current spherically symmetric background~\cite{LeTiec:2020bos,Charalambous:2023jgq},
\be
	k_{\ell}^{\text{Love}}\left(\omega\right) = \text{Re}\left\{k_{\ell}\left(\omega\right)\right\} \,,\quad k_{\ell}^{\text{diss}}\left(\omega\right) = \text{Im}\left\{k_{\ell}\left(\omega\right)\right\} \,.
\ee
A direct calculation using the leading-order near-zone equations of motion then shows that
\be
	k_{\ell}\left(\omega\right) = \frac{\Gamma(-2\hat{\ell}-1)\Gamma(\hat{\ell}+1)\Gamma(\hat{\ell}+1-i\beta_{+}\omega)}{\Gamma(2\hat{\ell}+1)\Gamma(-\hat{\ell})\Gamma(-\hat{\ell}-i\beta_{+}\omega)}\left(\frac{\rho_{+}-\rho_{-}}{\rho_{s}}\right)^{2\hat{\ell}+1} + \mathcal{O}\left(\beta_{+}^2\omega^2\right) \,,
\ee
where it is emphasized that these are accurate only up to $\mathcal{O}\left(\beta_{+}^2\omega^2\right)$ corrections, since we are working at leading-order in the near-zone expansion. A real/imaginary separation then reveals that the scalar Love numbers and scalar dissipation numbers of the black hole are given by
\be
	\begin{gathered}
		k_{\ell}\left(\omega\right) = k_{\ell}^{\text{Love}}\left(\omega\right) + i k_{\ell}^{\text{diss}}\left(\omega\right) \,, \\
		\ba
			k_{\ell}^{\text{Love}}\left(\omega\right) &= A_{\ell}\left(\omega\right)\tan\pi\hat{\ell}\cosh\pi\beta_{+}\omega + \mathcal{O}\left(\beta_{+}^2\omega^2\right) \,, \\
			k_{\ell}^{\text{diss}}\left(\omega\right) &= A_{\ell}\left(\omega\right)\sinh\pi\beta_{+}\omega + \mathcal{O}\left(\beta_{+}^3\omega^3\right) \,,
		\ea
	\end{gathered}
\ee
where $A_{\ell}\left(\omega\right)$ is the real, positive and no-where vanishing constant
\be
	A_{\ell}\left(\omega\right) = \frac{\Gamma^2(\hat{\ell}+1)\left|\Gamma(\hat{\ell}+1-i\beta_{+}\omega)\right|^2}{2\pi\Gamma(2\hat{\ell}+1)\Gamma(2\hat{\ell}+2)}\left(\frac{\rho_{+}-\rho_{-}}{\rho_{s}}\right)^{2\hat{\ell}+1} \,.
\ee
More importantly, we observe that static scalar Love numbers are exactly zero for the resonant conditions $\hat{\ell}\in\mathbb{N}$, while their diverging behavior for half-integer $\hat{\ell}$ is interpreted as a classical RG flow from the worldline EFT point of view~\cite{Kol:2011vg}. This is to be compared with expectations from Wilsonian naturalness arguments. More explicitly, power counting arguments show that the scalar Love numbers are expected to be non-zero and running for $2\hat{\ell}\in\mathbb{N}$~\cite{Charalambous:2022rre}. However, the vanishing of the static Love numbers for $\hat{\ell}\in\mathbb{N}$ raises naturalness concerns and calls upon an enhanced symmetry resolution~\cite{tHooft:1979rat,Porto:2016zng,Charalambous:2021mea}.

The resolution is the Love symmetry proposal~\cite{Charalambous:2021kcz,Charalambous:2022rre}, which is based on the important observation that the leading order near-zone equations of motion are equipped with an enhanced globally defined $\SL$ symmetry. In fact, as we will demonstrate now, the Love $\SL$ symmetry is identical to the $\SL$ isometry subgroup of the NNHE geometry.

At the level of the equations of motion, the leading-order near-zone wave operator exactly coincides with the NNHE wave operator in Eq.~\eqref{eq:KGnearNHE}. More importantly, the leading-order near-zone radial operator in Eq.~\eqref{eq:NZRN} is precisely the NNHE $\SL$ Casimir
\be
	\mathbb{O}_{\text{NZ}} = \mathcal{C}_2^{\SL_{\text{NNHE}}} \,.
\ee
In the spirit of this identification, we now realize that the corresponding Love symmetry generators that have been reported in association with the non-extremal black hole response problem~\cite{Charalambous:2021kcz,Charalambous:2022rre,Charalambous:2024tdj},
\be
	L_0^{\text{Love}} = -\beta_{+}\,\partial_{t} \,,\quad L_{\pm1}^{\text{Love}} = e^{\pm t/\beta_{+}}\left[\mp\sqrt{\Delta}\,\partial_{\rho} + \left(\partial_{\rho}\sqrt{\Delta}\right)\,\beta_{+}\,\partial_{t}\right] \,,
\ee
are exactly the $\SL$ Killing vectors of the NNHE geometry in Eq.~\eqref{eq:LoveGeneratorsRp},
\be
\label{eq:NNHE_love}
	\sla_{\text{Love}} = \sla_{\text{NNHE}} \,.
\ee
Note that the above statement 
relies on a non-trivial fact that the 
near-zone Klein-Gordon operator
matches the near-horizon Klein-Gordon operator. Equivalently, 
the near-horizon scaling limit does not 
alter the static Klein-Gordon 
equation. This enables one to use the 
near-horizon symmetries to describe 
the static solution even at asymptotic 
infinity, relevant for the Love number
matching. 

The associated NNHE geometry in Eq.~\eqref{eq:NNHEgeometryRN} is then what has been referred in Refs.~\cite{Charalambous:2022rre,Charalambous:2023jgq} as the associated ``subtracted'' geometries~\cite{Cvetic:2011dn,Cvetic:2011hp,Cvetic:2012tr}; effective non-extremal black hole geometries that preserve the black hole thermodynamics.

Supplementing on this near-zone/NNHE dictionary, the response coefficients $a_{\ell}$ that were encountered in the response problem of the asymptotically AdS black hole in Section~\ref{sec:nearNHEGeometricConstraints}, see Eq.~\eqref{eq:nearNHERCs}, now acquire an interpretation in terms of observables for asymptotically flat non-extremal black holes: they are the leading-order-in-near-zone response coefficients $k_{\ell}$ (up to an overall $\rho_{s}^{2\hat{\ell}+1}$ factor). Furthermore, we see how the vanishing of the static scalar Love numbers of the Reissner-Nordstr\"{o}m black hole whenever $\hat{\ell}\in\mathbb{N}$, originally proposed as a consequence of a Love symmetry highest-weight representation~\cite{Charalambous:2021kcz,Charalambous:2022rre}, has already been outputted as a geometric constraint of the NNHE enhanced $\SL$ isometry subgroup. In Section~\ref{sec:nearNHEGeometricConstraints}, this vanishing was attributed to the corresponding static perturbation being a zero-frequency QNM and a descendant in a highest-weight representation. Even though the algebraic arguments of the highest-weight representation being spanned by QNMs does not acquire a $1$-to-$1$ correspondence with the QNMs of the asymptotically flat Reissner-Nordstr\"{o}m black hole due to the different boundary conditions at large distances, the fact that the static scalar perturbation regular at the horizon belongs to a highest-weight representation remains true, with the highest-weight property predicting a polynomial form of the scalar field profile without any decaying (response) modes. A rigorous interpretation of the other elements in the highest-weight representation in terms of observables in asymptotically flat spacetimes is still to be determined, but there have been proposals~\cite{Ivanov:2022qqt,Charalambous:2022rre,Charalambous:2023jgq} that relate these other states with total-transmission modes~\cite{Hod:2013fea,Cook:2016fge,Cook:2016ngj}.

It is also interesting to remark here that the near-zone Love symmetry generators can recover the NHE $\SL$ Killing vectors in the extremal limit. This is achieved via a Winger-like contraction as already noted in Refs.~\cite{Charalambous:2022rre,Charalambous:2023jgq}, namely,
\be\ba
	\xi_{+1} &= \lambda^{-1}\lim_{T_{H}\rightarrow0}\left(-2\pi T_{H}L_0\right) = \lambda^{-1}\partial_{t} \,, \\
	\xi_{0} &= \lim_{T_{H}\rightarrow0}\frac{L_{+1}-L_{-1}}{2} = t\partial_{t} - \left(\rho-\rho_{+}\right)\partial_{\rho} \,, \\
	\xi_{-1} &= \lambda\lim_{T_{H}\rightarrow0}\frac{L_{+1}+L_{-1}+2L_0}{2\pi T_{H}} = \lambda\left[\left(\left(\frac{r_{+}}{\hat{d}}\frac{\rho_{+}}{\rho-\rho_{+}}\right)^2 + t^2\right)\partial_{t} - 2t\left(\rho-\rho_{+}\right)\partial_{\rho}\right] \,,
\ea\ee
which indeed match those of Eq.~\eqref{eq:NHESL2R_RN} upon identifying $\lambda$ with the scaling parameter in Eq.~\eqref{eq:NHECoordsRN}.


\section{Love numbers and Love symmetries of non-dilatonic black $p$-branes}
\label{sec:WVEFTandLNs}

We will now formulate the response problem of black $p$-branes and compute, in particular, the scalar Love numbers of non-dilatonic $p$-branes. Similar to the case of the Reissner-Nordstr\"{o}m black hole, we will see that a tower of magic zeroes emerges in the response problem of the black $p$-brane, under static and homogeneous massless scalar perturbations. We will interpret these vanishings as selection rules dictated by near-zone symmetries. For the case of black strings ($p=1$) these Love symmetry vector fields generate the expected $2$-dimensional conformal group, $SO\left(2,2;\mathbb{R}\right)$. For $p\ge2$, however, one does not find a near-zone $\left(p+1\right)$-dimensional conformal group, $SO\left(p+1,2;\mathbb{R}\right)$. Nevertheless, one can still interpret these vanishings from a particular ``static'' $\SL$ symmetry that is not a symmetry of the near-zone equations of motion but has the special property of outputting exact results for static and homogeneous perturbations of the black $p$-brane. This behavior is ultimately attributed to the fact that static and homogeneous perturbations of black $p$-branes in $D$ spacetime dimensions obey the same equations of motion as static perturbations of the $\left(D-p\right)$-dimensional Reissner-Nordstr\"{o}m black hole~\cite{Rodriguez:2023xjd}. We will finish this section with an analogous study of the scalar response problem of an extremal black $p$-brane. For the extremal configuration, we will find an even larger class of vanishing Love numbers, that also contain the dynamical case of perturbations that follow light-like dispersion relations. These vanishings will be addressed in the next section by the enhanced isometries that emerge in the near-horizon throat of the extremal configuration.

\subsection{Background geometry of non-dilatonic $p$-branes}

The Reissner-Nordstr\"{o}m black hole is a black object that is charged under the $1$-form Maxwell field $A_{\mu}$. In this spirit, a black $p$-brane is an extended black object, whose worldvolume has dimension
\be
	d = p+1 \,,
\ee
that is charged under a $d$-form gauge field $A_{\mu_1\mu_2\dots\mu_{d}}$. We will focus here to black $p$-branes with no (or constant) dilaton, such that the resulting geometries do not exhibit a curvature singularity at the inner horizon and, hence, admit a smooth extremal limit~\cite{Gibbons:1987ps,Duff:1991pe,Duff:1993ye,Gibbons:1994vm}. Nevertheless, the ``usual'' self-dual black $3$-brane of $D=10$ supergravity, as well as the black $2$-brane and its dual black $5$-brane in $D=11$ spacetime dimensions, belong to the this class of non-dilatonic black $p$-branes.

The action that describes a non-dilatonic black $p$-brane charged under a $d$-form gauge field consists of a bulk piece and a worldvolume piece~\cite{Duff:1993ye,Horowitz:1991cd,Galtsov:2005thm},
\be\label{eq:pBraneAction}
	S = I_{D}\left(d\right) + S_{d} \,.
\ee 
The bulk action for the gravitational and $d$-form force fields reads
\be\label{eq:pBraneActionBulk}
	I_{D}\left(d\right) = \frac{1}{2\kappa^2}\int d^{D}x\,\sqrt{-g}\left[ R - \frac{1}{2\left(d+2\right)!}F_{d+1}^2 \right] \,,
\ee 
while, for the worldvolume action, we take the leading order EFT action describing the static infinitely thin brane with tension $T_{d}$,
\be\label{eq:pBraneActionWV}
	\begin{aligned}
		S_{d} = T_{d}\int d^{d}\sigma&\bigg\{\sqrt{-\gamma}\left[-\frac{1}{2}\gamma^{ab}g_{\mu\nu}\partial_{a}X^{\mu}\partial_{b}X^{\nu}  + \frac{d-2}{2}\right] \\
		&-\frac{1}{d!}\epsilon^{a_1\dots a_{d}}\partial_{a_1}X^{\mu_1}\dots\partial_{a_{d}}X^{\mu_{d}}A_{\mu_1\dots\mu_{d}} \bigg\} \,.
	\end{aligned}
\ee
The static point particle corresponds to the limit $p=0$ ($d=1$) of the above action. In what follows we will also make extended use of the dual worldvolume dimensionality,
\be
	\hat{d} = D-2-d = D-3-p \,.
\ee

The solution of the field equations following from the above action for a charged static black $p$-brane with spherical, $SO\,(\hat{d}+2)$, symmetry along the transverse directions and Euclidean, $ISO\left(d-1\right)$, symmetry along the longitudinal directions is given by~\cite{Duff:1993ye,Horowitz:1991cd,Galtsov:2005thm}
\be\label{eq:pBraneGeoNonDilatonic}
	\begin{gathered}
		\star F_{\hat{d}+1} = Q_{d}\epsilon_{\hat{d}+1} \,, \\
		ds^2 = f_{-}^{2/d}\left[-\frac{f_{+}}{f_{-}}dt^2 + d\mathbf{x}^2\right] + \frac{dr^2}{f_{+}f_{-}} + r^2d\Omega_{\hat{d}+1}^2 \,,
	\end{gathered}
\ee
where
\be
	f_{\pm}\left(r\right) = 1 - \left(\frac{r_{\pm}}{r}\right)^{\hat{d}} \,.
\ee
The inner and outer horizons are related to the charge, $Q_{d}$, and the ADM mass per unit volume, $M_{d}$, of the $p$-brane according to~\cite{Lu:1993vt}
\be
	M_{d} = \frac{\text{Vol}(\mathbb{S}^{\hat{d}})}{2\kappa^2}\left[(\hat{d}+1)\,r_{+}^{\hat{d}}-r_{-}^{\hat{d}}\right] \,,\quad Q_{d} = \hat{d}\left(r_{+}r_{-}\right)^{\hat{d}/2} \,.
\ee
The inverse surface gravity of the black $p$-brane at the outer horizon is given by
\be\label{eq:betappbrane}
	\beta_{+} \equiv \kappa_{+}^{-1} = \frac{2r_{+}}{\hat{d}}\left(\frac{r_{+}^{\hat{d}}}{r_{+}^{\hat{d}}-r_{-}^{\hat{d}}}\right)^{\frac{1}{d}} \,,
\ee
while it will also be useful to introduce the quantity $\beta_{-}$, obtained from a $r_{+} \leftrightarrow r_{-}$ exchange in $\beta_{+}$,
\be\label{eq:betampbrane}
	\beta_{-} \equiv \frac{2r_{-}}{\hat{d}}\left(\frac{r_{-}^{\hat{d}}}{r_{+}^{\hat{d}}-r_{-}^{\hat{d}}}\right)^{\frac{1}{d}} \,.
\ee
As opposed to the black hole case ($p=0$), $\beta_{-}$ above does not have any interpretation as an inverse surface gravity at the inner horizon.

The isometry group for the non-extremal black $p$-brane is $\mathbb{R}_{t}\times ISO\left(d-1\right)\times SO\,(\hat{d}+2)$, which gets enhanced to $ISO\left(d-1,1\right)\times SO\,(\hat{d}+2)$ in the extremal case, where
\be
	Q_{d} = \frac{2\kappa^2}{\text{Vol}(\mathbb{S}^{\hat{d}})}M_{d} \Rightarrow r_{-} = r_{+} = \left(\frac{2\kappa^2}{\text{Vol}(\mathbb{S}^{\hat{d}})}\frac{M_{d}}{\hat{d}}\right)^{1/\hat{d}} \,.
\ee

\subsection{Membrane (world-volume) effective field theory}

The world-volume EFT has been studied in details in Refs.~\cite{Emparan:2009cs,Emparan:2009at}. Note that, in contrast to Refs.~\cite{Emparan:2009cs,Emparan:2009at}, in this work we focus on infinitely long branes that are described by a single scale, set by a membrane mass, and hence our discussion will be a simple generalization of the worldline EFT~\cite{Goldberger:2004jt,Porto:2005ac} for black holes~\cite{Goldberger:2005cd,Kol:2007rx,Kol:2011vg,Goldberger:2020fot,Hui:2020xxx,Charalambous:2021mea,Ivanov:2022hlo}, see also Refs.~\cite{Porto:2016pyg,Levi:2018nxp} for comprehensive reviews on the world-line formulation of gravitational dynamics for a binary system. Here we give some basic overview that allows us to define Love numbers as Wilson coefficients of the classical world-volume EFT.

Consider a general case of $p$-branes in $D$ dimensions. The brane is at rest w.r.t an external observer. The brane is described by a world-volume action with coordinates $\sigma^{a}$. The long-wavelength degrees of freedom that capture dynamics of the brane and its back-reaction on the bulk geometry are collective coordinates $X^{I}\left(\sigma^{a}\right)$, which capture the center of mass displacements (in directions transverse to the world-volume), and other external long wavelength fields, e.g. the dilaton $\phi$, the $d$-form potential $A_{\mu_1...\mu_{d}}$, and the metric fluctuations $g^{\rm (long)}_{\m\n}$. The EFT description requires $\d_{a}X^{I}$ to vary slowly (the velocity can be large, at the same time, i.e. $X^{I}$ can vary fast). In order to preserve gauge invariance, it is convenient to consider all spacetime coordinates $X^{\mu}\left(\sigma^{a}\right)$ as long-wavelengths fields. Note that $X^{a}$ are gauge fields in this picture. This embedding allows us to define an induced metric,
\be 
	\g_{ab}=g^{\rm (long)}_{\m\n}\d_{a} X^{\m}\d_{b}X^{\n} \,.
\ee
In the EFT, we build a full solution of the Einstein equations by expanding over the Minkowski space in the direction orthogonal to the brane. This means that $g^{\rm (long)}_{\m\n}=\eta_{\mu\nu} + h_{\mu\nu}$, with $h_{\mu\nu}$ capturing the long-distance metric perturbations.
The dynamics of an infinitely long thin $p$-brane of tension $T_{d}$ are described by the Dirac action
\be
	S_{\rm Dirac}= -T_{d} \int d^{d}\sigma~\sqrt{-\det\left(g_{\m\n}\d_a X^\m \d_b X^\n\right)} \,.
\ee
The leading (in the infinitely-thin brane approximation) coupling to the $d$-form field is given by
\be
	S_{\text{d-form}}= \frac{1}{d!}\int d^{d}\sigma \varepsilon^{a_1\dots a_{d}} \partial_{a_1}X^{\mu_1}\dots\partial_{a_{d}}X^{\mu_{d}} A_{\mu_1\dots\mu_{d}}\,,
\ee
where $\varepsilon^{a_1\dots a_{d}}$ is the Levi-Civita antisymmetric symbol. To obtain the leading order long-distance description of a brane at rest, we fix static gauge
\be
	X^{a}\left(\sigma^{b}\right)= \sigma^{a} \,,
\ee
place the brane at the origin of the coordinate system, and work in its rest frame, so $\d_{a}X^{I} = 0$. At zeroth order the induced brane metric is flat $\g_{ab}=\eta_{ab}$. The brane action then takes the form
\be 
	\begin{split}
		S_{\rm Dirac}&= -T_{d} \int d^{d}\sigma \,.
	\end{split}
\ee
Now we can consider adding a small metric perturbation, $g_{\m\n}=\eta_{\m\n}+h_{\m\n}$, $|h_{\m\n}|\ll 1$. In Newtonian gauge we consider only two scalar components, corresponding to perturbations of $h_{00}$ and the trace of the spatial orthogonal metric, $h_{ii}$. Adding the Einstein-Hilbert part to the Dirac action, we can then find the usual static gravitational potential in the bulk,
\be
	h_{00}\left(r\right) \propto \frac{T_{d}}{r^{D-p-3}}\,.
\ee
The calculation that we have done is exactly the same as the calculation of the Newtonian potential in a $D-p$ dimensional spacetime. In full analogy, one can also easily compute the flux of the $d$-form.

Let us now take into account corrections to the infinitely thin brane approximation, which capture effects of the finite thickness and consequently, response to external fields. In what follows, we consider only the scalar sector of the theory, i.e. polarization effects induced by a test scalar field.
At leading order, these are captured by irrelevant operators on the world-volume that are quadratic in the multipole moments of the scalar field,
\be\label{eq:LoveFT}
	S_{\rm FT}^{\text{Static Love}} = \sum_{\ell=0}^\infty \frac{\lambda_\ell R^{2\ell+\hat{d}}}{2\ell!} \int d^{d}\sigma~\mathcal{E}_{L}\mathcal{E}^{L} \,,
\ee
where the characteristic length scale $R$ is related to the $p$-brane tension according to
\be
    T_{d} = R^{\hat{d}}
\ee
in our current geometrized units with $G=c=1$. The introduction of the scale $R$ also emphasizes that we are employing a power counting prescription using off-shell fields, i.e. in powers of $\left(\frac{R}{r}\right)$ with $r$ a transversal radial distance, as opposed to the power counting rules of on-shell fields, i.e. in powers of, for instance, $\omega r$ with $\omega$ the frequency of the field. The multipole moments $\mathcal{E}_{L}$ of the scalar field carry a multi-index $L\equiv I_1\dots I_{\ell}$ of the transversal spatial indices, and are defined in terms of $\ell$ derivatives projected onto the transversal spatial slices via the codimension-$d$ projector
\be
    P^{\mu\nu} = g^{\mu\nu} - \partial_{a}X^{\mu}\partial_{b}X^{\nu}\gamma^{ab}
\ee
according to
\be
    \mathcal{E}_{L} = P^{\nu_1}_{\langle\mu_1}\dots P^{\nu_{\ell}}_{\mu_{\ell}\rangle}\nabla_{\nu_1}\dots\nabla_{\nu_{\ell}}\Phi \,,
\ee
with $\langle\dots\rangle$ denoting the STF part with respect to the spatial directions.

The Wilson coefficients $\lambda_{\ell}$ in Eq.~\eqref{eq:LoveFT} encode the conservative response of the $p$-brane under static and homogeneous scalar field perturbations and define the static and homogeneous scalar Love numbers. Dynamical and non-homogeneous Love numbers can also be defined in a similar manner, by introducing first the covariant longitudinal derivatives
\be
    D_{a} = \partial_{a}X^{\mu}\,\nabla_{\mu} \,.
\ee
These are the world-volume equivalents of the $u^\mu \nabla_\m$ derivatives in the world-line EFT. Since the internal Poincar\'{e} symmetry is broken, one needs to distinguish between non-static effects, captured from $D_0$, and non-homogeneities along the spatial directions of the world-volume, captured from actions of $D_{i}$. Then, for instance, the first few dynamical and non-homogeneous scalar Love numbers are defined from the following irrelevant finite-size operators
\be\ba\label{eq:LoveFTDyn0}
	S_{\rm FT}^{\text{Dynamical Love}} &\supset \sum_{\ell=0}^\infty\frac{\lambda_{\ell}^{\left(2,0\right)} R^{2\ell+\hat{d}+2}}{2\ell!} \int d^{d}\sigma~D_0\mathcal{E}_{L}D_0\mathcal{E}^{L} \\
    &+\quad \sum_{\ell=0}^\infty\frac{\lambda_{\ell}^{\left(0,2\right)} R^{2\ell+\hat{d}+2}}{2\ell!} \int d^{d}\sigma~\delta^{ij}D_{i}\mathcal{E}_{L}D_{j}\mathcal{E}^{L} \,, \\
    &+\quad \sum_{\ell=0}^\infty\frac{\lambda_{\ell}^{\left(2,2\right)} R^{2\ell+\hat{d}+4}}{2\ell!} \int d^{d}\sigma~\delta^{ij}D_0D_{i}\mathcal{E}_{L}D_0D_{j}\mathcal{E}^{L} \,.
\ea\ee
Following the same pattern, one construct the rest of the infinite tower of Wilson coefficients $\lambda_{\ell}^{\left(2n,2m\right)}$, where we have made use of the $\mathbb{R}_{t}\times ISO\left(p\right)$ isometries of the background and also used the fact that the above local operators must be time-reversal, $\sigma^0\rightarrow-\sigma^0$, and world-volume parity, $\sigma^{i}\rightarrow-\sigma^{i}$, symmetric. Equivalently, and more explicitly, we may work in Fourier space along the world-volume directions and repackage all the dynamical and homogeneous Love numbers into a single phase space ``Wilson function'' $\lambda_{\ell}\left(k\right)$, with $k^{a}=\left(\omega,\mathbf{k}\right)$ the $d$-momentum along the world-volume, according to
\be
    \lambda_{\ell}\left(k\right) = \sum_{n,m=0}^{\infty}\left(-1\right)^{n+m}\omega^{2n}\mathbf{k}^{2m}\lambda_{\ell}^{\left(2n,2m\right)}R^{2\left(n+m\right)}
\ee
such that,
\be\label{eq:LoveFTDynFull}
    S_{\rm FT}^{\text{dynamical Love}} = \sum_{\ell=0}^\infty R^{2\ell+\hat{d}} \int \frac{d^{d}k}{\left(2\pi\right)^{d}} \frac{\lambda_{\ell}\left(k\right)}{2\ell!}\mathcal{E}_{L}\left(-k\right)\mathcal{E}^{L}\left(k\right) \,.
\ee
Our EFT description is by construction valid for modes with
\be
	\omega R \ll 1 \,,\quad \left|\mathbf{k}\right| R\ll 1 \,.
\ee

\subsection{Scalar Love numbers of non-extremal black $p$-branes}
\label{sec:MatchingNZ}

The Wilson coefficients in Eq.~\eqref{eq:LoveFT} capture the response of a brane to external scalar fields. They are scalar equivalents of the static tidal Love numbers. To see this, let us consider a bulk dilaton sourced by a static current $J$,
\be\label{eq:bulkscal}
	S_{\rm bulk}=\int d^{D}x~\left[-\frac{1}{2}(\d \phi)^2 + J\phi \right] \,.
\ee

We will now demonstrate how to compute these scalar Love numbers defined within the world-volume EFT, via a matching condition. We will employ the off-shell ``Newtonian matching''~\cite{Kol:2011vg,Charalambous:2021mea,Charalambous:2023jgq}, which has the advantage of entering at the level of the equations of motion and, hence, allows to probe consequences of approximate enhanced symmetries, such as the near-zone Love symmetries. Alternatively, one can also perform on-shell matchings onto scattering observables, see, e.g., Refs~\cite{Ivanov:2022hlo,Ivanov:2022qqt,Saketh:2023bul}.

From the world-volume EFT point of view, the Newtonian matching consists of computing the EFT $1$-point function in a background of a $2^{\ell}$-polar Newtonian source,
\be
	\phi\left(k,\mathbf{x}\right) = \bar{\phi}\left(k,\mathbf{x}\right) + \delta	\phi\left(k,\mathbf{x}\right) \,,\quad \bar{\phi}\left(k,\mathbf{x}\right) = \bar{\phi}_{L}\left(k\right)x^{L} \,,
\ee
where we are working in Fourier space along the world-volume longitudinal directions and we have adopted the multi-index notation $L \equiv I_1\dots I_{\ell}$, $x^{L}\equiv x^{I_1}\dots x^{I_{\ell}}$, with $\bar{\phi}_{L}\left(k\right)$ being a constant STF tensor that captures the strength of the scalar source. We remark that in this subsection, and \textit{only here}, $x^{I}$ will denote the spatial directions transverse to the world-volume and are not to be confused with the spatial directions $x^{i}$ longitudinal to the world-volume we are using throughout the rest of the paper. Using spherical coordinates in the directions orthogonal to the brane, this can be rewritten as $\bar\phi\propto r^{\ell}Y_{\ell \mathbf{m}}\left(\theta\right)$, where $Y_{\ell \mathbf{m}}(\theta)$ are the $\left(D-2-p\right)$-dimensional scalar spherical harmonics. The symbol $\mathbf{m}$ here is the multi-index denoting the whole set of higher-dimensional orbital and magnetic quantum numbers\footnote{More explicitly, the ``azimuthal'' multi-index entering the scalar spherical harmonics $Y_{\ell\mathbf{m}}$ on $\mathbb{S}^{n}$ is $\mathbf{m} = \ell_2,\dots,\ell_{n-2},m_{\text{mag}}$, with $\ell\ge\ell_1\ge\ell_2\ge\dots\ge\ell_{n-3}\ge\left|m_{\text{mag}}\right|$. In other words, the scalar spherical harmonics on $\mathbb{S}^{n}$ are labelled by $n-1$ orbital numbers (one for each of the $n-1$ polar angles $\theta_1,\dots\theta_{n-1}\in\left[0,\pi\right]$) and one magnetic quantum number amounting for the single azimuthal angle $\phi\equiv\theta_{n}\in\left[0,2\pi\right)$.}. Adding now action~\eqref{eq:LoveFTDynFull} to \eqref{eq:bulkscal} and computing corrections at linear order in the Newtonian background source, we have the following total expression diagrammatically,
\be
	\vev{\delta\phi\left(k,\mathbf{x}\right)} =
	\vcenter{\hbox{\begin{tikzpicture}
			\begin{feynman}
				\vertex (a0);
				\vertex[right=0.6cm of a0] (gblobaux);
				\vertex[left=0.00cm of gblobaux, blob] (gblob){};
				\vertex[below=1cm of a0] (p1);
				\vertex[above=1cm of a0] (p2);
				\vertex[right=1cm of p1] (a1);
				\vertex[right=1cm of p2] (a2);
				\vertex[right=0.69cm of p2] (a22){$\times$};
				\diagram*{
					(p1) -- [double,double distance=0.5ex] (p2),
					(a1) -- (gblob) -- (a2),
				};
			\end{feynman}
	\end{tikzpicture}}} =
	\underbrace{\vcenter{\hbox{\begin{tikzpicture}
				\begin{feynman}
					\vertex[dot] (a0);
					\vertex[below=1cm of a0] (p1);
					\vertex[above=1cm of a0] (p2);
					\vertex[right=0.4cm of a0, blob] (gblob){};							
					\vertex[right=1.5cm of p1] (b1);
					\vertex[right=1.5cm of p2] (b2);
					\vertex[right=1.19cm of p2] (b22){$\times$};
					\vertex[above=0.7cm of a0] (g1);
					\vertex[above=0.4cm of a0] (g2);
					\vertex[right=0.05cm of a0] (gdtos){$\vdots$};
					\vertex[below=0.7cm of a0] (gN);
					\diagram*{
						(p1) -- [double,double distance=0.5ex] (p2),
						(g1) -- [photon] (gblob),
						(g2) -- [photon] (gblob),
						(gN) -- [photon] (gblob),
						(b1) -- (gblob) -- (b2),
					};
				\end{feynman}
	\end{tikzpicture}}}}_{\text{``source''}} + 
	\underbrace{\vcenter{\hbox{\begin{tikzpicture}
				\begin{feynman}
					\vertex[dot] (a0);
					\vertex[left=0.00cm of a0] (lambda){$\lambda_{\ell}\left(k\right)$};
					\vertex[below=1.6cm of a0] (p1);
					\vertex[above=0.4cm of a0] (p2);
					\vertex[right=1.5cm of p1] (b1);
					\vertex[right=1.5cm of p2] (b2);
					\vertex[right=1.19cm of p2] (b22){$\times$};
					\vertex[below=0.4cm of a0] (g1);
					\vertex[below=0.3cm of g1] (g2);
					\vertex[below=0.14cm of g2] (gdotsaux);
					\vertex[right=0.00cm of gdotsaux] (gdtos){$\vdots$};
					\vertex[below=0.5cm of gdotsaux] (gN);
					\vertex[below=0.7cm of a0] (gblobaux);
					\vertex[right=0.4cm of gblobaux, blob] (gblob){};
					\diagram*{
						(p1) -- [double,double distance=0.5ex] (p2),
						(b2) -- (a0) -- (gblob) -- (b1),
						(g1) -- [photon] (gblob),
						(g2) -- [photon] (gblob),
						(gN) -- [photon] (gblob),
					};
				\end{feynman}
	\end{tikzpicture}}}}_{\text{``response''}} \,.
\ee
Our diagrammatic notation above is as follows: The double line represents the world-volume, straight lines represent propagators of the scalar field, a cross (``$\times$'') represents an insertion of the background field $\bar{\phi}$ and wavy lines represent graviton propagators, whose interactions with the world-volume come from the minimal Dirac action and capture relativistic corrections. The second equality serves to demonstrate the ``source''/``response'' split of the scalar field profile, disentangling relativistic corrections to the Newtonian source from actual response effects~\cite{Charalambous:2021mea,Ivanov:2022hlo}. Choosing to work in the body-centered frame, the world-volume EFT $1$-point function acquires the characteristic bi-monomial form in the Newtonian limit,
\be\ba\label{eq:EFTresp}
    \vev{\delta\phi\left(k,\mathbf{x}\right)} &\rightarrow
	\vcenter{\hbox{\begin{tikzpicture}
			\begin{feynman}
				\vertex[dot] (a0);
				\vertex[below=1cm of a0] (p1);
				\vertex[above=1cm of a0] (p2);						
				\vertex[right=1cm of p1] (b1);
				\vertex[right=1cm of p2] (b2);
				\vertex[right=0.69cm of p2] (b22){$\times$};
				\diagram*{
					(p1) -- [double,double distance=0.5ex] (p2),
					(b1) -- (b2),
				};
			\end{feynman}
	\end{tikzpicture}}} + 
	\vcenter{\hbox{\begin{tikzpicture}
			\begin{feynman}
				\vertex[dot] (a0);
				\vertex[left=0.00cm of a0] (lambda){$\lambda_{\ell}\left(k\right)$};
				\vertex[below=1cm of a0] (p1);
				\vertex[above=1cm of a0] (p2);
				\vertex[right=1.2cm of p1] (b1);
				\vertex[right=1.2cm of p2] (b2);
				\vertex[right=0.89cm of p2] (b22){$\times$};
				\diagram*{
					(p1) -- [double,double distance=0.5ex] (p2),
					(b2) -- (a0) -- (b1),
				};
			\end{feynman}
	\end{tikzpicture}}} \\
	&= \left[1 + \frac{2^{\ell-2}\Gamma\left(\ell+\frac{\hat{d}}{2}\right)}{\pi^{(\hat{d}+2)/2}}\lambda_{\ell}\left(k\right)\left(\frac{R}{r}\right)^{2\ell+\hat{d}}\right]\bar{\phi}_{L}\left(k\right)x^{L} \,.
\ea\ee
The correction $\propto \lambda_\ell$ behaves like an induced $\ell$'th multipole moment in ($D-1-p$)-dimensional space, and hence the Wilson coefficient $\lambda_\ell$ can be indeed interpreted as a scalar response coefficient, or scalar Love number (SLN). For $p=0$ \eqref{eq:EFTresp} coincides with the response of a $D$-dimensional Schwarzschild black hole. From the EFT point of view we expect $\lambda_\ell =\mathcal{O}(1)$~\cite{tHooft:1979rat,Porto:2016zng}.

Before moving on, we note that in the EFT there can also be a dilaton curvature coupling. However, it can be forbidden by requiring the shift symmetry $\phi\to \phi+\text{const}$. Even if it is present, the dilaton coupling does not play any role at zeroth order in metric perturbations where the induced metric is flat.

We now move on to the microscopic theory sides, microscopic here referring to the full theory at the purely classical level. More explicitly, for the case of massless scalar perturbations, the microscopic equations of motion consist of the Klein-Gordon equation
\be\label{eq:KGfullp}
	\Box_{\text{full}}\psi = \frac{\hat{d}^2}{r^2}\left[\partial_{\rho}\,\Delta\,\partial_{\rho} + \frac{r^2\rho^{2/d}\Delta_{-}^{\left(d-2\right)/d}}{\hat{d}^2}\left(-\frac{1}{\Delta_{+}}\partial_{t}^2 + \frac{1}{\Delta_{-}}\delta^{ij}\partial_{i}\partial_{j}\right) + \frac{1}{\hat{d}^2}\Delta_{\mathbb{S}^{\hat{d}+1}}\right]\psi = j_{\psi} \,,
\ee
where we have focused to the background geometry of interest, i.e. the geometry of the non-dilatonic $p$-brane, see Eq.~\eqref{eq:pBraneGeoNonDilatonic}, and we have also introduced the transformed radial variable
\be
	\rho \equiv r^{\hat{d}} \,,
\ee
such that
\be
	\Delta_{\pm} \equiv \rho f_{\pm} = \rho-\rho_{\pm} \quad\text{and}\quad \Delta\equiv\Delta_{+}\Delta_{-} = \left(\rho-\rho_{+}\right)\left(\rho-\rho_{-}\right) \,.
\ee

To match onto the world-volume EFT, one should also employ an expansion of the equations of motion within the regime where the EFT is accurate. This is the near-zone expansion and amounts to working in the region where both the frequency and the momentum along the longitudinal directions are small compared to the inverse distance from the event horizon,
\be
	\omega\left(r-r_{+}\right) \ll 1 \,,\quad \left|\mathbf{k}\right|\left(r-r_{+}\right) \ll 1 \,,
\ee
and also small compared to the inverse (transversal) size of the black string, i.e. $\omega r_{+}\ll1$ and $\left|\mathbf{k}\right|r_{+}\ll1$; these last conditions allow to glue together the near-zone and far-zone regions and probe observables defined in the world-volume EFT to scattering observables defined in the asymptotic region. Furthermore, sources of the scalar field perturbations are taken to reside in the far-zone region, with their effect within the near-zone region being captured by proper asymptotic boundary conditions, namely, after expanding into monochromatic scalar spherical harmonic modes $\psi_{\omega k\ell\mathbf{m}}$, the asymptotic boundary conditions to be imposed are
\be
	\psi_{\omega\mathbf{k}\ell\mathbf{m}} \sim e^{-i\omega t}e^{i\mathbf{k}\cdot\mathbf{x}}\bar{\mathcal{E}}_{\ell\mathbf{m}}\left(\omega,\mathbf{k}\right)r^{\ell}\left[1+k_{\ell}\left(\omega,\mathbf{k}\right)\left(\frac{R}{r}\right)^{2\ell+\hat{d}}\right] \quad\text{as $r\rightarrow\infty$} \,.
\ee
These boundary conditions at large distances are exactly what want to match onto the world-volume EFT $1$-point function. More explicitly, the scalar Love numbers, i.e., the conservative part of the response coefficients $k_{\ell}\left(\omega,\mathbf{k}\right)$ entering at the level of the microscopic calculation, are related to the Wilson coefficients $\lambda_{\ell}\left(\omega,\mathbf{k}\right)$ according to
\be
	k_{\ell}^{\text{Love}}\left(\omega,\mathbf{k}\right) = \frac{2^{\ell-2}\Gamma\left(\ell+\frac{\hat{d}}{2}\right)}{\pi^{(\hat{d}+2)/2}}\lambda_{\ell}\left(\omega,\mathbf{k}\right) \,.
\ee
As a reminder, the Love (dissipative) part of the response coefficient is identified with the part that is even (odd) under time-reversal, $\omega\rightarrow-\omega$, and world-volume parity, $\mathbf{k}\rightarrow-\mathbf{k}$, transformations. For the current static and transversely spherically symmetric background configuration, this is equivalent to the real/imaginary split of the response coefficient~\cite{LeTiec:2020bos,Charalambous:2023jgq},
\be
	k_{\ell}^{\text{Love}}\left(\omega,\mathbf{k}\right) = \text{Re}\left\{k_{\ell}\left(\omega,\mathbf{k}\right)\right\} \,,\quad k_{\ell}^{\text{diss}}\left(\omega,\mathbf{k}\right) = \text{Im}\left\{k_{\ell}\left(\omega,\mathbf{k}\right)\right\} \,.
\ee

To perform an explicit calculation, let us choose the following near-zone expansion of the equations of motion
\be\label{eq:NZsplitp}
	\Box_{\text{full}}\psi = \frac{\hat{d}^2}{r^2}\left[\mathbb{O}_{\text{NZ}} + \frac{1}{\hat{d}^2}\Delta_{\mathbb{S}^{\hat{d}+1}} + \epsilon V_1\right]\psi
\ee
with $\epsilon$ a formal expansion parameter and
\be\ba\label{eq:RadialNZp}
	\mathbb{O}_{\text{NZ}} &= \partial_{\rho}\,\Delta\,\partial_{\rho} + \frac{\rho_{+}-\rho_{-}}{4}\beta_{+}^2\left(-\frac{1}{\Delta_{+}}\,\partial_{t}^2 + \frac{1}{\Delta_{-}}\,\delta^{ij}\partial_{i}\partial_{j}\right) \,, \\
	V_1 &= \frac{r^2\rho^{2/d}\Delta_{-}^{\left(d-2\right)/d}-r_{+}^2\rho_{+}^{2/d}\left(\rho_{+}-\rho_{-}\right)^{\left(d-2\right)/d}}{\hat{d}^2}\left(-\frac{1}{\Delta_{+}}\,\partial_{t}^2+\frac{1}{\Delta_{-}}\,\delta^{ij}\partial_{i}\partial_{j}\right) \,.
\ea\ee
Then, the leading-order near-zone equations of motion are of Fuchsian type and can be solved in terms of Euler's hypergeometric functions. More specifically, looking at monochromatic modes and expanding into spherical harmonics on $\mathbb{S}^{\hat{d}+1}$,
\be
	\psi\left(t,\mathbf{x},\rho,\theta\right) = \sum_{\ell,\mathbf{m}}\psi_{\omega\mathbf{k}\ell\mathbf{m}}\left(t,\mathbf{x},\rho\right)Y_{\ell\mathbf{m}}\left(\theta\right) \,,\quad \psi_{\ell\mathbf{m}}\left(t,x,\rho\right) = e^{i\left(\mathbf{k}\cdot\mathbf{x}-\omega t\right)}R_{\omega\mathbf{k}\ell\mathbf{m}}\left(\rho\right) \,,
\ee
the solution that is ingoing at the event horizon is given by
\be\ba
	{}&\psi_{\omega k\ell\mathbf{m}} = e^{i\left(\mathbf{k}\cdot\mathbf{x}-\omega t\right)}\bar{R}_{\ell\mathbf{m}}\left(\omega,\mathbf{k}\right)\left(\frac{\rho-\rho_{+}}{\rho-\rho_{-}}\right)^{-i\beta_{+}\omega/2} \\
	&\times\left(\frac{\rho-\rho_{-}}{\rho_{+}-\rho_{-}}\right)^{-i\beta_{+}\varpi_{\pm}}{}_2F_1\left(\hat{\ell}+1-i\beta_{+}\varpi_{\pm},-\hat{\ell}-i\beta_{+}\varpi_{\pm};1-i\beta_{+}\omega;-\frac{\rho-\rho_{+}}{\rho_{+}-\rho_{-}}\right) \,.
\ea\ee
In the expression above, the integration constants $\bar{R}_{\ell\mathbf{m}}\left(\omega,k\right)$ are proportional to the transmission amplitudes, and we have also introduced the rescaled orbital number
\be
	\hat{\ell} \equiv \frac{\ell}{\hat{d}} = \frac{\ell}{D-p-3} \,,
\ee
as well as the phase space parameters
\be\label{eq:OmegaPMp}
	\varpi_{\pm} \equiv \frac{\omega\mp \left|\mathbf{k}\right|}{2} \,.
\ee
The fact that we can use either $\varpi_{+}$ or $\varpi_{-}$ in the scalar field solution above follows from Euler's transformation, ${}_2F_1\left(a,b;c;z\right)=\left(1-z\right)^{c-a-b}{}_2F_1\left(c-a,c-b;c;z\right)$.

Looking at the large-distance behavior of this solution that is ingoing at the event horizon, one then deduces the following expression for the scalar response coefficients
\be\label{eq:RCspBrane}
\begin{split}
	k_{\ell}\left(\omega,\mathbf{k}\right) = &\frac{\Gamma(-2\hat{\ell}-1)\Gamma(\hat{\ell}+1-i\beta_{+}\varpi_{+})\Gamma(\hat{\ell}+1-i\beta_{+}\varpi_{-})}{\Gamma(2\hat{\ell}+1)\Gamma(-\hat{\ell}-i\beta_{+}\varpi_{+})\Gamma(-\hat{\ell}-i\beta_{+}\varpi_{-})}\left(\frac{\rho_{+}-\rho_{-}}{R^{\hat{d}}}\right)^{2\hat{\ell}+1} \\
 &+ \mathcal{O}\left(\beta_{+}^2\omega^2,\beta_{+}^2\mathbf{k}^2\right) \,,
 \end{split}
\ee
with the $\mathcal{O}\left(\beta_{+}^2\omega^2,\beta_{+}^2\mathbf{k}^2\right)$ corrections arising from subleading orders in the near-zone expansion. As a result the scalar Love numbers and scalar dissipation numbers of the non-dilatonic black $p$-brane are given by
\be
	\begin{gathered}
		k_{\ell}^{\text{diss}}\left(\omega,\mathbf{k}\right) = A_{\ell}\left(\omega,\mathbf{k}\right)\sinh\pi\beta_{+}\omega + \mathcal{O}\left(\beta_{+}^3\omega^3,\beta_{+}^3\left|\mathbf{k}\right|^3\right) \,, \\
		\ba
			k_{\ell}^{\text{Love}}\left(\omega,\mathbf{k}\right) &= A_{\ell}\left(\omega,\mathbf{k}\right) \times \bigg\{\tan\pi\hat{\ell}\cosh\pi\beta_{+}\varpi_{+}\cosh\pi\beta_{+}\varpi_{-} \\
			&-\cot\pi\hat{\ell}\sinh\pi\beta_{+}\varpi_{+}\sinh\pi\beta_{+}\varpi_{-}\bigg\} + \mathcal{O}\left(\beta_{+}^2\omega^2,\beta_{+}^2\left|\mathbf{k}\right|^2\right) \,,
		\ea
	\end{gathered}
\ee
where $A_{\ell}\left(\omega,k\right)$ are the real constants
\be
	A_{\ell}\left(\omega,\mathbf{k}\right) = \frac{\left|\Gamma(\hat{\ell}+1-i\beta_{+}\varpi_{+})\right|^2\left|\Gamma(\hat{\ell}+1-i\beta_{+}\varpi_{-})\right|^2}{2\pi\Gamma(2\hat{\ell}+1)\Gamma(2\hat{\ell}+2)}\left(\frac{\rho_{+}-\rho_{-}}{R^{\hat{d}}}\right)^{2\hat{\ell}+1} \,.
\ee

Despite the approximate nature of the calculation, it gives us exact results for static and homogeneous perturbations,
\be
	k_{\ell}^{\text{Love}}\left(\omega=0,\mathbf{k}=0\right) = \frac{\Gamma^4(\hat{\ell}+1)}{2\pi\Gamma(2\hat{\ell}+1)\Gamma(2\hat{\ell}+2)}\left(\frac{\rho_{+}-\rho_{-}}{R^{\hat{d}}}\right)^{2\hat{\ell}+1}\tan\pi\hat{\ell} \,.
\ee
The behavior of the static and homogeneous scalar Love numbers of the black $p$-brane are then strongly dependent on the orbital number of the perturbation and follow the exact same pattern as the static scalar Love numbers of a Reissner-Nordstr\"{o}m black hole with shifted spacetime dimensionality $D\rightarrow D-p$: they are non-zero and non-running for generic $\hat{\ell}$, they are logarithmically running for half-integer $\hat{\ell}$ and they are vanishing for integer $\hat{\ell}$~\cite{Kol:2011vg,Hui:2020xxx,Charalambous:2024tdj,Hadad:2024lsf}.

\subsection{Love symmetry for black strings}
\label{sec:LoveSymmetryp1}

The only seemingly fine-tuned behavior is the class of $\hat{\ell}\in\mathbb{N}$ for which the static and homogeneous scalar Love numbers of the black $p$-brane vanish. We will now focus on the case of black strings ($p=1$). As with the case of the Reissner-Nordstr\"{o}m black hole, an enhanced conformal Love symmetry emerges in the near-zone~\cite{Charalambous:2021kcz,Charalambous:2022rre,Charalambous:2023jgq,Charalambous:2024tdj}. For the black string, this turns out to be an $SO\left(2,2;\mathbb{R}\right)$ symmetry. The six vector fields $\left\{L_0^{\text{Love}},L_{01}^{\text{Love}},L_{+1,0}^{\text{Love}},L_{-1,0}^{\text{Love}},L_{+1,1}^{\text{Love}},L_{-1,1}^{\text{Love}}\right\}$ generating the Love $SO\left(2,2;\mathbb{R}\right)$ symmetry associated with this near-zone truncation are, using $D \times SO\left(1,1;\mathbb{R}\right)$ as the stabilizer of the group, with $\mathfrak{d} = \left\{\tilde{L}_0\right\}$ and $\mathfrak{so}\left(1,1;\mathbb{R}\right)=\left\{\tilde{L}_{01}\right\}$,
\be\ba\label{eq:LoveS022p1}
	L_0^{\text{Love}} &= -\beta_{+}\,\partial_{t} \,,\quad L_{01}^{\text{Love}} = +\beta_{+}\partial_{x} \,, \\
	L_{\pm1,0}^{\text{Love}} &= e^{\pm t/\beta_{+}}\left[\cosh\frac{x}{\beta_{+}}\left(\mp2\sqrt{\Delta}\,\partial_{\rho} + \sqrt{\frac{\rho-\rho_{-}}{\rho-\rho_{+}}}\,\beta_{+}\partial_{t}\right) \pm \sinh\frac{x}{\beta_{+}}\sqrt{\frac{\rho-\rho_{+}}{\rho-\rho_{-}}}\,\beta_{+}\partial_{x}\right] \,, \\
	L_{\pm1,1}^{\text{Love}} &= e^{\pm t/\beta_{+}}\left[\sinh\frac{x}{\beta_{+}}\left(\mp2\sqrt{\Delta}\,\partial_{\rho} + \sqrt{\frac{\rho-\rho_{-}}{\rho-\rho_{+}}}\,\beta_{+}\partial_{t}\right) \pm \cosh\frac{x}{\beta_{+}}\sqrt{\frac{\rho-\rho_{+}}{\rho-\rho_{-}}}\,\beta_{+}\partial_{x}\right] \,.
\ea\ee
These indeed satisfy the $SO\left(2,2;\mathbb{R}\right)$ algebra
\be\ba
	\left[L^{\text{Love}}_{\pm1,0},L^{\text{Love}}_0\right] &= \pm L^{\text{Love}}_{\pm1,0} \,,\quad \left[L^{\text{Love}}_{\pm1,1},L^{\text{Love}}_0\right] = \pm L^{\text{Love}}_{\pm1,1} \,, \\
	\left[L^{\text{Love}}_{\pm1,0},L^{\text{Love}}_{01}\right] &= -L^{\text{Love}}_{\pm1,1} \,,\quad \left[L^{\text{Love}}_{\pm1,1},L^{\text{Love}}_{01}\right] = -L^{\text{Love}}_{\pm1,0} \,, \\
	\left[L^{\text{Love}}_{\pm1,0},L^{\text{Love}}_{\mp1,0}\right] &= \pm2L^{\text{Love}}_0 \,,\quad \left[L^{\text{Love}}_{\pm1,1},L^{\text{Love}}_{\mp1,1}\right] = \mp2L^{\text{Love}}_0 \,,\quad \left[L^{\text{Love}}_{\pm1,0},L^{\text{Love}}_{\mp1,1}\right] = 2L^{\text{Love}}_{01} \,,
\ea\ee
with all other commutators being vanishing. Equivalently, using the Lie algebra isomorphism $\mathfrak{so}\left(2,2;\mathbb{R}\right) \cong \sla_{\left(+\right)}\oplus\sla_{\left(-\right)}$, with 
\[ \sla_{\left(\pm\right)} = \left\{L^{\left(\pm\right)\text{Love}}_{-1},L^{\left(\pm\right)\text{Love}}_{0},L^{\left(\pm\right)\text{Love}}_{+1}\right\}\,,
\]
that is, using $D_{\left(+\right)}\times D_{\left(-\right)}$ as the stabilizer of the group, with $\mathfrak{d}_{\left(\pm\right)} = \left\{L_0^{\left(\pm\right)\text{Love}}\right\}$, the six vector fields above can be rearranged into two commuting sets of $\SL$ vector fields,
\be\label{eq:LoveSL2R2p1}
	\begin{gathered}
		\ba
			L_0^{\left(\sigma\right)\text{Love}} &= \frac{L^{\text{Love}}_0 -\sigma L^{\text{Love}}_{01}}{2} = -\frac{\beta_{+}}{2}\left(\partial_{t}+\sigma\partial_{x}\right) \,, \\
			L_{\pm1}^{\left(\sigma\right)\text{Love}} &= \frac{L^{\text{Love}}_{\pm1,0} \pm\sigma L^{\text{Love}}_{\pm,1}}{2} = e^{\pm\left(t+\sigma x\right)/\beta_{+}}\left[\mp\sqrt{\Delta}\,\partial_{\rho} + \sqrt{\frac{\rho-\rho_{-}}{\rho-\rho_{+}}}\,\frac{\beta_{+}}{2}\partial_{t} + \sqrt{\frac{\rho-\rho_{+}}{\rho-\rho_{-}}}\,\frac{\beta_{+}}{2}\sigma\partial_{x}\right] \,,
		\ea \\\\
		\left[L_{m}^{\left(\sigma\right)\text{Love}},L_{n}^{\left(\sigma^{\prime}\right)\text{Love}}\right] = \left(m-n\right)L_{m+n}^{\left(\sigma\right)\text{Love}}\delta_{\sigma,\sigma^{\prime}} \,,\quad m,n\in\left\{-1,0,+1\right\} \,,\quad \sigma,\sigma^{\prime} \in\left\{+,-\right\} \,.
	\end{gathered}
\ee

The Casimir operator for the Love symmetry generators is then given by
\be\ba
	\mathcal{C}_2^{SO\left(2,2;\mathbb{R}\right)_{\text{Love}}} &= \frac{1}{4}\bigg[\big(L_0^{\text{Love}}\big)^2 - \frac{1}{2}\left(L^{\text{Love}}_{+1,0}L^{\text{Love}}_{-1,0}+L^{\text{Love}}_{-1,0}L^{\text{Love}}_{+1,0}\right) \\
	&\quad\quad+ \frac{1}{2}\left(L^{\text{Love}}_{+1,1}L^{\text{Love}}_{-1,1}+L^{\text{Love}}_{-1,1}L^{\text{Love}}_{+1,1}\right) + \big(L^{\text{Love}}_{01}\big)^2 \bigg] \\
	&= \frac{1}{2}\left(\mathcal{C}_2^{\SL_{\left(+\right)\text{Love}}} + \mathcal{C}_2^{\SL_{\left(-\right)\text{Love}}}\right) \\
	&= \partial_{\rho}\,\Delta\,\partial_{\rho} + \frac{\rho_{+}-\rho_{-}}{4}\beta_{+}^2\left(-\frac{1}{\Delta_{+}}\,\partial_{t}^2 + \frac{1}{\Delta_{-}}\partial_{x}^2\right) \,,
\ea\ee
and we immediately see that this indeed matches with the near-zone radial operator $\mathbb{O}_{\text{NZ}}$ in Eq.~\eqref{eq:RadialNZp} for the case of black strings,
\be
	\mathbb{O}_{\text{NZ}} = \mathcal{C}_2^{SO\left(2,2;\mathbb{R}\right)_{\text{Love}}} \quad \text{for $p=1$} \,.
\ee
We also note here that the Casimirs of the two commuting $\SL$'s turn out to be the same, $\mathcal{C}_2^{\SL_{\left(+\right)\text{Love}}} = \mathcal{C}_2^{\SL_{\left(-\right)\text{Love}}} = \mathcal{C}_2^{SO\left(2,2;\mathbb{R}\right)_{\text{Love}}}$. An important property of the $SO\left(2,2;\mathbb{R}\right)$ Love generators above is that they are regular at both the future and the past event horizon, as can be seen by employing advanced/retarded null coordinates respectively.

Let us now solve the near-zone equations of motion algebraically, using the enhanced near-zone $SO\left(2,2;\mathbb{R}\right)$ Love symmetry. The key observation is that the monochromatic modes $\psi_{\omega k\ell\mathbf{m}}\left(t,x,\rho\right)$ furnish representations of the Love $SO\left(2,2;\mathbb{R}\right)$ algebra. This is most easily written in the $\sla_{\left(+\right)}\oplus\sla_{\left(-\right)}$ basis of the $SO\left(2,2;\mathbb{R}\right)$ algebra, spanned by the two sets of $3$ vectors $\left\{L_{m}^{\left(+\right)\text{Love}}\right\}\cup\left\{L_{m}^{\left(-\right)\text{Love}}\right\}$ given in Eq.~\eqref{eq:LoveSL2R2p1},
\be
	L_0^{\left(\pm\right)\text{Love}}\psi_{\omega k\ell\mathbf{m}} = i\beta_{+}\varpi_{\pm}\,\psi_{\omega k\ell\mathbf{m}} \quad\text{and}\quad \mathcal{C}_2^{\SL_{\left(\pm\right)\text{Love}}}\psi_{\omega k\ell\mathbf{m}} = \hat{\ell}\,(\hat{\ell}+1)\,\psi_{\omega k\ell\mathbf{m}} \,,
\ee
where we have used the fact that the two $\SL$ Casimirs coincide. From a $\text{CFT}_2$ perspective, the scaling dimension and the $2$-dimensional spin are nothing else than the frequency and wavenumber magnitude of the near-zone scalar field perturbations, $\Delta_{\text{CFT}_2} = h_{\left(+\right)}+h_{\left(-\right)} = i\beta_{+}\omega$ and $J_{\text{CFT}_2} = h_{\left(+\right)}-h_{\left(-\right)} = -i\beta_{+}k$.

We now look at the branch of vectors $\left\{\upsilon_{-\hat{\ell},n}^{\left(+\right)}\right\}$ that furnish a highest-weight representation of $\SL_{\left(+\right)\text{NNHE}}$ with highest-weight $h_{\left(+\right)}=-\hat{\ell}$. These are constructed as descendants,
\be
	\upsilon_{-\hat{\ell},n}^{\left(+\right)} = \left(L_{-1}^{\left(+\right)\text{Love}}\right)^{n}\upsilon_{-\hat{\ell},0}^{\left(+\right)} \,,\quad n\in\mathbb{N} \,,
\ee
of the primary vector $\upsilon_{-\hat{\ell},0}^{\left(+\right)}$, defined according to
\be
	L_{+1}^{\left(+\right)\text{Love}}\upsilon_{-\hat{\ell},0}^{\left(+\right)} = 0 \,,\quad L_0^{\left(+\right)\text{Love}}\upsilon_{-\hat{\ell},0}^{\left(+\right)} = -\hat{\ell}\,\upsilon_{-\hat{\ell},0}^{\left(+\right)} \,,
\ee
as well as the condition that it also belongs to a representation of $\SL_{\left(-\right)\text{NNHE}}$, i.e. $L_0^{\left(-\right)\text{Love}}\upsilon_{-\hat{\ell},0}^{\left(+\right)} = \left(\text{const.}\right)\upsilon_{-\hat{\ell},0}^{\left(+\right)}$, though not necessarily a highest-weight one. One then finds
\be
	\upsilon_{-\hat{\ell},0}^{\left(+\right)} = e^{ik\left(x-t\right)}\left(\frac{\rho-\rho_{+}}{\rho-\rho_{-}}\right)^{-i\beta_{+}k/2}\left[-e^{+2t/\beta_{+}}\left(\rho-\rho_{+}\right)\right]^{\hat{\ell}} \,.
\ee
The $\SL_{\left(-\right)\text{Love}}$ weight for this state is
\be
	L_0^{\left(-\right)\text{Love}} \upsilon_{-\hat{\ell},0}^{\left(+\right)} = (-\hat{\ell}+i\beta_{+}k)\,\upsilon_{-\hat{\ell},0}^{\left(+\right)} \,,
\ee
and it is easy to see that this highest-weight state of $\SL_{\left(+\right)\text{Love}}$ is in general \textit{not} a highest-weight state of $\SL_{\left(-\right)\text{Love}}$, except in the special case $k=0$, i.e. for perturbations that are homogeneous along the extended dimension of the black string.

Employing advanced null coordinates, it is then realized that $\upsilon_{-\hat{\ell},0}^{\left(+\right)}$ is regular at the future event horizon and, furthermore, it solves the leading order near-zone radial problem with frequency
\be
	\omega_{-\hat{\ell},0}^{\left(+\right)} = +k + i4\pi T_{H}\hat{\ell} \,,
\ee
where we have restored the explicit dependence on the Hawking temperature, using the fact that $\beta_{+}^{-1} \equiv \kappa_{+} = 2\pi T_{H}$. Regularity of the Love $SO\left(2,2;\mathbb{R}\right)$ vector fields then implies that all its descendants $\upsilon_{-\hat{\ell},n}^{\left(+\right)}$ are also solutions of the leading order near-zone radial problem that are regular at the future event horizon, but with frequencies
\be
	\omega_{-\hat{\ell},n}^{\left(+\right)} = +k - i4\pi T_{H}(n-\hat{\ell}) \,.
\ee
Following the same prescription, it is straightforward to realize that the set of states $\left\{\upsilon_{-\hat{\ell},n}^{\left(-\right)}\right\}$ that furnish a highest-weight representation of $\SL_{\left(-\right)\text{Love}}$ of highest-weight $h_{\left(-\right)}=-\hat{\ell}$, but that do not in general belong to a highest-weight representation of $\SL_{\left(+\right)\text{Love}}$ are also regular solutions of the near-zone equation of motion, with frequencies following the dispersion relation $\omega_{-\hat{\ell},n}^{\left(-\right)} = -k - i4\pi T_{H}(n-\hat{\ell})$.

As a last comment, the special case $k=0$, of perturbations that are homogeneous along the black string, is degenerate in that these states belong to a highest-weight representation of both $\SL_{\left(+\right)\text{Love}}$ and $\SL_{\left(-\right)\text{Love}}$. This should not be a surprise since it is for this value of $k$ that the two branches of frequencies coincide, $\omega_{n\ell}^{\left(+\right)}\left(k=0\right) = \omega_{n\ell}^{\left(-\right)}\left(k=0\right)$. In fact, these states are algebraically special, being purely imaginary. Furthermore, static and homogeneous perturbations of the non-dilatonic black string belong to this degenerate highest-weight representation of $SO\left(2,2;\mathbb{R}\right)$ if and only if $\hat{\ell}$ is an integer. More specifically, the static and homogeneous massless scalar perturbation of orbital number $\ell$ is identified as the $\hat{\ell}$'th descendant. This observation outputs the selection rule
\be
	k_{\ell}\left(\omega=0,k=0\right)\bigg|_{\hat{\ell}\in\mathbb{N}} = 0 \,,
\ee
which is an \textit{exact} result, since static and homogeneous perturbations survive beyond the near-zone regime, i.e. the action of the near-zone scalar wave operator coincides with the action of the full scalar wave operator when acting on static and homogeneous perturbations of the black string. The above selection rule of vanishing static and homogeneous scalar Love numbers of the non-dilatonic black string is a direct consequence of the highest-weight property, namely, the annihilation condition $\left(L_{+1}^{\left(\sigma\right)\text{Love}}\right)^{\hat{\ell}+1}\psi_{\omega=0,k=0,\ell\mathbf{m}}\left(\rho\right) \propto \partial_{\rho}^{\hat{\ell}+1}\psi_{\omega=0,k=0,\ell\mathbf{m}}\left(\rho\right)$ implies that the corresponding perturbations are purely polynomial in $\rho$, with no decaying response component. More generally, for integer $\hat{\ell}$, the $\hat{\ell}$'th descendant follows a light-like dispersion, with $\left|\omega\right| = \left|k\right|$. However, such near-zone modes do not survive beyond the near-zone regime for $k\ne0$. Nevertheless, as we will see in the next section, near-horizon modes with light-like dispersion relations turn out to do survive in the far-horizon region in the extremal case. In the next section, we will also see that the states spanning the aforementioned representations of the Love $SO\left(2,2;\mathbb{R}\right)$ have another interesting property: they can be realized as QNMs of the NNHE geometry of the black string.

\subsection{Homogeneous $\SL$ symmetry for $\left(p\ge2\right)$-branes}
\label{sec:HomSL2R}

We will close this part of near-zone symmetries addressing the vanishing static and homogeneous scalar Love numbers of non-extremal black $p$-branes by supplementing with the case of $\left(p\ge2\right)$-branes. Unfortunately, there is no near-zone Love symmetry that spans the algebra of the $p\ge2$ conformal group, $SO\left(p+1,2;\mathbb{R}\right)$. We will geometrically prove this statement in the form of a ``no-go theorem'' in the next section. Here, we will instead show that one can construct a ``homogeneous'' $\SL$ which emerges only in the reduced solution space of homogeneous ($\mathbf{k}=\mathbf{0}$) perturbations and has the special property of having an exact action on static and homogeneous perturbations.

Let us then consider the full equations of motion for the special case of homogeneous scalar perturbations $\psi_{\text{hom}}$ of the black $p$-brane, see Eq.~\eqref{eq:KGfullp} with $\partial_{i}\psi_{\text{hom}} = 0$,
\be
	\Box_{\text{full}}\psi_{\text{hom}} = \frac{\hat{d}^2}{r^2}\left[\partial_{\rho}\,\Delta\,\partial_{\rho} - \frac{r^2\rho^{2/d}\Delta_{-}^{2\left(d-1\right)/d}}{\hat{d}^2\Delta}\,\partial_{t}^2 + \frac{1}{\hat{d}^2}\Delta_{\mathbb{S}^{\hat{d}+1}}\right]\psi_{\text{hom}} \,,
\ee
and let us choose the following near-zone truncation for this reduced solution space
\be
	\Box_{\text{full}}\psi_{\text{hom}} = \frac{\hat{d}^2}{r^2}\left[\mathbb{O}_{\text{NZ}}^{\text{hom}} + \frac{1}{\hat{d}^2}\Delta_{\mathbb{S}^{\hat{d}+1}} + \epsilon V_1^{\text{hom}}\right]\psi_{\text{hom}}
\ee
with
\be\ba\label{eq:RadialNZHomp}
	\mathbb{O}_{\text{NZ}}^{\text{hom}} &= \partial_{\rho}\,\Delta\,\partial_{\rho} - \frac{\left(\rho_{+}-\rho_{-}\right)^2}{4\Delta}\,\beta_{+}^2\partial_{t}^2 \,, \\
	V_1^{\text{hom}} &= \frac{r^2\rho^{2/d}\Delta_{-}^{2\left(d-1\right)/d}-r_{+}^2\rho_{+}^{2/d}\left(\rho_{+}-\rho_{-}\right)^{2\left(d-1\right)/d}}{\hat{d}^2\Delta}\,\partial_{t}^2 \,.
\ea\ee

One then realizes that the leading order near-zone equations of motion for such homogeneous scalar perturbations of the black $p$-brane are functionally identical to the near-zone equations of motion for scalar perturbations of a $\left(D-p\right)$-dimensional Reissner-Nordstr\"{o}m black hole, see Eq.~\eqref{eq:NZRN}; an observation that was already remarked in Ref.~\cite{Rodriguez:2023xjd} for the case of rotating black strings and here extended to non-rotating black $p$-branes. One can then straightforwardly write down the following set of $3$ vector fields $\left\{L_{-1}^{\text{hom}},L_0^{\text{hom}},L_{+1}^{\text{hom}}\right\}$,
\be
	L_0^{\text{hom}} = -\beta_{+}\,\partial_{t} \,,\quad L_{\pm1}^{\text{hom}} = e^{\pm t/\beta_{+}}\left[\mp\sqrt{\Delta}\,\partial_{\rho} + \left(\partial_{\rho}\sqrt{\Delta}\right)\beta_{+}\,\partial_{t}\right] \,,
\ee
which satisfies the $\SL$ algebra,
\be
	\left[L_{m}^{\text{hom}},L_{n}^{\text{hom}}\right] = \left(m-n\right)L_{m+n}^{\text{hom}} \,,\quad m,n\in\left\{-1,0,+1\right\} \,,
\ee
and whose Casimir element precisely reproduces the reduced leading order near-zone radial operator,
\be
	\mathcal{C}_2^{\SL_{\text{hom}}} = \mathbb{O}_{\text{NZ}}^{\text{hom}} \,.
\ee
Monochromatic and homogeneous perturbations of the black $p$-brane can then be seen to form representations of this $\sla_{\text{hom}}\equiv\text{span}\left\{L_{m}^{\text{hom}}|m=-1,0,+1\right\}$. Following the exact same representation theory arguments laid down in the previous section around the near-zone Love $\SL$ symmetry of scalar perturbations of the Reissner-Nordstr\"{o}m black hole, the vanishing of the static and homogeneous scalar Love numbers of black $p$-branes whenever $\hat{\ell}$ is an integer emerges as the selection rule that the corresponding perturbation belongs to a highest-weight representation of $\SL_{\text{hom}}$,
\be
	\left(L_{+1}^{\text{hom}}\right)^{\hat{\ell}+1}\psi_{\omega=0,\mathbf{k}=\mathbf{0},\ell\mathbf{m}}\bigg|_{\hat{\ell}\in\mathbb{N}} = 0 \Rightarrow k_{\ell}\left(\omega=0,\mathbf{k}=\mathbf{0}\right)\bigg|_{\hat{\ell}\in\mathbb{N}} = 0 \,.
\ee

\subsection{Scalar Love numbers of extremal black $p$-branes}
\label{sec:ExtremalLNs}

We will finish this section by supplementing with a study of what happens at extremality, namely, we will compute the scalar Love numbers of an extremal black $p$-brane. The full equations of motion in the extremal case read, after expanding the scalar field into its spherical harmonics modes,
\be
    \mathbb{O}_{\text{full}}\psi_{\ell\mathbf{m}} = \hat{\ell}(\hat{\ell}+1)\,\psi_{\ell\mathbf{m}} \,,
\ee
with the full radial operator given by
\be
    \mathbb{O}_{\text{full}} = \partial_{\rho}\left(\rho-\rho_{+}\right)^2\partial_{\rho} + \frac{r^2}{\hat{d}^2}\left(\frac{\rho}{\rho-\rho_{+}}\right)^{2/d}\eta^{ab}\partial_{a}\partial_{b} \,.
\ee
These cannot be solved analytically for generic perturbations. Nevertheless, one can identify the exceptional case of perturbations that follow a light-like dispersion relation,
\be\label{eq:LightLikeDispEOM}
    \mathbb{O}_{\text{full}}\psi_{\omega^2=\mathbf{k}^2;\ell\mathbf{m}} = \partial_{\rho}\left(\rho-\rho_{+}\right)^2\partial_{\rho}\psi_{\omega^2=\mathbf{k}^2;\ell\mathbf{m}} \,,
\ee
for which the full equations of motion can be solved analytically everywhere. Namely, the solution that is regular at the extremal horizon is given by
\be
	\psi_{\omega^2=\mathbf{k}^2;\ell\mathbf{m}} = e^{i\left(\mathbf{k}\cdot\mathbf{x}-\omega t\right)}\bar{R}_{\ell\mathbf{m}}\left(\omega^2=\mathbf{k}^2\right) \left(\rho-\rho_{+}\right)^{\hat{\ell}} \,.
\ee
In particular, the regularity condition at the horizon has eliminated the decaying branch $\propto\left(\rho-\rho_{+}\right)^{-\hat{\ell}-1}$ and, consequently, the scalar response coefficients of the extremal black $p$-brane are exactly zero for perturbations that follow light-like dispersion relations,
\be
	k_{\ell}\left(\omega^2=\mathbf{k}^2\right) = 0 \quad\text{for extremal black $p$-branes} \,.
\ee
More importantly, this result is true for \textit{any} orbital number $\ell$, not just those for which $\hat{\ell}$ is an integer as with what happens in the non-extremal case and, furthermore, it implies vanishing not only of the conservative Love numbers, but also of the dissipation coefficients for these particular perturbation modes.

At last, let us consider the more generic response problem of the extremal black $p$-brane, without choosing specific values for the longitudinal momentum of the perturbation. For the sake of this, we will employ a near-zone expansion adapted to the current extremal configuration. Since we want to preserve the near-horizon behavior, this near-zone expansion will, in fact, also be an expansion around modes that are light-like-ly dispersed. Indeed, let us consider the following near-zone split of the radial operator,
\be\label{eq:NZOpExtremalp}
    \begin{gathered}
        \mathbb{O}_{\text{full}} = \mathbb{O}_{\text{NZE}} + \epsilon\,V_1^{\text{NZE}} \,, \\
        \ba
            \mathbb{O}_{\text{NZE}} &= \partial_{\rho}\left(\rho-\rho_{+}\right)^2
            \partial_{\rho}
            + \frac{r_{+}^2}{\hat{d}^2}\left(\frac{\rho_{+}}{\rho-\rho_{+}}\right)^{2/d}\eta^{ab}\partial_{a}\partial_{b} \,, \\
            V_1^{\text{NZE}} &= \frac{r^2\rho^{2/d}-r_{+}^2\rho_{+}^{2/d}}{\hat{d}^2\left(\rho-\rho_{+}\right)^{2/d}}\eta^{ab}\partial_{a}\partial_{b} \,.
        \ea
    \end{gathered}
\ee
One then sees that the corrections in $V_1^{\text{NZE}}$ are of the form $\mathcal{O}\left(k_{\perp}\left(r-r_{+}\right)\right)$ when acting on monochromatic perturbations, where
\be
	k_{\perp} \equiv \sqrt{-\eta^{ab}k_{a}k_{b}} = \sqrt{\omega^2-\mathbf{k}_{\parallel}^2}
\ee
is the transverse frequency, $\mathbf{k}_{\parallel} = \mathbf{k}$ being the spatial momentum along the world-volume. As such, we can extract the scalar response coefficients of the extremal black $p$-brane up to $\mathcal{O}\left(k^2r_{+}^2\right)$ corrections just by solving
\be
	\mathbb{O}_{\text{NZE}}\psi_{\omega\mathbf{k}\ell\mathbf{m}} = \hat{\ell}(\hat{\ell}+1)\psi_{\omega\mathbf{k}\ell\mathbf{m}} \,.
\ee
Imposing ingoing boundary conditions at the horizon we find the following solution for the scalar perturbations of the extremal black $p$-brane at leading order
\be
	\psi_{\omega\mathbf{k}\ell\mathbf{m}} = e^{i\left(\mathbf{k}\cdot\mathbf{x}-\omega t\right)}\bar{R}_{\ell\mathbf{m}}\left(\omega,\mathbf{k}\right)\sqrt{\frac{\rho-\rho_{+}}{\rho_{+}}}H^{\left(2\right)}_{d\left(\hat{\ell}+\frac{1}{2}\right)}\left(\frac{d}{\hat{d}}k_{\perp}r_{+}\left(\frac{\rho_{+}}{\rho-\rho_{+}}\right)^{1/d}\right) \,.
\ee
Expanding the Hankel function around small arguments, i.e. large radial distances, we then find
\be
	k_{\ell}\left(\omega,\mathbf{k}\right) = -\frac{\pi}{\Gamma(\tilde{\ell})\Gamma(\tilde{\ell}+1)\sin\pi\tilde{\ell}}\left(i\frac{d}{2\hat{d}}k_{\perp}r_{+}\right)^{2\tilde{\ell}}  + \mathcal{O}\left(k_{\perp}^2r_{+}^2\right)
\ee
where
\be
    \tilde{\ell} \equiv d\left(\hat{\ell}+\frac{1}{2}\right) \,.
\ee
For extremal black $p$-branes with world-volume dimensionality $d\ge2$, one then sees that the vanishing response coefficients persist up to $\mathcal{O}\left(k_{\perp}^2r_{+}^2\right)$ that are not included in the near-zone approximation,
\be
    k_{\ell}\left(\omega,\mathbf{k}\right) = 0+\mathcal{O}\left(k_{\perp}^2r_{+}^2\right) \quad\text{for extremal $\left(p\ge1\right)$-branes} \,.
\ee
On the other hand, for the special case of extremal black holes ($p=0$), one instead finds that there is a non-vanishing response if $\hat{\ell}<\frac{1}{2}$,
\be
    k_{\ell}\left(\omega,\mathbf{k}\right)\bigg|_{p=0,\hat{\ell}<\frac{1}{2}} = -\frac{\pi e^{i\pi\hat{\ell}}}{\Gamma(\hat{\ell}+\frac{1}{2})\Gamma(\hat{\ell}+\frac{3}{2})\cos\pi\hat{\ell}}\left(-\frac{1}{2\hat{d}}\omega r_{+}\right)^{2\hat{\ell}+1} + \mathcal{O}\left(\omega^2r_{+}^2\right) \,.
\ee
In four spacetime dimensions, this tells us that the $s$-wave mode exhibits a purely dissipative response, and hence the vanishing of the scalar Love numbers of extremal four-dimensional black holes also persists up to order $\mathcal{O}\left(\omega^2r_{+}^2\right)$. Note that 
for general $\hat \ell$
the above response appears 
to be non-analytic in 
$\omega$ for modes with $\ell<\frac{D-3}{2}$. 
This does not contradict 
EFT for the imaginary part of the 
response which could 
be non-analytic due to 
gapless dissipative modes~\cite{Goldberger:2004jt,Goldberger:2005cd,Goldberger:2020fot}.
We believe that 
the apparent non-analyticity 
in $\omega$ of the real part 
of the response
is an artifact 
of the leading order near-zone approximation,
which is known to 
produce a wrong answer for the 
conservative 
dynamical Love numbers, 
see e.g.~\cite{Charalambous:2021mea}.


As we will see in the next section, the vanishing responses we have encountered for the current extremal black $p$-brane solution will be addressed by another type of extended symmetries: the enhanced $SO\left(d;2\mathbb{R}\right)$ isometry subgroup that emerges in the near-horizon throat of the extremal configuration. The resulting Casimir operator of the NHE algebra, which will be by construction adapted to a near-horizon expansion of the equations of motion, will turn out to have the unexpected property of also being identical to the near-zone operator $\mathbb{O}_{\text{NZE}}$ of Eq.~\eqref{eq:NZOpExtremalp}.


\section{Enhanced near-horizon symmetries of 
near-extremal 
non-dilatonic black $p$-branes}
\label{sec:NNHEsympbrane}

In last two sections, we have demonstrated that enhanced conformal symmetries associated with the scalar response problem of black holes and black strings emerge in the near-zone region and that they impose physical selection rules, namely, they dictate the vanishing of static and homogeneous scalar Love numbers. At leading order in the near-zone expansion, the response problem of the non-dilatonic black hole and black string can, in fact, be restated in terms of effective geometries, with the near-zone symmetries manifesting as isometries of these effective geometries. For black holes, we have showed in Section~\ref{sec:NearHorisonSymmetriesRN} that the corresponding effective geometry turned out to be identical with a certain near-horizon limit, what we called the Maldacena limit~\cite{Maldacena:1998uz,Maldacena:1997re,Horowitz:1998pq}, itself being directly related to the NNHE geometry.

Here, we will extend these results to higher world-volume dimensionalities, starting with the well-established result of an enhanced spacetime isometry manifesting in the near-horizon region of strictly extremal black $p$-branes, identified to be the $\left(p+1\right)$-dimensional conformal symmetry group, $SO\left(p+1,2;\mathbb{R}\right)$. There, we will also draw a relation between these NHE Killing vectors and the vector field generating the near-zone Love $SO\left(2,2;\mathbb{R}\right)$ symmetry for the case of black strings. We will will furthermore contrast the behavior of the scalar Love numbers of extremal black $p$-branes with the physical geometric constraints implied from the NHE Killing vectors. We will then study the Maldacena limit of the non-extremal non-dilatonic black $p$-brane and explore the relation of the resulting NNHE geometry with the enhanced near-zone symmetries. As we will see, it is only for black holes and black strings ($p=0,1$) that an identification between the near-zone Love symmetries and the NNHE isometries exist, while, for $p\ge2$, we will show that there cannot be an underlying pure $\text{AdS}_{p+2}$ structure equipped with a non-degenerate horizon, neither in the near-zone nor in the NNHE region.

\subsection{Extremal limit - Enhanced isometries in near-horizon throat}
Let us begin with the strictly extremal black $p$-brane, for which $r_{+}=r_{-}$ and, thus, the full geometry is described by the line element
\be
	ds^2 = f_{+}^{2/d}\eta_{ab}dx^{a}dx^{b} + \frac{dr^2}{f_{+}^2} + r^2 d\Omega_{\hat{d}+1}^2 \,,\quad f_{+} = 1-\left(\frac{r_{+}}{r}\right)^{\hat{d}}\,.
\ee
The full source-less equation of motion for linear massless scalar perturbations of the black $p$-brane then reads, after expanding into spherical harmonic modes,
\be\label{eq:EOMfullExtremalp}
	\begin{gathered}
		\Box_{\text{full}}\psi = 0 \Rightarrow \mathbb{O}_{\text{full}}\psi_{\ell\mathbf{m}} = \hat{\ell}\,(\hat{\ell}+1)\,\psi_{\ell\mathbf{m}} \,, \\
		\mathbb{O}_{\text{full}} = \partial_{\rho}\left(\rho-\rho_{+}\right)^2\partial_{\rho} + \frac{r^2}{\hat{d}^2}\left(\frac{\rho}{\rho-\rho_{+}}\right)^{2/d}\eta^{ab}\partial_{a}\partial_{b} \,.
	\end{gathered}
\ee
The first thing to observe is that the isometry group of the full geometry has now been enhanced from $\mathbb{R}_{t}\times ISO\left(d-1\right)\times SO\,(\hat{d}+2)$ to $ISO\left(d-1,1\right)\times SO\,(\hat{d}+2)$. An even larger enhancement enters in the near-horizon throat, reached by performing the change of coordinates,
\be
	\tilde{y} = \frac{d}{\hat{d}}\frac{r_{+}}{\lambda}\left(\frac{r^{\hat{d}}-r_{+}^{\hat{d}}}{r_{+}^{\hat{d}}}\right)^{1/d} \,,\quad \tilde{x}^{a} = \lambda x^{a} \,,
\ee
and taking the scaling limit $\lambda\rightarrow0$. The resulting NHE geometry,
\be\label{eq:NHEmetricp}
	ds_{\text{NHE}}^2 = \frac{\tilde{y}^2}{b^2} \eta_{ab}d\tilde{x}^{a}d\tilde{x}^{b} + b^2\frac{d\tilde{y}^2}{\tilde{y}^2} + r_{+}^2d\Omega_{\hat{d}+1}^2 \,,
\ee
is an $\text{AdS}_{d+1}\times\mathbb{S}^{\hat{d}+1}$ manifold, with the $\text{AdS}_{d+1}$ radius being
\be\label{eq:NHEAdSRadiusp}
	b = \frac{d}{\hat{d}}r_{+} \,.
\ee
The $\text{AdS}_{d+1}$ Killing vectors are given by
\be
	\begin{gathered}
		\xi_0 = \tilde{x}^{a}\tilde{\partial}_{a} - \tilde{y}\,\partial_{\tilde{y}} \,,\quad \xi_{ab} = \tilde{x}_{a}\tilde{\partial}_{b}-\tilde{x}_{b}\tilde{\partial}_{a} \,,\quad \xi_{+1,a} = \tilde{\partial}_{a} \,, \\
		\xi_{-1,a} = \left(\frac{b^4}{\tilde{y}^2} + \tilde{x}^2 \right)\tilde{\partial}_{a} - 2\tilde{x}_{a}\tilde{x}^{b}\tilde{\partial}_{b} + 2\tilde{x}_{a}\tilde{y}\,\partial_{\tilde{y}}\,,
	\end{gathered}
\ee
with $\tilde{x}^2\equiv\eta_{ab}\tilde{x}^{a}\tilde{x}^{b}$, and satisfy the $SO\left(d,2;\mathbb{R}\right)$ algebra,
\be\ba
	\left[\xi_{\pm1,a},\xi_0\right] &= \pm \xi_{\pm1,a} \,, \\
	\left[\xi_{\pm1,a},\xi_{bc}\right] &= \eta_{ab}\xi_{\pm1,c} - \eta_{ac}\xi_{\pm1,b} \,, \\
	\left[\xi_{\pm1,a},\xi_{\mp1,b}\right] &= 2\left(\mp\eta_{ab}\xi_0 + \xi_{ab}\right) \,, \\
	\left[\xi_{ab},\xi_{cd}\right] &= -\left(\eta_{ac}\xi_{bd}-\eta_{ad}\xi_{bc} - \eta_{bc}\xi_{ad}+\eta_{bd}\xi_{ac}\right) \,.
\ea\ee

In the original $\left(x^{a},\rho\right)$ coordinates, the $SO\left(d,2;\mathbb{R}\right)$ Killing vectors are written as
\be
	\begin{gathered}
		\xi_0 = x^{a}\partial_{a} - d\left(\rho-\rho_{+}\right)\partial_{\rho} \,,\quad \xi_{ab} = x_{a}\partial_{b}-x_{b}\partial_{a} \,,\quad \xi_{+1,a} = \lambda^{-1}\partial_{a} \,, \\
		\xi_{-1,a} = \lambda\left\{\left[\left(\frac{d}{\hat{d}}r_{+}\left(\frac{\rho_{+}}{\rho-\rho_{+}}\right)^{1/d}\right)^2 + x^2\right]\partial_{a} - 2x_{a}x^{b}\partial_{b} + 2d\,x_{a}\left(\rho-\rho_{+}\right)\partial_{\rho} \right\} \,.
	\end{gathered}
\ee
The Casimir of this NHE $SO\left(d,2;\mathbb{R}\right)$ algebra can be worked out to be
\be\ba\label{eq:CasimirSOd2p}
	\mathcal{C}_2^{SO\left(d,2;\mathbb{R}\right)_{\text{NHE}}} &= \frac{1}{d^2}\left[\xi_0^2 + \frac{1}{2}\eta^{ab}\left(\xi_{+1,a}\xi_{-1,b}+\xi_{-1,a}\xi_{+1,b}\right) - \frac{1}{2}\xi_{ab}\xi^{ab}\right] \\
	&= \frac{1}{d^2}\left[\frac{1}{\tilde{y}^{d-1}}\partial_{\tilde{y}}\,\tilde{y}^{d+1}\partial_{\tilde{y}} + \frac{b^4}{\tilde{y}^2}\eta^{ab}\tilde{\partial}_{a}\tilde{\partial}_{b}\right] \\
	&= \partial_{\rho}\left(\rho-\rho_{+}\right)^2\partial_{\rho} + \frac{r_{+}^2}{\hat{d}^2}\left(\frac{\rho_{+}}{\rho-\rho_{+}}\right)^{2/d}\eta^{ab}\partial_{a}\partial_{b} \,,
\ea\ee
and one can see that the full equations of motion in the near-horizon scaling limit are precisely this Casimir operator of the enhanced NHE algebra,
\be\label{eq:NHVsFullRadial}
	\mathbb{O}_{\text{full}} = \mathcal{C}_2^{SO\left(d,2;\mathbb{R}\right)_{\text{NHE}}} + \mathcal{O}\left(\lambda\right) \,.
\ee

\subsection{Geometric constraints of NHE algebra on scalar Love numbers of extremal black $p$-branes}

Let us now explore the implications of these enhanced NHE isometries for the scalar response problem of the black $p$-brane. To begin with, the full equations of motion Eq.~\eqref{eq:EOMfullExtremalp} cannot be solved analytically everywhere. Nevertheless, the NHE Casimir element provides with an approximation that has the remarkable property of being exact when it acts on static and homogeneous perturbations. In fact, this remains true for the non-static perturbations that follow a light-like dispersion relation,
\be\label{eq:LightLikeDispEOM}
	 \mathcal{C}_2^{SO\left(d,2;\mathbb{R}\right)_{\text{NHE}}}\psi_{\omega^2=\mathbf{k}^2;\ell\mathbf{m}} = \mathbb{O}_{\text{full}}\psi_{\omega^2=\mathbf{k}^2;\ell\mathbf{m}} = \partial_{\rho}\left(\rho-\rho_{+}\right)^2\partial_{\rho}\psi_{\omega^2=\mathbf{k}^2;\ell\mathbf{m}} \,,
\ee
and, hence, the full equations of motion $\mathbb{O}_{\text{full}}\psi_{\omega^2=\mathbf{k}^2;\ell\mathbf{m}} = \hat{\ell}(\hat{\ell}+1)\,\psi_{\omega^2=\mathbf{k}^2;\ell\mathbf{m}}$ can be solved analytically everywhere as we have also demonstrated in Section~\ref{sec:ExtremalLNs}, with the final result being the vanishing of the corresponding scalar response coefficients.

We will now demonstrate that the NHE $SO\left(d,2;\mathbb{R}\right)_{\text{NHE}}$ Killing vector address these vanishings. To begin with, let us focus to the simplest case of black strings ($p=1$). The $SO\left(2,2;\mathbb{R}\right)_{\text{NHE}}$ Killing vectors read
\be\ba\label{eq:NHESO22R_p1}
	\xi_0 &= t\partial_{t} + x\partial_{x} - 2\left(\rho-\rho_{+}\right)\partial_{\rho} \,,\quad \xi_{01} = -t\partial_{x} - x\partial_{t} \,, \\
	\xi_{+1,0} &= \lambda^{-1}\partial_{t} \,,\quad \xi_{+1,1} = \lambda^{-1}\partial_{x} \,, \\
	\xi_{-1,0} &= \lambda\left\{\left[\left(\frac{2r_{+}}{\hat{d}}\right)^2\frac{\rho_{+}}{\rho-\rho_{+}} + t^2+x^2\right]\partial_{t} + 2t\,x\partial_{x} - 4t\left(\rho-\rho_{+}\right)\partial_{\rho}\right\} \,, \\
	\xi_{-1,1} &= \lambda\left\{\left[\left(\frac{2r_{+}}{\hat{d}}\right)^2\frac{\rho_{+}}{\rho-\rho_{+}} - t^2-x^2\right]\partial_{x} - 2x\,t\partial_{t} + 4x\left(\rho-\rho_{+}\right)\partial_{\rho}\right\} \,,
\ea\ee
or, in the $\sla_{\left(+\right)}\oplus\sla_{\left(-\right)}$ basis,
\be\ba
	\xi_0^{\left(\pm\right)} &= \frac{1}{2}\left(t\pm x\right)\left(\partial_{t}\pm\partial_{x}\right) - \left(\rho-\rho_{+}\right)\partial_{\rho} \,,\quad \xi_{+1}^{\left(\pm\right)} = \lambda^{-1}\frac{1}{2}\left(\partial_{t}\pm\partial_{x}\right) \,, \\
	\xi_{-1}^{\left(\pm\right)} &= \lambda\left[\frac{2r_{+}^2}{\hat{d}^2}\frac{\rho_{+}}{\rho-\rho_{+}}\left(\partial_{t}\mp\partial_{x}\right) + \frac{1}{2}\left(t\pm x\right)^2\left(\partial_{t}\pm\partial_{x}\right) - 2\left(t\pm x\right)\left(\rho-\rho_{+}\right)\partial_{\rho}\right] \,.
\ea\ee
Let us now explore particular representations of this NHE $\SL_{\left(+\right)}\times\SL_{\left(-\right)}$ algebra, namely, those representations which are highest-weight with respect to only one of the $\SL$ factors. For instance, let us look at the primary vector $\upsilon^{\left(+\right)}_{-\hat{\ell},0}$ of weight $h_{\left(+\right)}=-\hat{\ell}$ with respect to $\SL_{\left(+\right)\text{NHE}}$, defined by
\be
	\xi_{+1}^{\left(+\right)}\upsilon^{\left(+\right)}_{-\hat{\ell},0} = 0 \,,\quad \xi_0^{\left(+\right)}\upsilon^{\left(+\right)}_{-\hat{\ell},0}=-\hat{\ell}\,\upsilon^{\left(+\right)}_{-\hat{\ell},0} \,.
\ee
The solution is
\be
	\upsilon^{\left(+\right)}_{-\hat{\ell},0} = f\left(t-x\right)\left(\rho-\rho_{+}\right)^{\hat{\ell}} \,,
\ee
with $f\left(t-x\right)$ some arbitrary function of the combination $t-x$. The pure polynomial form in $\rho-\rho_{+}$ dictates the vanishing of the corresponding scalar response coefficients for such near-horizon modes of the extremal black string, this time for any orbital number $\ell$, not just for integer $\hat{\ell}$. The question remains of whether these near-horizon modes are physical, in the sense that they satisfy the full equations of motion in the background of the extremal black string. This is indeed the case, as demonstrated explicitly in Eq.~\eqref{eq:LightLikeDispEOM}, since the $-\partial_{t}^2+\partial_{x}^2 \propto \xi_{+1}^{\left(-\right)}\xi_{+1}^{\left(+\right)}$ term annihilates $\upsilon^{\left(+\right)}_{-\hat{\ell},0}$ by definition. Therefore, the primary vector $\upsilon^{\left(+\right)}_{-\hat{\ell},0}$ we just constructed from $\SL_{\left(+\right)\text{NHE}}$ representation theory arguments is, in fact, a solution to the \textit{full} equations motion. We remark, however, that this primary vector does not in general belong to a representation of $\SL_{\left(-\right)\text{NHE}}$, since $\xi_0^{\left(-\right)}\upsilon^{\left(+\right)}_{-\hat{\ell},0} \ne \left(\text{const.}\right)\upsilon^{\left(+\right)}_{-\hat{\ell},0}$.

In a plane wave basis for the massless scalar perturbations, $\psi_{\omega k\ell\mathbf{m}}\left(t,x,\rho\right) = e^{i\left(kx-\omega t\right)}R_{\omega k \ell\mathbf{m}}\left(\rho\right)$, the highest-weight property $\xi_{+1}^{\left(+\right)}\upsilon^{\left(+\right)}_{-\hat{\ell},0} = 0$ is precisely translated to the light-like dispersion $\omega = +k$. The same steps can be followed to construct a primary vector $\upsilon^{\left(-\right)}_{-\hat{\ell},0}$ that belongs to a highest-weight representation of $\SL_{\left(-\right)\text{NHE}}$, but, evidently, does not belong to a representation of $\SL_{\left(+\right)\text{NHE}}$, which is identified with a solution to the full equations of motion for monochromatic perturbation modes with dispersion relation $\omega=-k$. The highest-weight property again dictates a purely polynomial form in the $\rho-\rho_{+}$, hence outputting vanishing response coefficients for these modes.

As a last remark for the case of black strings, it is interesting to note that the Killing vectors generating the NHE $SO\left(2,2;\mathbb{R}\right)$ can also be recovered from the near-zone Love $SO\left(2,2;\mathbb{R}\right)$ generators we encountered in Section~\ref{sec:LoveSymmetryp1}, see Eq.~\eqref{eq:LoveS022p1} or Eq.~\eqref{eq:LoveSL2R2p1}. This is achieved by taking the extremal limit in the form of a Wigner-like contraction as follows
\be\ba
	\xi_0 &= +\lim_{T_{H}\rightarrow0}\frac{L_{+1,0}-L_{-1,0}}{2} \,, \\
	\xi_{01} &= -\lim_{T_{H}\rightarrow0}\frac{L_{+1,1}+L_{-1,1}}{2} \,, \\
	\xi_{+1,0} &= -\lambda^{-1}\lim_{T_{H}\rightarrow0}2\pi T_{H}L_0 \,, \\
	\xi_{+1,1} &= +\lambda^{-1}\lim_{T_{H}\rightarrow0}2\pi T_{H}L_{01} \,, \\
	\xi_{-1,0} &= +\lambda\lim_{T_{H}\rightarrow0}\frac{L_{+1,0}+L_{-1,0}+2L_0}{2\pi T_{H}} \,, \\
	\xi_{-1,1} &= -\lambda\lim_{T_{H}\rightarrow0}\frac{L_{+1,1}-L_{-1,1}-2L_{01}}{2\pi T_{H}} \,,
\ea\ee
after identifying $\lambda$ with the scaling parameter.

For the case of extremal black $\left(p\ge2\right)$-branes, we can follow similar algebraic arguments to address the vanishings of the static and homogeneous scalar Love numbers. In particular, consider the following specific highest-weight representation of the $SO\left(d,2;\mathbb{R}\right)_{\text{NHE}}$ algebra, with the primary vector defined by
\be
	\xi_{+1,a}\upsilon_{-d\,\hat{\ell},0} = 0 \,,\quad \xi_0\upsilon_{-d\,\hat{\ell},0} = -d\hat{\ell}\,\upsilon_{-d\,\hat{\ell},0} \Rightarrow \upsilon_{-d\,\hat{\ell},0} = \left(\rho-\rho_{+}\right)^{\hat{\ell}} \,.
\ee
This is precisely the static and homogeneous perturbation that is regular at the horizon and its purely polynomial form is exactly the condition that dictates the vanishing of the corresponding response coefficients, just as we extracted from the explicit calculation.

Last, let us remark an interesting property regarding the more generic response problem of the extremal black $p$-brane, without choosing specific values for the longitudinal momentum of the perturbation. The full equations of motion Eq.~\eqref{eq:EOMfullExtremalp} cannot be solved analytically, but we may use the enhanced NHE isometries and expand around the near-horizon region. Interestingly, for the current background geometry, which arises as a solution for the action give in Eqs.~\eqref{eq:pBraneAction}-\eqref{eq:pBraneActionWV} whose bulk gravitational degrees of freedom are general-relativistic, this near-horizon expansion turns out to be identical with a near-zone expansion. This fact comes from the observation that the $\mathcal{O}\left(\lambda\right)$ corrections in the near-horizon expansion of the full equations of motion, Eq.~\eqref{eq:NHVsFullRadial}, are of the form $\mathcal{O}\left(k_{\perp}\left(r-r_{+}\right)\right)$ when acting on monochromatic perturbations, with $k_{\perp}^2 \equiv -\eta^{ab}k_{a}k_{b} = \omega^2-\mathbf{k}^2$ the trasnverse frequency. In fact, the near-zone operator $\mathbb{O}_{\text{NZE}}$ in Eq.~\eqref{eq:NZOpExtremalp}, that we have employed when computing the scalar Love numbers of the extremal black $p$-brane in Section~\ref{sec:ExtremalLNs}, is identical to the Casimir operator of NHE algebra
\be
    \mathbb{O}_{\text{NZE}} = \mathcal{C}_2^{SO\left(d,2;\mathbb{R}\right)_{\text{NHE}}} \,.
\ee
This is a rather unexpected result that appears to be specific to genera-relativistic gravitational effects and be practically traced to the fact that $g^{rr}$ is a perfect a square in the extremal limit, instead of some generic function with a double root at the degenerate horizon. Physically, it implies that there exist near-horizon modes that survive in the far-horizon region, which at no-way needed to be the case. We will leave the better understanding of this accidental effect for future development.


\subsection{Contraction of homogeneous Love symmetry in the extremal limit}

In Section~\ref{sec:HomSL2R}, we have constructed a ``homogeneous Love symmetry'', i.e., an $\SL$ symmetry of the near-zone equations of motions that acts on massless scalar perturbations that are homogeneous with respect to the spatial directions of the black $p$-brane. In the extremal limit, one can perform a contraction of the homogeneous Love symmetry generators as follows
\be\ba
    \zeta_{+1}^{\text{hom}} &= \tilde{\lambda}^{-1}\lim_{T_{H}\rightarrow0}\left(-2\pi T_{H}L_0^{\text{hom}}\right) = \tilde{\lambda}^{-1}\partial_{t} \,, \\
	\zeta_{0}^{\text{hom}} &= \lim_{T_{H}\rightarrow0}\frac{L_{+1}^{\text{hom}}-L_{-1}^{\text{hom}}}{2} = t\partial_{t} - \left(\rho-\rho_{+}\right)\partial_{\rho} \,, \\
	\zeta_{-1}^{\text{hom}} &= \tilde{\lambda}\lim_{T_{H}\rightarrow0}\frac{L_{+1}^{\text{hom}}+L_{-1}^{\text{hom}}+2L_0^{\text{hom}}}{2\pi T_{H}} = \tilde{\lambda}\left[\left(\left(\frac{r_{+}}{\hat{d}}\frac{\rho_{+}}{\rho-\rho_{+}}\right)^2 + t^2\right)\partial_{t} - 2t\left(\rho-\rho_{+}\right)\partial_{\rho}\right] \,,
\ea\ee
with $\tilde{\lambda}$ some arbitrary constant. The resulting vectors fields satisfy an $\SL$ algebra,
\be
    \left[\zeta_{m}^{\text{hom}},\zeta_{n}^{\text{hom}}\right] = \left(m-n\right)\zeta_{m+n}^{\text{hom}} \,,\quad m,n\in\left\{-1,0,+1\right\} \,,
\ee
which also tells us that the parameter $\tilde{\lambda}$ is algebraically irrelevant as it enters automorphically. The above set of vector fields has the special property of surviving in the following near-horizon limit
\be
    \hat{y} = \frac{1}{\hat{d}}\frac{r_{+}}{\lambda}\frac{\rho-\rho_{+}}{\rho_{+}} \,,\quad \tilde{t} = \lambda t \,,\quad \lambda\rightarrow0
\ee
once one identifies $\tilde{\lambda}$ with the scaling parameter $\lambda$.

In fact, the vector fields $\zeta_{m}^{\text{hom}}$ exactly match with the generators of the enhanced $\SL$ symmetry of the near-horizon throat of a $\left(D-p\right)$-dimensional extremal Reissner-Nordstr\"{o}m black hole, see Eq.~\eqref{eq:NHESL2R_RN}. However, the set $\left\{\zeta_{m}^{\text{hom}}\right\}$ does not span any isometry subgroup of the near-horizon throat of the $D$-dimensional extremal black $p$-brane or any of its $x^{i}=\text{const.}$ subspaces.
This can be traced back to the 
fact that the metric component $g_{tt}$
for $p>1$ does not reduce to
the AdS$_2$ one in the extremal 
near-horizon scaling limit. 

Nevertheless, it is an isometry of the $\left(t=\text{const.},x^{i}=\text{const.}\right)$ subspace of the near-horizon throat, i.e.
\be
    \mathcal{L}_{\zeta_{m}^{\text{hom}}}g_{\tilde{y}\tilde{y}}^{\text{NHE}} = 0 \,,
\ee
where $g_{\tilde{y}\tilde{y}}^{\text{NHE}}$ is the $\tilde{y}\tilde{y}$-component of the NHE metric in Eq.~\eqref{eq:NHEmetricp}. As a result, the $\SL$ algebra generated by $\left\{\zeta_{m}^{\text{hom}}\right\}$ has an exact action on static and homogeneous perturbations of the near-horizon throat. In fact, its Casimir operator has an exact action on static and homogeneous perturbations of the full extremal black $p$-brane geometry
\be
    \mathcal{C}_2\left(\zeta_{m}^{\text{hom}}\right)\psi_{\omega=0,\mathbf{k}=\mathbf{0}} = \partial_{\rho}\left(\rho-\rho_{+}\right)^2\partial_{\rho}\psi_{\omega=0,\mathbf{k}=\mathbf{0}} = \mathbb{O}_{\text{full}}\psi_{\omega=0,\mathbf{k}=\mathbf{0}} \,.
\ee
As such, one can study algebraic constraints on static and homogeneous perturbations of the extremal black $p$-brane, just by studying representations of this $\SL$ algebra generated by $\zeta_{m}^{\text{hom}}$. It is not hard to see that static and homogeneous perturbations of the extremal black $p$-brane are primary vectors of a highest-weight representation of this $\SL$ algebra of weight $-\hat{\ell}$,
\be
    \zeta_{+1}^{\text{hom}}\psi_{\omega=0,\mathbf{k}=\mathbf{0},\ell\mathbf{m}} = 0 \,,\quad \zeta_{0}^{\text{hom}}\psi_{\omega=0,\mathbf{k}=\mathbf{0},\ell\mathbf{m}} = -\hat{\ell}\,\psi_{\omega=0,\mathbf{k}=\mathbf{0},\ell\mathbf{m}} \,.
\ee
This implies a purely polynomial solution, up to an overall integration constant,
\be
    \psi_{\omega=0,\mathbf{k}=\mathbf{0},\ell\mathbf{m}} = \left(\rho-\rho_{+}\right)^{\hat{\ell}} \,,
\ee
from which one immediately infers the vanishing of the static and homogeneous scalar Love numbers of the extremal black $p$-brane.

The existence of the vector fields $\zeta_{m}^{\text{hom}}$ should not come as a surprise. It can be traced back to the observation that the near-zone equations of motion for homogeneous perturbations of a black $p$-brane are functionally identical to the near-zone equations of motion for perturbations of $\left(D-p\right)$-dimensional black hole, as already remarked in Section~\ref{sec:HomSL2R}.
At the moment, it is not 
clear if this symmetry
can be robustly interpreted 
from the geometric point of view.
However, this might be related to the observation that AdS$_{p+2}$ can be viewed as a warped geometry of AdS$_2\times\mathbb{S}^{p}$, see e.g. Ref.~\cite{Cvetic:2020axz}.


\subsection{Near-extremal near-horizon region: equivalence to SAdS$_{p+2}$}
We now look at the non-extremal black $p$-brane and study its NIH geometry. Using the coordinate $\rho=r^{\hat{d}}$, the full geometry in Eq.~\eqref{eq:pBraneGeoNonDilatonic} reads
\be
	\begin{gathered}
		ds^2 = \left[H\left(\rho\right)\right]^{1/d}\left[-Z\left(\rho\right)dt^2 + d\mathbf{x}^2\right] + \frac{r^2}{\hat{d}^2\rho^2}\frac{d\rho^2}{H\left(\rho\right)Z\left(\rho\right)} + r^2d\Omega_{\hat{d}+1}^2 \,, \\
		Z\left(\rho\right) = \frac{f_{+}}{f_{-}} \,,\quad H\left(\rho\right) = f_{-}^{2} \,.
	\end{gathered}
\ee
To obtain the NIH region, we first introduce the variable
\be
	z = \frac{d}{\hat{d}}r_{-}\left(\frac{\rho_{-}}{\rho-\rho_{-}}\right)^{1/d} \,
\ee
and take the Maldacena limit, that is, we take the near-inner horizon limit $z\rightarrow\infty$ while keeping $Z = \frac{f_{+}}{f_{-}}$ fixed. The resulting NIH geometry is found to be
\be\label{eq:pg2NHgeometry_x}
	ds_{\text{NIH}}^2 = \left(\frac{d}{\hat{d}}r_{-}\right)^2\frac{1}{z^2}\left[-Zdt^2 + d\mathbf{x}^2 + \frac{dz^2}{Z}\right] + r_{-}^2d\Omega_{\hat{d}+1}^2 \,,\quad Z = 1 - \left(\frac{2z}{d\beta_{-}}\right)^{d} \,,
\ee
where $\beta_{-}$ is given in Eq.~\eqref{eq:betampbrane}. Fixed points on the sphere, charted by the $\left(t,\mathbf{x},z\right)$ coordinates,
\be\label{eq:NIHGeometryp}
	ds_{\text{NIH}}^2\bigg|_{\mathbb{S}^{\hat{d}+1}} = \frac{b^2}{z^2}\left[-Zdt^2 + d\mathbf{x}^2 + \frac{dz^2}{Z}\right] \,,\quad Z = 1 - \left(\frac{z}{z_{\text{h}}}\right)^{d} \,,
\ee
now lie on a $\text{SAdS}_{p+2}$ spacetime in Poincar\'{e} coordinates, whose AdS radius and event horizon location read
\be
	b = \frac{d}{\hat{d}}r_{-} \quad\text{and}\quad z_{\text{h}} = \frac{d}{2}\beta_{-} \,.
\ee

Switching back to $\left(t,\mathbf{x},\rho,\theta\right)$ coordinates, the NIH geometry becomes
\be 
	ds^2_{\text{NIH}} =\left(\frac{\Delta_-}{\r_-}\right)^{2/d}\left[-\frac{\Delta_{+}}{\Delta_-}dt^2 + d\mathbf{x}^2\right] +\frac{r_-^{2}}{\hat{d}^2}\left[\frac{d\rho^2}{\Delta_+\Delta_-} + \hat{d}^2\,d\Omega_{\hat{d}+1}^2\right] \,.
\ee 
If we now treat this as a background geometry and consider massless scalar perturbations $\psi$, the linearized scalar wave operator reads
\be 
	\Box_{\text{NIH}}\psi = \frac{\hat{d}^2}{r_{-}^2}\left[\partial_{\rho}\,\Delta\,\partial_{\rho} + \frac{r_{-}^2\rho_{-}^{2/d}\Delta_{-}^{\left(d-2\right)/d}}{\hat{d}^2}\left(-\frac{1}{\Delta_{+}}\partial_{t}^2 + \frac{1}{\Delta_{-}}\,\delta^{ij}\partial_{i}\partial_{j}\right) + \frac{1}{\hat{d}^2}\Delta_{\mathbb{S}^{\hat{d}+1}}\right]\psi \,.
\ee
This is to be contrasted to the full equations of motion. Eq.~\eqref{eq:KGfullp} and, in particular, to the near-zone truncation employed in Section~\ref{sec:MatchingNZ}. Comparing with the leading order near-zone operator in Eqs.~\eqref{eq:NZsplitp}-\eqref{eq:RadialNZp}, we see that it does not quite match the above wave operator. Nevertheless, the two operators do become identical if one takes the NNHE limit in the following way
\be\ba
	{}&\frac{r_{-}^2\rho_{-}^{2/d}\Delta_{-}^{\left(d-2\right)/d}}{\hat{d}^2}\left(-\frac{1}{\Delta_{+}}\partial_{t}^2 + \frac{1}{\Delta_{-}}\,\delta^{ij}\partial_{i}\partial_{j}\right) \\
	&\quad\xrightarrow{\text{NNHE}} \frac{r_{+}^2\rho_{+}^{2/d}\left(\rho_{+}-\rho_{-}\right)^{\left(d-2\right)/d}}{\hat{d}^2}\left(-\frac{1}{\Delta_{+}}\partial_{t}^2 + \frac{1}{\Delta_{-}}\,\delta^{ij}\partial_{i}\partial_{j}\right) \,,
\ea\ee
the crucial point being that we keep the terms inside the brackets intact. This operator is associated with the following NNHE geometry
\be
	ds^2_{\text{NNHE}} = \frac{\left(\rho_{+}-\rho_{-}\right)^{\left(2-d\right)/d}}{\rho_{+}^{2/d}}\left[-\Delta_{+}dt^2 + \Delta_{-}d\mathbf{x}^2\right] + \frac{r_{+}^{2}}{\hat{d}^2}\left[\frac{d\rho^2}{\Delta_+\Delta_-} + \hat{d}^2\,d\Omega_{\hat{d}+1}^2\right]
\ee
and precisely corresponds to the effective black $p$-brane geometry that reproduces the near-zone equations of motion.

At this stage, it is still not clear whether the NIH or the NNHE geometries above enjoy enhanced symmetries that will prove relevant to the full scalar response problem of the non-extremal black $p$-brane. As we will demonstrate right away, however, the case of black strings ($p=1$) follows a pattern similar to the Reissner-Nordstr\"{o}m black hole paradigm we analyzed in Section~\ref{sec:NearHorisonSymmetriesRN}, that is, the NIH/NNHE geometries are equipped with enhanced conformal symmetries.

\subsection{NNHE region for black strings: equivalence to AdS$_3$}
\label{sec:nearNHEGeometryp1}

As remarked in Ref.~\cite{Satoh:1998sg}, the black string geometry can be brought into the form of a BTZ black hole after a coordinate transformation, modulo periodic identifications. This in turn has been demonstrated in Ref~\cite{Carlip:1995qv} to be isomorphic to a patch of pure $\text{AdS}_3$. Adapting these considerations to our analysis, the coordinate transformations
\be\label{eq:SAdS3toAdS3}
	\tilde{\tau}\,\tilde{r} = b^2\frac{z_{\text{h}}}{z}\cosh\frac{x}{z_{\text{h}}} \,,\quad \tilde{\chi}\,\tilde{r} = b^2\frac{z_{\text{h}}}{z}\sinh\frac{x}{z_{\text{h}}} \,,\quad \tilde{\tau}^2 - \tilde{\chi}^2 - \frac{b^4}{\tilde{r}^2} = L^2e^{2t/z_{\text{h}}} \,,
\ee
with $L$ some arbitrary length scale, shows that the NIH geometry of the black string, Eq.~\eqref{eq:NIHGeometryp} with $p=1$, is precisely isomorphic to the Poincar\'{e} patch of pure $\text{AdS}_3$,
\be
	ds_{\text{NIH}}^2\bigg|_{\mathbb{S}^{\hat{d}+1}} = \frac{\tilde{r}^2}{b^2}\left(-d\tilde{\tau}^2+d\tilde{\chi}^2\right) + b^2\frac{d\tilde{r}^2}{\tilde{r}^2} \,.
\ee

The six Killing vectors generating the $SO\left(2,2;\mathbb{R}\right)$ isometry group of this $\text{AdS}_3$ geometry are then
\be\ba\label{eq:SO22generatorsNIH}
	\tilde{L}_0 &= \tilde{\tau}\partial_{\tilde{\tau}} + \tilde{\chi}\partial_{\tilde{\chi}} - \tilde{r}\partial_{\tilde{r}} \,,\quad \tilde{L}_{01} = -\tilde{\tau}\partial_{\tilde{\chi}} - \tilde{\chi}\partial_{\tilde{\tau}} \,, \\
	\tilde{L}_{+1,0} &= \partial_{\tilde{\tau}} \,,\quad \tilde{L}_{+1,1} = \partial_{\tilde{\chi}} \,, \\
	\tilde{L}_{-1,0} &= \left(\frac{b^4}{\tilde{r}^2} + \tilde{\tau}^2 + \tilde{\chi}^2\right)\partial_{\tilde{\tau}} + 2\tilde{\tau}\tilde{\chi}\partial_{\tilde{\chi}} - 2\tilde{\tau}\tilde{r}\partial_{\tilde{r}} \,, \\
	\tilde{L}_{-1,1} &= \left(\frac{b^4}{\tilde{r}^2} - \tilde{\tau}^2 - \tilde{\chi}^2\right)\partial_{\tilde{\chi}} - 2\tilde{\chi}\tilde{\tau}\partial_{\tilde{\tau}} + 2\tilde{\chi}\tilde{r}\partial_{\tilde{r}} \,.
\ea\ee
or, in the $\sla_{\left(+\right)}\oplus\sla_{\left(-\right)}$ basis,
\be
	\begin{gathered}
		\ba
		\tilde{L}_0^{\left(\pm\right)} &= \frac{\tilde{L}_0 \mp \tilde{L}_{01}}{2} = \frac{1}{2}\left(\tilde{\tau}\pm\tilde{\chi}\right)\left(\partial_{\tilde{\tau}}\pm\partial_{\tilde{\chi}}\right) - \frac{1}{2}\tilde{r}\partial_{\tilde{r}} \,, \\
		\tilde{L}_{+1}^{\left(\pm\right)} &= \frac{\tilde{L}_{+1,0} \pm \tilde{L}_{+1,1}}{2} = \frac{1}{2}\left(\partial_{\tilde{\tau}}\pm\partial_{\tilde{\chi}}\right) \,, \\
		\tilde{L}_{-1}^{\left(\pm\right)} &= \frac{\tilde{L}_{-1,0} \mp \tilde{L}_{-1,1}}{2} = \frac{b^4}{2\tilde{r}^2}\left(\partial_{\tilde{\tau}}\mp\partial_{\tilde{\chi}}\right) + \frac{1}{2}\left(\tilde{\tau}\pm\tilde{\chi}\right)^2\left(\partial_{\tilde{\tau}}\pm\partial_{\tilde{\chi}}\right) - \left(\tilde{\tau}\pm\tilde{\chi}\right)\tilde{r}\partial_{\tilde{r}} \,,
		\ea \\
		\left[\tilde{L}_{m}^{\left(\sigma\right)},\tilde{L}_{n}^{\left(\sigma^{\prime}\right)}\right]=\left(m-n\right)\tilde{L}_{m+n}^{\left(\sigma\right)}\delta_{\sigma,\sigma^{\prime}} \,,\quad m,n\in\left\{-1,0,+1\right\} \,,\quad \sigma,\sigma^{\prime} \in\left\{+,-\right\} \,.
	\end{gathered}
\ee
The Casimir operator for the Killing vectors is then given by
\be\ba
	\tilde{\mathcal{C}}_2^{SO\left(2,2;\mathbb{R}\right)_{\text{NIH}}} &= \frac{1}{4}\bigg[\tilde{L}_0^2 - \frac{1}{2}\left(\tilde{L}_{+1,0}\tilde{L}_{-1,0}+\tilde{L}_{-1,0}\tilde{L}_{+1,0}\right) \\
	&\quad\quad+ \frac{1}{2}\left(\tilde{L}_{+1,1}\tilde{L}_{-1,1}+\tilde{L}_{-1,1}\tilde{L}_{+1,1}\right) + \tilde{L}_{01}^2 \bigg] \\
	&= \frac{1}{2}\left(\tilde{\mathcal{C}}_2^{\SL_{\left(+\right)\text{NIH}}} + \tilde{\mathcal{C}}_2^{\SL_{\left(-\right)\text{NIH}}}\right) \\
	&= \frac{1}{4}\left[\frac{1}{\tilde{r}}\partial_{\tilde{r}}\,\tilde{r}^3\,\partial_{\tilde{r}} + \frac{b^4}{\tilde{r}^2}\left(-\partial_{\tilde{\tau}}^2+\partial_{\tilde{\chi}}^2\right)\right] \,,
\ea\ee
and we also note here that the Casimirs of the two commuting $\SL$'s turn out to be the same, $\tilde{\mathcal{C}}_2^{\SL_{\left(+\right)\text{NIH}}} = \tilde{\mathcal{C}}_2^{\SL_{\left(-\right)\text{NIH}}} = \tilde{\mathcal{C}}_2^{SO\left(2,2;\mathbb{R}\right)_{\text{NIH}}}$.

In summary, the full isometry group of the NIH geometry is $\SL_{\left(+\right)}\times\SL_{\left(-\right)}\times SO\,(\hat{d}+2)$, with $\SL_{\left(\pm\right)}$ generated by $\tilde{L}_{m}^{\left(\pm\right)}$ and the commuting $SO\,(\hat{d}+2)$ amounting for the spherical symmetry of the geometry.

In the original $\left(t,x,\rho\right)$ coordinates, the full NIH geometry becomes
\be\label{eq:NIHgeometryp1}
	\begin{gathered}
		ds_{\text{NIH}}^2 = -\frac{\Delta_{+}}{\rho_{-}}dt^2 + \frac{\Delta_{-}}{\rho_{-}}dx^2 + \frac{r_{-}^2}{\hat{d}^2}\left[\frac{d\rho^2}{\Delta} + \hat{d}^2\,d\Omega_{\hat{d}+1}^2\right] \,, \\
		\Delta_{\pm} = \rho-\rho_{\pm} \,,\quad \Delta \equiv \Delta_{+}\Delta_{-} = \left(\rho-\rho_{+}\right)\left(\rho-\rho_{-}\right) \,,
	\end{gathered}
\ee
the $\text{AdS}_3$ Killing vectors read\footnote{The vector fields written here that span $\mathfrak{so}\left(2,2;\mathbb{R}\right)$ have been automorphically redefined for future convenience. More specifically, we performed the redefinitions $\tilde{L}_0\rightarrow-\tilde{L}_0$, $\tilde{L}_{01}\rightarrow-\tilde{L}_{01}$, $\tilde{L}_{\pm1,0}\rightarrow\tilde{L}_{\mp1,0}$, $\tilde{L}_{\pm1,1}\rightarrow-\tilde{L}_{\mp1,1}$ in Eq.~\eqref{eq:SO22generatorsNIH}, plus rescalings of the form $\tilde{L}_{\pm1,0}\rightarrow \lambda^{\pm1}\tilde{L}_{\pm1,0}$ and $\tilde{L}_{\pm1,1}\rightarrow \lambda^{\pm1}\tilde{L}_{\pm1,1}$ with the \textit{same} global factor to remove any dependence on the arbitrary length scale $L$ and signs associated with the covered spacetime patch; both types of redefinitions preserve the current basis used to write down the $SO\left(2,2;\mathbb{R}\right)$ algebra.}
\be\ba\label{eq:LoveGeneratorsp1RmSO22}
	\tilde{L}_0 &= -\beta_{-}\,\partial_{t} \,,\quad \tilde{L}_{01} = +\beta_{-}\partial_{x} \,, \\
	\tilde{L}_{\pm1,0} &= e^{\pm t/\beta_{-}}\left[\cosh\frac{x}{\beta_{-}}\left(\mp2\sqrt{\Delta}\,\partial_{\rho} + \sqrt{\frac{\rho-\rho_{-}}{\rho-\rho_{+}}}\,\beta_{-}\partial_{t}\right) \pm \sinh\frac{x}{\beta_{-}}\sqrt{\frac{\rho-\rho_{+}}{\rho-\rho_{-}}}\,\beta_{-}\partial_{x}\right] \,, \\
	\tilde{L}_{\pm1,1} &= e^{\pm t/\beta_{-}}\left[\sinh\frac{x}{\beta_{-}}\left(\mp2\sqrt{\Delta}\,\partial_{\rho} + \sqrt{\frac{\rho-\rho_{-}}{\rho-\rho_{+}}}\,\beta_{-}\partial_{t}\right) \pm \cosh\frac{x}{\beta_{-}}\sqrt{\frac{\rho-\rho_{+}}{\rho-\rho_{-}}}\,\beta_{-}\partial_{x}\right] \,,
\ea\ee
or, in the $\sla_{\left(+\right)}\oplus\sla_{\left(-\right)}$ basis,
\be\ba\label{eq:LoveGeneratorsp1RmSL2R2}
	\tilde{L}_0^{\left(\sigma\right)} &= -\frac{\beta_{-}}{2}\left(\partial_{t}+\sigma\partial_{x}\right) \,, \\
	\tilde{L}_{\pm1}^{\left(\sigma\right)} &= e^{\pm\left(t+\sigma x\right)/\beta_{-}}\left[\mp\sqrt{\Delta}\,\partial_{\rho} + \sqrt{\frac{\rho-\rho_{-}}{\rho-\rho_{+}}}\,\frac{\beta_{-}}{2}\partial_{t} + \sqrt{\frac{\rho-\rho_{+}}{\rho-\rho_{-}}}\,\frac{\beta_{-}}{2}\sigma\partial_{x}\right] \,,
\ea\ee
with $\sigma\in\left\{-,+\right\}$, while the corresponding $SO\left(2,2;\mathbb{R}\right)$ Casimir becomes
\be\label{eq:SO22NIHCasimir}
	\tilde{\mathcal{C}}_2^{SO\left(2,2;\mathbb{R}\right)_{\text{NIH}}} = \partial_{\rho}\,\Delta\,\partial_{\rho} + \frac{\rho_{+}-\rho_{-}}{4}\beta_{-}^2\left(-\frac{1}{\Delta_{+}}\,\partial_{t}^2 + \frac{1}{\Delta_{-}}\,\partial_{x}^2\right) \,.
\ee
Working in advanced or retarded null coordinates, one can then see that all generators are regular at both the future and past inner horizons, but are singular at the outer horizon.

As with the $p=0$ case of black holes, we now consider the NNHE limit, consisting of taking $r_{-}\rightarrow r_{+}$, while keeping the inverse surface gravity at the event horizon finite, $\beta_{+} = \text{finite}$, plus remembering to keep the function $Z = \frac{f_{+}}{f_{-}}$ fixed. We find
\be\label{eq:NNHEgeometryp1}
	ds_{\text{NNHE}}^2 = -\frac{\Delta_{+}}{\rho_{+}}dt^2 + \frac{\Delta_{-}}{\rho_{+}}dx^2 + \frac{r_{+}^2}{\hat{d}^2}\left[\frac{d\rho^2}{\Delta} + \hat{d}^2\,d\Omega_{\hat{d}+1}^2\right] \,.
\ee
The corresponding $\text{AdS}_3$ Killing vectors are given by
\be\ba\label{eq:LoveGeneratorsp1RpSO22}
	L_0 &= -\beta_{+}\,\partial_{t} \,,\quad L_{01} = +\beta_{+}\partial_{x} \,, \\
	L_{\pm1,0} &= e^{\pm t/\beta_{+}}\left[\cosh\frac{x}{\beta_{+}}\left(\mp2\sqrt{\Delta}\,\partial_{\rho} + \sqrt{\frac{\rho-\rho_{-}}{\rho-\rho_{+}}}\,\beta_{+}\partial_{t}\right) \pm \sinh\frac{x}{\beta_{+}}\sqrt{\frac{\rho-\rho_{+}}{\rho-\rho_{-}}}\,\beta_{+}\partial_{x}\right] \,, \\
	L_{\pm1,1} &= e^{\pm t/\beta_{+}}\left[\sinh\frac{x}{\beta_{+}}\left(\mp2\sqrt{\Delta}\,\partial_{\rho} + \sqrt{\frac{\rho-\rho_{-}}{\rho-\rho_{+}}}\,\beta_{+}\partial_{t}\right) \pm \cosh\frac{x}{\beta_{+}}\sqrt{\frac{\rho-\rho_{+}}{\rho-\rho_{-}}}\,\beta_{+}\partial_{x}\right] \,,
\ea\ee
or, in the $\sla_{\left(+\right)}\oplus\sla_{\left(-\right)}$ basis,
\be\ba\label{eq:LoveGeneratorsp1RpSL2R2}
	L_0^{\left(\sigma\right)} &= -\frac{\beta_{+}}{2}\left(\partial_{t}+\sigma\partial_{x}\right) \,,\quad \sigma\in\left\{+,-\right\} \,, \\
	L_{\pm1}^{\left(\sigma\right)} &= e^{\pm\left(t+\sigma x\right)/\beta_{+}}\left[\mp\sqrt{\Delta}\,\partial_{\rho} + \sqrt{\frac{\rho-\rho_{-}}{\rho-\rho_{+}}}\,\frac{\beta_{+}}{2}\partial_{t} + \sqrt{\frac{\rho-\rho_{+}}{\rho-\rho_{-}}}\,\frac{\beta_{+}}{2}\sigma\partial_{x}\right] \,,
\ea\ee
and the corresponding $SO\left(2,2;\mathbb{R}\right)$ Casimir is given by
\be\label{eq:LoveCasimirRpSO22}
	\mathcal{C}_2^{SO\left(2,2;\mathbb{R}\right)_{\text{NNHE}}} = \partial_{\rho}\,\Delta\,\partial_{\rho} + \frac{\rho_{+}-\rho_{-}}{4}\beta_{+}^2\left(-\frac{1}{\Delta_{+}}\,\partial_{t}^2 + \frac{1}{\Delta_{-}}\,\partial_{x}^2\right) \,.
\ee
This way, we have obtained an enhanced near-horizon symmetry group whose generators satisfy the opposite regularity behaviors near the horizons: they are now regular at both the future and the past event horizons, but singular at the inner horizon, thus, permitting an investigation of geometric constraints on the near-horizon physics.

Consider then the problem of massless scalar perturbations $\psi$ of the NNHE geometry for the black string in Eq.~\eqref{eq:NNHEgeometryp1}. As with the case of the Reissner-Nordstr\"{o}m black hole, the problem is completely integrable, as can be seen from the complete separability of the linearized massless Klein-Gordon equation,
\be\ba\label{eq:KGnearNHEp1}
	\Box_{\text{NNHE}}\psi &= \frac{\hat{d}^2}{r_{+}^2}\left[\partial_{\rho}\,\Delta\,\partial_{\rho} + \frac{\rho_{+}-\rho_{-}}{4}\beta_{+}^2\left(-\frac{1}{\Delta_{+}}\,\partial_{t}^2 + \frac{1}{\Delta_{-}}\,\partial_{x}^2\right) + \frac{1}{\hat{d}^2}\Delta_{\mathbb{S}^{\hat{d}+1}}\right]\psi \\
	&= \frac{\hat{d}^2}{r_{+}^2}\left[\mathcal{C}_2^{SO\left(2,2;\mathbb{R}\right)_{\text{NNHE}}} + \frac{1}{\hat{d}^2}\Delta_{\mathbb{S}^{\hat{d}+1}}\right]\psi \,.
\ea\ee

Solving the angular problem by expanding into scalar spherical harmonic modes on $\mathbb{S}^{\hat{d}+1}$~\cite{Hui:2020xxx,Chodos:1983zi,Higuchi:1986wu},
\be
	\psi\left(t,x,\rho,\theta\right) = \sum_{\ell,\mathbf{m}}\psi_{\ell\mathbf{m}}\left(t,x,\rho\right)Y_{\ell\mathbf{m}}\left(\theta\right) \,,
\ee
the radial problem reduces to
\be
	\mathcal{C}_2^{SO\left(2,2;\mathbb{R}\right)_{\text{NNHE}}}\psi_{\ell\mathbf{m}} = \hat{\ell}\,(\hat{\ell}+1)\,\psi_{\ell\mathbf{m}} \,,\quad \hat{\ell} \equiv \frac{\ell}{\hat{d}} \,.
\ee

From the explicit form of the radial operator, the indicial powers at large $\rho$ can be seen to be
\be\label{eq:nearNHERCsp1}
	\psi_{\ell\mathbf{m}} \sim \mathcal{E}_{\ell\mathbf{m}}\left[\rho^{\hat{\ell}} + \frac{a_{\ell}}{\rho^{\hat{\ell}+1}}\right] \quad\text{as $\rho\rightarrow\infty$} \,,
\ee
which coincide with the boundary conditions imposed in asymptotically $\text{AdS}_3$ spacetimes for operators with scaling dimensions $\Delta_{+} = -\hat{\ell}$ and $\Delta_{-} = \hat{\ell}+1$. The non-normalizable, $\propto \rho^{-\Delta_{+}}=r^{\ell}$, mode then corresponds to a source, while the normalizable, $\propto \rho^{-\Delta_{-}} = r^{-\ell-\hat{d}}$, mode corresponds to the response, interpreting the integration ``constants'' $a_{\ell}$ as ``response coefficients''.

Comparing with the response problem of the non-extremal black string we encountered in Section~\ref{sec:MatchingNZ}, we then realize that these response coefficients from the perspective of the NNHE geometry are in fact exactly the same as the black string scalar Love numbers at leading order in the near-zone approximation. More interestingly, the Love $SO\left(2,2;\mathbb{R}\right)$ symmetry that we found emerging in the near-zone region is identical to the enhanced isometry subgroup of the NNHE geometry. Indeed, a simple inspection of the effective geometry associated with the employed near-zone truncation of the equations of motion in Eq.~\eqref{eq:NZsplitp} shows that it is identical with the NNHE geometry in Eq.~\eqref{eq:NNHEgeometryp1} or, alternatively, from the fact that the near-zone radial operator $\mathbb{O}_{\text{NZ}}$ in Eq.~\eqref{eq:RadialNZp} precisely coincides with the Casimir element, Eq.~\eqref{eq:LoveCasimirRpSO22}, of the enhanced $SO\left(2,2;\mathbb{R}\right)$ isometry subgroup of the NNHE geometry. Even more explicitly, the Love symmetry generators we have written down in Eq.~\eqref{eq:LoveS022p1} and Eq.~\eqref{eq:LoveSL2R2p1} are identical to the Killing vectors of the NNHE geometry in Eqs.~\eqref{eq:LoveGeneratorsp1RpSO22}-\eqref{eq:LoveGeneratorsp1RpSL2R2}. In conclusion, we have just demonstrated that
\be
	\mathfrak{so}\left(2,2;\mathbb{R}\right)_{\text{Love}} = \mathfrak{so}\left(2,2;\mathbb{R}\right)_{\text{NNHE}} \quad \text{for black strings} \,.
\ee

The upshot of this identification is that can now give an alternative interpretation of the states spanning the highest-weight multiplets of $\SL_{\left(+\right)\text{Love}}\times\SL_{\left(-\right)\text{Love}}$ from the perspective of the NNHE geometry. As with the Reissner-Nordstr\"{o}m black hole case, these states are precisely the QNMs of the NNHE black string geometry. Indeed, the spectrum of these states that we have extracted in Section~\ref{sec:LoveSymmetryp1},
\be
	\omega_{n\ell}^{\left(\pm\right)} \equiv \omega_{-\hat{\ell},n}^{\left(\pm\right)} = \pm k - i4\pi T_{H}(n-\hat{\ell}) \,,
\ee
are roots of the NNHE response coefficients $a_{\ell}$, i.e. the leading order scalar response coefficients of the non-extremal black string, see Eq.\eqref{eq:RCspBrane}. Compared to the case of the Reissner-Nordstr\"{o}m black hole, there is an extra factor of $2$ in the dumping parameter, in accordance with results reported in Ref.~\cite{Berti:2009kk}, namely, the QNMs frequencies of the BTZ black hole whose relation to the NNHE geometry has already been remarked in Section~\ref{sec:nearNHEGeometryp1}.

\subsection{NNHE region for black $\left(p\ge2\right)$-branes: inequivalence to AdS$_{p+2}$}
\label{sec:nearNHEGeometryp}

Inspired by the results around the $p=0$ black hole and the $p=1$ black string, the question now arises of whether $\text{SAdS}_{p+2}$ is isomorphic to the Poincar\'{e} patch of $\text{AdS}_{p+2}$. A minimum indication that this cannot be true revolves around the fact that the isometries of $\text{SAdS}_{p+2}$ and pure $\text{AdS}_{p+2}$ do not match for $p\ge2$. We will now demonstrate this non-isomorphism in the form of a ``no-go theorem''.

Consider a generic $\mathbb{R}^{1,p}\times SO(\hat{d}+2)$-symmetric $p$-brane geometry, which can always be brought to the form
\be
	ds^2 = H_{ab}\left(r\right)dx^{a}dx^{b} + \frac{dr^2}{f\left(r\right)} + R^2\left(r\right)d\Omega_{\hat{d}+1}^2 \,.
\ee
The question we wish to answer is whether the submanifold charted by the $\left(x^{a},r\right)$ coordinates can be maximally symmetric and admit a non-degenerate horizon. Let us focus to the worldvolume directions. The relevant components of the Riemann tensor are
\be
	R_{abcd} = -\frac{1}{4}f\left(H_{ac}^{\prime}H_{bd}^{\prime}-H_{ad}^{\prime}H_{bc}^{\prime}\right) \,.
\ee
One set of the conditions for the submanifold to be maximally symmetric is that
\be
	R_{abcd} = \lambda\left(H_{ac}H_{bd}-H_{ad}H_{bc}\right) \,,
\ee
with $\lambda$ the constant curvature of the submanifold. This gives rise to $\frac{d^2\left(d^2-1\right)}{12}$ independent constraints for the $\frac{d\left(d+1\right)}{2}$ longitudinal metric components $H_{ab}$. This system is not over-constrained only for $d\le2$, i.e. for black holes ($p=0$) and black strings ($p=1$). Let us now focus to the case relevant for our analysis, where the longitudinal isometry subgroup is enhanced to $\mathbb{R}_{t}\times ISO\left(d-1\right)$. Then, the generic line element can without loss of generality be rewritten as
\be
	ds^2 = F\left(r\right)\left[-Z\left(r\right)dt^2 + d\mathbf{x}^2\right] + \frac{dr^2}{H\left(r\right)Z\left(r\right)} + R^2\left(r\right)d\Omega_{\hat{d}+1}^2 \,,
\ee
where the functions $F\left(r\right)$, $Z\left(r\right)$, $H\left(r\right)$ and $R\left(r\right)$ are left arbitrary at the moment. We will investigate whether the maximally symmetric condition for the submanifold charted by the $\left(t,\mathbf{x},r\right)$ coordinates can have a non-degenerate horizon, i.e. whether $Z\left(r\right)$ can have a simple root at some radial position, while the other functions remain finite there. The conditions we want to impose are that $R_{\alpha\beta\gamma\delta} = \lambda\left(g_{\alpha\gamma}g_{\beta\delta}-g_{\alpha\delta}h_{\beta\gamma}\right)$, where the Greek indices from the beginning of the alphabet, $\alpha,\beta,\gamma,\delta\,\dots$, label $\left(t,\mathbf{x},r\right)$. For $d=1$ ($p=0$), the only condition that needs to be imposed is
\be
	R_{trtr} = \lambda\,g_{tt}g_{rr} \Rightarrow \frac{\left(FZ\right)^{\prime\prime}}{2FZ} + \frac{\left(FZ\right)^{\prime}}{2FZ}\left(\frac{\left(HZ\right)^{\prime}}{2HZ} - \frac{\left(FZ\right)^{\prime}}{2FZ}\right) = -\frac{\lambda}{HZ} \,.
\ee
From a simple residue analysis, we see that if $Z\left(r\right)$ has a root at some radial position $r=r_{\text{h}}$, then the orders of the poles on the two sides of the above equation are the same. For $d=2$ ($p=1$), the Riemann tensor along the worldvolume directions starts becoming non-trivial, with its single component imposing the further constraint
\be
	R_{titj} = \lambda\,g_{tt}g_{ij} \Rightarrow \frac{F^{\prime}}{2F}\frac{\left(FZ\right)^{\prime}}{2FZ} = -\frac{\lambda}{HZ} \,.
\ee
The orders of the poles on the two sides of this additional constraint are again of the same order, hence allowing the possibility of a non-degenerate horizon for the case of black strings. Last, for $d\ge3$ ($p\ge2$), the Riemann tensor induced on the longitudinal spatial slices starts becoming non-trivial and imposes one further constraint,
\be
	R_{ijkl} = \lambda \left(g_{ik}g_{jl}-g_{il}g_{jk}\right) \Rightarrow \frac{F^{\prime2}}{4F^2} = -\frac{\lambda}{HZ} \,.
\ee
We see then that $Z\left(r\right)$ cannot have a simple root at some $r=r_{\text{h}}$ while having $F\left(r_{\text{h}}\right) \ne 0$ and $H\left(r_{\text{h}}\right)\ne0$. As a result, we have a no-go theorem that shows that non-degenerate black $p$-branes with isometry subgroup $\mathbb{R}_{t}\times ISO\left(d-1\right)$ along the worldvolume directions cannot be mapped to a pure maximally symmetric spacetime, as opposed to what happens when $p=0$ or $p=1$, for which cases the SAdS$_{p+2}$ black holes/strings are diffeomoprhic to the Poincar\'{e} patch of pure AdS$_{p+2}$.


\section{Discussion and Open Questions}
\label{sec:concl}

In this work we have studied the Love numbers and 
Love symmetries of non-dilatonic black $p$-branes. 
Our results shed new light on the 
origin of the Love symmetry and 
its connection to the underlying background 
geometry. Our main results are summarized 
in Section~\ref{sec:summ}. Let us now list some 
remarks and open questions, and discuss future 
research directions suggested by our study.

\paragraph{Geometrization of Love symmetries for rotating black holes}
In this work, we have focused to spherically symmetric and static black holes. However, Love symmetries have been demonstrated to exist also for Kerr black holes~\cite{Charalambous:2021kcz,Charalambous:2022rre}, as well as for some higher-dimensional rotating black holes~\cite{Charalambous:2023jgq,Gray:2024qys}. For the four-dimensional case of Kerr black holes, its NNHE geometry is still SAdS$_2$, as was found in Refs.~\cite{Bredberg2009,Kapec:2019hro}, and is, hence, still isomorphic to a patch of pure AdS$_2$. Nevertheless, the Love symmetry generators that were introduced in Refs.~\cite{Charalambous:2021kcz,Charalambous:2022rre} are not Killing vectors of the NNHEK geometry of Ref.~\cite{Bredberg2009}. This may not come as a surprise though, once one realizes that the Love $\SL$ symmetry for the Kerr black is only one particular $\SL$ subalgebra of an infinitely extended $\SL\ltimes\hat{U}\left(1\right)$ structure~\cite{Charalambous:2021kcz,Charalambous:2022rre}, which naturally aligns with the spirit of the Kerr/CFT correspondence~\cite{Guica:2008mu,Lu:2008jk,Castro:2010fd,Krishnan:2010pv,Chen:2010zwa,Lowe:2011aa}\footnote{See also Refs.~\cite{Aminov:2020yma,Bonelli:2021uvf,Consoli:2022eey,Bautista:2023sdf,Arnaudo:2024rhv,Arnaudo:2024bbd} for recent gauge/gravity dictionaries that exploit CFT$_2$ tools for studying Kerr black hole perturbations.}. It is then expected that this infinite extension provides with $\SL$ subalgebras which precisely generate the NNHEK geometry of Ref.~\cite{Bredberg2009}. We leave this prospect, as well as its generalization to higher-dimensional rotating black holes and black strings, for future investigation.

\paragraph{Physicality of near-horizon modes of extremal black $p$-branes}
One interesting observation that emerges at extremality is the existence of physical near-horizon modes, i.e. of modes that live in the near-horizon AdS$_2$ throat but solve the full equations of motion. This corresponds to perturbations satisfying a light-like dispersion relation ($\omega^2=\mathbf{k}^2$) with respect to the extremal black $p$-brane worldvolume directions. Technically, this can be traced to the fact that the degenerate horizon is located at the roots of a metric function that is a perfect square in suitable coordinates, and appears to be a special characteristic of general-relativistic gravitational interactions. Instead, higher-derivative gravitational interactions would in general result in, for instance, static spherically symmetric extremal configurations where the degenerate horizon radius $r_{\text{h}}$ is the root of a metric component of the form $g^{rr} = f\left(r\right)\left(\frac{r-r_{\text{h}}}{r_{\text{h}}}\right)^2$, for some radial function $f\left(r\right)$ that does not vanish on the horizon; general-relativistic gravitation interactions imply that this $f\left(r\right)$ is such that there precisely exists a transformation $r=r\left(\rho\right)$ for which $g^{\rho\rho} = \frac{1}{r^2}\left(\rho-\rho_{\text{h}}\right)^2$. The fact that such physical near-horizon modes exist immediately implies that the extremal near-horizon throat is smoothly connected to the far-horizon region, and is to be combined with the recent investigations on the singular nature of extremal horizons when higher-derivative corrections are switched on~\cite{Horowitz:2022mly,Horowitz:2023xyl,Horowitz:2024dch,Horowitz:2024kcx}.

\paragraph{Geometrization of Love symmetries from accidental symmetries of extremal black holes}
Ultimately, the aforementioned ``quadratic polynomial'' behavior of the metric discriminant function is responsible for the existence of near-zone symmetries, such as the Love symmetry, in the non-extremal configuration as well~\cite{Charalambous:2024tdj}. In this work, we have attributed this geometrization of near-zone symmetries to the fact that they are isometries of the NNHE geometry of the non-dilatonic black hole/string configuration, as prescribed in Section~\ref{sec:nearNHEAdS2} and Section~\ref{sec:nearNHEGeometryp1}. For the case of black holes, it was recently remarked that four-dimensional general-relativistic extremal black holes are equipped with an ``accidental symmetry''~\cite{Hadar:2020kry,Porfyriadis:2021psx,deCesare:2024csp,Banerjee:2024zix} whose action on the NHE geometry precisely transforms it into the NNHE geometry. In this spirit, one may infer the existence of the non-extremal Love symmetries as a remnant of the extremal near-horizon throat isometries, propagated at non-extremality via the accidental symmetries of Refs.~\cite{Porfyriadis:2021psx,Banerjee:2024zix}. It would be interesting to better understand this connection and seek its extension to higher-dimensional black holes/strings.

\paragraph{Vanishing of extremal $p$-brane Love numbers}
In Section~\ref{sec:ExtremalLNs}, we have computed the scalar Love numbers of extremal black $p$-branes. As opposed to the non-extremal situation, the static and homogeneous scalar Love numbers of the extremal configuration vanish for \textit{any} orbital number $\ell$ of the perturbation and worldvolume and spacetime dimensionalities $p$ and $D$; in fact, this result remains true for any perturbations satisfying the light-like dispersion relation, $\omega^2=\mathbf{k}^2$. This remarkably rigid behavior of extremal black $p$-branes is reminiscent of the extremal black hole Meissner effect~\cite{King:1975,Bicak:1985,Bicak:2006hs}, see also Ref.~\cite{Giribet:2023qhz}, and proposes its according extension: the extremal black $p$-brane near-horizon throat expels external fields that follow a light-like dispersion relation. Regardless, the vanishing of the extremal black $p$-brane Love numbers suggests a holographic interpretation. A primary step towards this is to obtain a better understanding of the symmetry structure of extremal geometries. Related to this, let it be noted that the vanishing of the extremal black hole Love numbers in four spacetime dimensions was recently reinterpreted from discrete conformal symmetries of the equations of motion~\cite{Kehagias:2024yzn}, notably, the discrete conformal isometry of the four-dimensional extremal Reissner-Nordstr\"{o}m black, also known as the Couch-Torrence inversion~\cite{Couch:1984}, which maps the horizon onto null infinity and vice versa.

\noindent\textit{Acknowledgments ---} We are grateful to Wentao Cui,
Walter Goldberger, 
and 
Hong Liu
for useful
discussions. 
P.C. is supported by the European Research Council (ERC) Project 101076737 -- CeleBH. 
Views and opinions expressed are however those of the author only and do not necessarily reflect those of the European Union or the European Research Council. Neither the European Union nor the granting authority can be held responsible for them.
SD is supported in part by the NSF grant PHY-2210349, and by the IBM Einstein Fellow Fund at the IAS. 
\newpage
\appendix

\bibliographystyle{JHEP}
\bibliography{short_new}

\providecommand{\href}[2]{#2}\begingroup\raggedright\begin{thebibliography}{100}

\bibitem{LIGOScientific:2016aoc}
{\scshape LIGO Scientific, Virgo} collaboration, B.~P. Abbott et~al.,
  \emph{{Observation of Gravitational Waves from a Binary Black Hole Merger}},
  \href{https://doi.org/10.1103/PhysRevLett.116.061102}{\emph{Phys. Rev. Lett.}
  {\bfseries 116} (2016) 061102}
  [\href{https://arxiv.org/abs/1602.03837}{{\ttfamily 1602.03837}}].

\bibitem{LIGOScientific:2017vwq}
{\scshape LIGO Scientific, Virgo} collaboration, B.~P. Abbott et~al.,
  \emph{{GW170817: Observation of Gravitational Waves from a Binary Neutron
  Star Inspiral}},
  \href{https://doi.org/10.1103/PhysRevLett.119.161101}{\emph{Phys. Rev. Lett.}
  {\bfseries 119} (2017) 161101}
  [\href{https://arxiv.org/abs/1710.05832}{{\ttfamily 1710.05832}}].

\bibitem{Chatziioannou2020}
K.~Chatziioannou, \emph{Neutron-star tidal deformability and equation-of-state
  constraints}, \href{https://doi.org/10.1007/s10714-020-02754-3}{\emph{General
  Relativity and Gravitation} {\bfseries 52} (2020) 109}.

\bibitem{Chia:2024bwc}
H.~S. Chia, Z.~Zhou and M.~M. Ivanov, \emph{{Bring the Heat: Tidal Heating
  Constraints for Black Holes and Exotic Compact Objects from the
  LIGO-Virgo-KAGRA Data}},  \href{https://arxiv.org/abs/2404.14641}{{\ttfamily
  2404.14641}}.

\bibitem{Cardoso:2017cfl}
V.~Cardoso, E.~Franzin, A.~Maselli, P.~Pani and G.~Raposo, \emph{{Testing
  strong-field gravity with tidal Love numbers}},
  \href{https://doi.org/10.1103/PhysRevD.95.084014}{\emph{Phys. Rev. D}
  {\bfseries 95} (2017) 084014}
  [\href{https://arxiv.org/abs/1701.01116}{{\ttfamily 1701.01116}}].

\bibitem{Franzin:2017mtq}
E.~Franzin, V.~Cardoso, P.~Pani and G.~Raposo, \emph{{Testing strong gravity
  with gravitational waves and Love numbers}},
  \href{https://doi.org/10.1088/1742-6596/841/1/012035}{\emph{J. Phys. Conf.
  Ser.} {\bfseries 841} (2017) 012035}.

\bibitem{Cardoso:2018ptl}
V.~Cardoso, M.~Kimura, A.~Maselli and L.~Senatore, \emph{{Black Holes in an
  Effective Field Theory Extension of General Relativity}},
  \href{https://doi.org/10.1103/PhysRevLett.121.251105}{\emph{Phys. Rev. Lett.}
  {\bfseries 121} (2018) 251105}
  [\href{https://arxiv.org/abs/1808.08962}{{\ttfamily 1808.08962}}].

\bibitem{Katagiri:2023yzm}
T.~Katagiri, H.~Nakano and K.~Omukai, \emph{{Stability of relativistic tidal
  response against small potential modification}},
  \href{https://doi.org/10.1103/PhysRevD.108.084049}{\emph{Phys. Rev. D}
  {\bfseries 108} (2023) 084049}
  [\href{https://arxiv.org/abs/2304.04551}{{\ttfamily 2304.04551}}].

\bibitem{Xie:2022brn}
Y.~Xie, D.~Chatterjee, G.~Holder, D.~E. Holz, S.~Perkins, K.~Yagi et~al.,
  \emph{{Breaking bad degeneracies with Love relations: Improving
  gravitational-wave measurements through universal relations}},
  \href{https://doi.org/10.1103/PhysRevD.107.043010}{\emph{Phys. Rev. D}
  {\bfseries 107} (2023) 043010}
  [\href{https://arxiv.org/abs/2210.09386}{{\ttfamily 2210.09386}}].

\bibitem{Yagi:2013bca}
K.~Yagi and N.~Yunes, \emph{{I-Love-Q}},
  \href{https://doi.org/10.1126/science.1236462}{\emph{Science} {\bfseries 341}
  (2013) 365} [\href{https://arxiv.org/abs/1302.4499}{{\ttfamily 1302.4499}}].

\bibitem{Yagi:2013awa}
K.~Yagi and N.~Yunes, \emph{{I-Love-Q Relations in Neutron Stars and their
  Applications to Astrophysics, Gravitational Waves and Fundamental Physics}},
  \href{https://doi.org/10.1103/PhysRevD.88.023009}{\emph{Phys. Rev. D}
  {\bfseries 88} (2013) 023009}
  [\href{https://arxiv.org/abs/1303.1528}{{\ttfamily 1303.1528}}].

\bibitem{Pani:2015tga}
P.~Pani, \emph{{I-Love-Q relations for gravastars and the approach to the
  black-hole limit}},
  \href{https://doi.org/10.1103/PhysRevD.95.049902}{\emph{Phys. Rev. D}
  {\bfseries 92} (2015) 124030}
  [\href{https://arxiv.org/abs/1506.06050}{{\ttfamily 1506.06050}}].

\bibitem{Yagi:2015pkc}
K.~Yagi and N.~Yunes, \emph{{Binary Love Relations}},
  \href{https://doi.org/10.1088/0264-9381/33/13/13LT01}{\emph{Class. Quant.
  Grav.} {\bfseries 33} (2016) 13LT01}
  [\href{https://arxiv.org/abs/1512.02639}{{\ttfamily 1512.02639}}].

\bibitem{Uchikata:2016qku}
N.~Uchikata, S.~Yoshida and P.~Pani, \emph{{Tidal deformability and I-Love-Q
  relations for gravastars with polytropic thin shells}},
  \href{https://doi.org/10.1103/PhysRevD.94.064015}{\emph{Phys. Rev. D}
  {\bfseries 94} (2016) 064015}
  [\href{https://arxiv.org/abs/1607.03593}{{\ttfamily 1607.03593}}].

\bibitem{Yagi:2016qmr}
K.~Yagi and N.~Yunes, \emph{{Approximate Universal Relations among Tidal
  Parameters for Neutron Star Binaries}},
  \href{https://doi.org/10.1088/1361-6382/34/1/015006}{\emph{Class. Quant.
  Grav.} {\bfseries 34} (2017) 015006}
  [\href{https://arxiv.org/abs/1608.06187}{{\ttfamily 1608.06187}}].

\bibitem{Fang:2005qq}
H.~Fang and G.~Lovelace, \emph{{Tidal coupling of a Schwarzschild black hole
  and circularly orbiting moon}},
  \href{https://doi.org/10.1103/PhysRevD.72.124016}{\emph{Phys. Rev. D}
  {\bfseries 72} (2005) 124016}
  [\href{https://arxiv.org/abs/gr-qc/0505156}{{\ttfamily gr-qc/0505156}}].

\bibitem{Damour:2009vw}
T.~Damour and A.~Nagar, \emph{{Relativistic tidal properties of neutron
  stars}}, \href{https://doi.org/10.1103/PhysRevD.80.084035}{\emph{Phys. Rev.
  D} {\bfseries 80} (2009) 084035}
  [\href{https://arxiv.org/abs/0906.0096}{{\ttfamily 0906.0096}}].

\bibitem{Binnington:2009bb}
T.~Binnington and E.~Poisson, \emph{{Relativistic theory of tidal Love
  numbers}}, \href{https://doi.org/10.1103/PhysRevD.80.084018}{\emph{Phys. Rev.
  D} {\bfseries 80} (2009) 084018}
  [\href{https://arxiv.org/abs/0906.1366}{{\ttfamily 0906.1366}}].

\bibitem{Kol:2011vg}
B.~Kol and M.~Smolkin, \emph{{Black hole stereotyping: Induced gravito-static
  polarization}}, \href{https://doi.org/10.1007/JHEP02(2012)010}{\emph{JHEP}
  {\bfseries 02} (2012) 010} [\href{https://arxiv.org/abs/1110.3764}{{\ttfamily
  1110.3764}}].

\bibitem{LeTiec:2020spy}
A.~Le~Tiec and M.~Casals, \emph{{Spinning Black Holes Fall in Love}},
  \href{https://doi.org/10.1103/PhysRevLett.126.131102}{\emph{Phys. Rev. Lett.}
  {\bfseries 126} (2021) 131102}
  [\href{https://arxiv.org/abs/2007.00214}{{\ttfamily 2007.00214}}].

\bibitem{Chia:2020yla}
H.~S. Chia, \emph{{Tidal deformation and dissipation of rotating black holes}},
  \href{https://doi.org/10.1103/PhysRevD.104.024013}{\emph{Phys. Rev. D}
  {\bfseries 104} (2021) 024013}
  [\href{https://arxiv.org/abs/2010.07300}{{\ttfamily 2010.07300}}].

\bibitem{LeTiec:2020bos}
A.~Le~Tiec, M.~Casals and E.~Franzin, \emph{{Tidal Love Numbers of Kerr Black
  Holes}}, \href{https://doi.org/10.1103/PhysRevD.103.084021}{\emph{Phys. Rev.
  D} {\bfseries 103} (2021) 084021}
  [\href{https://arxiv.org/abs/2010.15795}{{\ttfamily 2010.15795}}].

\bibitem{Charalambous:2021mea}
P.~Charalambous, S.~Dubovsky and M.~M. Ivanov, \emph{{On the Vanishing of Love
  Numbers for Kerr Black Holes}},
  \href{https://doi.org/10.1007/JHEP05(2021)038}{\emph{JHEP} {\bfseries 05}
  (2021) 038} [\href{https://arxiv.org/abs/2102.08917}{{\ttfamily
  2102.08917}}].

\bibitem{DeLuca:2021ite}
V.~De~Luca and P.~Pani, \emph{{Tidal deformability of dressed black holes and
  tests of ultralight bosons in extended mass ranges}},
  \href{https://doi.org/10.1088/1475-7516/2021/08/032}{\emph{JCAP} {\bfseries
  08} (2021) 032} [\href{https://arxiv.org/abs/2106.14428}{{\ttfamily
  2106.14428}}].

\bibitem{DeLuca:2022xlz}
V.~De~Luca, A.~Maselli and P.~Pani, \emph{{Modeling frequency-dependent tidal
  deformability for environmental black hole mergers}},
  \href{https://doi.org/10.1103/PhysRevD.107.044058}{\emph{Phys. Rev. D}
  {\bfseries 107} (2023) 044058}
  [\href{https://arxiv.org/abs/2212.03343}{{\ttfamily 2212.03343}}].

\bibitem{Pani:2019cyc}
P.~Pani and A.~Maselli, \emph{{Love in Extrema Ratio}},
  \href{https://doi.org/10.1142/S0218271819440012}{\emph{Int. J. Mod. Phys. D}
  {\bfseries 28} (2019) 1944001}
  [\href{https://arxiv.org/abs/1905.03947}{{\ttfamily 1905.03947}}].

\bibitem{Datta:2019epe}
S.~Datta, R.~Brito, S.~Bose, P.~Pani and S.~A. Hughes, \emph{{Tidal heating as
  a discriminator for horizons in extreme mass ratio inspirals}},
  \href{https://doi.org/10.1103/PhysRevD.101.044004}{\emph{Phys. Rev. D}
  {\bfseries 101} (2020) 044004}
  [\href{https://arxiv.org/abs/1910.07841}{{\ttfamily 1910.07841}}].

\bibitem{Chakrabarti:2013lua}
S.~Chakrabarti, T.~Delsate and J.~Steinhoff, \emph{{New perspectives on neutron
  star and black hole spectroscopy and dynamic tides}},
  \href{https://arxiv.org/abs/1304.2228}{{\ttfamily 1304.2228}}.

\bibitem{Saketh:2023bul}
M.~V.~S. Saketh, Z.~Zhou and M.~M. Ivanov, \emph{{Dynamical tidal response of
  Kerr black holes from scattering amplitudes}},
  \href{https://doi.org/10.1103/PhysRevD.109.064058}{\emph{Phys. Rev. D}
  {\bfseries 109} (2024) 064058}
  [\href{https://arxiv.org/abs/2307.10391}{{\ttfamily 2307.10391}}].

\bibitem{Ivanov:2024sds}
M.~M. Ivanov, Y.-Z. Li, J.~Parra-Martinez and Z.~Zhou, \emph{{Gravitational
  Raman Scattering in Effective Field Theory: A Scalar Tidal Matching at
  O(G3)}}, \href{https://doi.org/10.1103/PhysRevLett.132.131401}{\emph{Phys.
  Rev. Lett.} {\bfseries 132} (2024) 131401}
  [\href{https://arxiv.org/abs/2401.08752}{{\ttfamily 2401.08752}}].

\bibitem{Gurlebeck:2015xpa}
N.~G\"urlebeck, \emph{{No-hair theorem for Black Holes in Astrophysical
  Environments}},
  \href{https://doi.org/10.1103/PhysRevLett.114.151102}{\emph{Phys. Rev. Lett.}
  {\bfseries 114} (2015) 151102}
  [\href{https://arxiv.org/abs/1503.03240}{{\ttfamily 1503.03240}}].

\bibitem{DeLuca:2023mio}
V.~De~Luca, J.~Khoury and S.~S.~C. Wong, \emph{{Nonlinearities in the tidal
  Love numbers of black holes}},
  \href{https://doi.org/10.1103/PhysRevD.108.024048}{\emph{Phys. Rev. D}
  {\bfseries 108} (2023) 024048}
  [\href{https://arxiv.org/abs/2305.14444}{{\ttfamily 2305.14444}}].

\bibitem{Riva:2023rcm}
M.~M. Riva, L.~Santoni, N.~Savi\'c and F.~Vernizzi, \emph{{Vanishing of
  nonlinear tidal Love numbers of Schwarzschild black holes}},
  \href{https://doi.org/10.1016/j.physletb.2024.138710}{\emph{Phys. Lett. B}
  {\bfseries 854} (2024) 138710}
  [\href{https://arxiv.org/abs/2312.05065}{{\ttfamily 2312.05065}}].

\bibitem{Combaluzier-Szteinsznaider:2024sgb}
O.~Combaluzier-Szteinsznaider, L.~Hui, L.~Santoni, A.~R. Solomon and S.~S.~C.
  Wong, \emph{{Symmetries of Vanishing Nonlinear Love Numbers of Schwarzschild
  Black Holes}},  \href{https://arxiv.org/abs/2410.10952}{{\ttfamily
  2410.10952}}.

\bibitem{Kehagias:2024rtz}
A.~Kehagias and A.~Riotto, \emph{{Black Holes in a Gravitational Field: The
  Non-linear Static Love Number of Schwarzschild Black Holes Vanishes}},
  \href{https://arxiv.org/abs/2410.11014}{{\ttfamily 2410.11014}}.

\bibitem{Iteanu:2024dvx}
S.~Iteanu, M.~M. Riva, L.~Santoni, N.~Savi\'c and F.~Vernizzi, \emph{{Vanishing
  of Quadratic Love Numbers of Schwarzschild Black Holes}},
  \href{https://arxiv.org/abs/2410.03542}{{\ttfamily 2410.03542}}.

\bibitem{Goldberger:2004jt}
W.~D. Goldberger and I.~Z. Rothstein, \emph{{An Effective field theory of
  gravity for extended objects}},
  \href{https://doi.org/10.1103/PhysRevD.73.104029}{\emph{Phys. Rev. D}
  {\bfseries 73} (2006) 104029}
  [\href{https://arxiv.org/abs/hep-th/0409156}{{\ttfamily hep-th/0409156}}].

\bibitem{Goldberger:2005cd}
W.~D. Goldberger and I.~Z. Rothstein, \emph{{Dissipative effects in the
  worldline approach to black hole dynamics}},
  \href{https://doi.org/10.1103/PhysRevD.73.104030}{\emph{Phys. Rev. D}
  {\bfseries 73} (2006) 104030}
  [\href{https://arxiv.org/abs/hep-th/0511133}{{\ttfamily hep-th/0511133}}].

\bibitem{Porto:2005ac}
R.~A. Porto, \emph{{Post-Newtonian corrections to the motion of spinning bodies
  in NRGR}}, \href{https://doi.org/10.1103/PhysRevD.73.104031}{\emph{Phys. Rev.
  D} {\bfseries 73} (2006) 104031}
  [\href{https://arxiv.org/abs/gr-qc/0511061}{{\ttfamily gr-qc/0511061}}].

\bibitem{Levi:2015msa}
M.~Levi and J.~Steinhoff, \emph{{Spinning gravitating objects in the effective
  field theory in the post-Newtonian scheme}},
  \href{https://doi.org/10.1007/JHEP09(2015)219}{\emph{JHEP} {\bfseries 09}
  (2015) 219} [\href{https://arxiv.org/abs/1501.04956}{{\ttfamily
  1501.04956}}].

\bibitem{Porto:2016pyg}
R.~A. Porto, \emph{{The effective field theorist\textquoteright{}s approach to
  gravitational dynamics}},
  \href{https://doi.org/10.1016/j.physrep.2016.04.003}{\emph{Phys. Rept.}
  {\bfseries 633} (2016) 1} [\href{https://arxiv.org/abs/1601.04914}{{\ttfamily
  1601.04914}}].

\bibitem{Levi:2018nxp}
M.~Levi, \emph{{Effective Field Theories of Post-Newtonian Gravity: A
  comprehensive review}},
  \href{https://doi.org/10.1088/1361-6633/ab12bc}{\emph{Rept. Prog. Phys.}
  {\bfseries 83} (2020) 075901}
  [\href{https://arxiv.org/abs/1807.01699}{{\ttfamily 1807.01699}}].

\bibitem{Charalambous:2021kcz}
P.~Charalambous, S.~Dubovsky and M.~M. Ivanov, \emph{{Hidden Symmetry of
  Vanishing Love Numbers}},
  \href{https://doi.org/10.1103/PhysRevLett.127.101101}{\emph{Phys. Rev. Lett.}
  {\bfseries 127} (2021) 101101}
  [\href{https://arxiv.org/abs/2103.01234}{{\ttfamily 2103.01234}}].

\bibitem{Charalambous:2022rre}
P.~Charalambous, S.~Dubovsky and M.~M. Ivanov, \emph{{Love symmetry}},
  \href{https://doi.org/10.1007/JHEP10(2022)175}{\emph{JHEP} {\bfseries 10}
  (2022) 175} [\href{https://arxiv.org/abs/2209.02091}{{\ttfamily
  2209.02091}}].

\bibitem{Hui:2021vcv}
L.~Hui, A.~Joyce, R.~Penco, L.~Santoni and A.~R. Solomon, \emph{{Ladder
  symmetries of black holes. Implications for Love numbers and no-hair
  theorems}}, \href{https://doi.org/10.1088/1475-7516/2022/01/032}{\emph{JCAP}
  {\bfseries 01} (2022) 032}
  [\href{https://arxiv.org/abs/2105.01069}{{\ttfamily 2105.01069}}].

\bibitem{Hui:2022vbh}
L.~Hui, A.~Joyce, R.~Penco, L.~Santoni and A.~R. Solomon, \emph{{Near-zone
  symmetries of Kerr black holes}},
  \href{https://doi.org/10.1007/JHEP09(2022)049}{\emph{JHEP} {\bfseries 09}
  (2022) 049} [\href{https://arxiv.org/abs/2203.08832}{{\ttfamily
  2203.08832}}].

\bibitem{Achour:2021dtj}
J.~B. Achour and E.~R. Livine, \emph{{Symmetries and conformal bridge in
  Schwarschild-(A)dS black hole mechanics}},
  \href{https://doi.org/10.1007/JHEP12(2021)152}{\emph{JHEP} {\bfseries 12}
  (2021) 152} [\href{https://arxiv.org/abs/2110.01455}{{\ttfamily
  2110.01455}}].

\bibitem{BenAchour:2022uqo}
J.~Ben~Achour, E.~R. Livine, S.~Mukohyama and J.-P. Uzan, \emph{{Hidden
  symmetry of the static response of black holes: applications to Love
  numbers}}, \href{https://doi.org/10.1007/JHEP07(2022)112}{\emph{JHEP}
  {\bfseries 07} (2022) 112}
  [\href{https://arxiv.org/abs/2202.12828}{{\ttfamily 2202.12828}}].

\bibitem{BenAchour:2022fif}
J.~Ben~Achour, E.~R. Livine, D.~Oriti and G.~Piani, \emph{{Schr\"odinger
  Symmetry in Gravitational Mini-Superspaces}},
  \href{https://doi.org/10.3390/universe9120503}{\emph{Universe} {\bfseries 9}
  (2023) 503} [\href{https://arxiv.org/abs/2207.07312}{{\ttfamily
  2207.07312}}].

\bibitem{BenAchour:2023dgj}
J.~Ben~Achour, E.~R. Livine and D.~Oriti, \emph{{Schr\"odinger symmetry of
  Schwarzschild-(A)dS black hole mechanics}},
  \href{https://doi.org/10.1103/PhysRevD.108.104028}{\emph{Phys. Rev. D}
  {\bfseries 108} (2023) 104028}
  [\href{https://arxiv.org/abs/2302.07644}{{\ttfamily 2302.07644}}].

\bibitem{Cvetic:2011dn}
M.~Cvetic and F.~Larsen, \emph{{Conformal Symmetry for Black Holes in Four
  Dimensions}}, \href{https://doi.org/10.1007/JHEP09(2012)076}{\emph{JHEP}
  {\bfseries 09} (2012) 076} [\href{https://arxiv.org/abs/1112.4846}{{\ttfamily
  1112.4846}}].

\bibitem{Cvetic:2011hp}
M.~Cvetic and F.~Larsen, \emph{{Conformal Symmetry for General Black Holes}},
  \href{https://doi.org/10.1007/JHEP02(2012)122}{\emph{JHEP} {\bfseries 02}
  (2012) 122} [\href{https://arxiv.org/abs/1106.3341}{{\ttfamily 1106.3341}}].

\bibitem{Cvetic:2012tr}
M.~Cvetic and G.~W. Gibbons, \emph{{Conformal Symmetry of a Black Hole as a
  Scaling Limit: A Black Hole in an Asymptotically Conical Box}},
  \href{https://doi.org/10.1007/JHEP07(2012)014}{\emph{JHEP} {\bfseries 07}
  (2012) 014} [\href{https://arxiv.org/abs/1201.0601}{{\ttfamily 1201.0601}}].

\bibitem{Bardeen:1999px}
J.~M. Bardeen and G.~T. Horowitz, \emph{{The Extreme Kerr throat geometry: A
  Vacuum analog of
  ${\mathrm{AdS}}_{2}{\ifmmode\times\else\texttimes\fi{}\mathrm{S}}^{2}$}},
  \href{https://doi.org/10.1103/PhysRevD.60.104030}{\emph{Phys. Rev. D}
  {\bfseries 60} (1999) 104030}
  [\href{https://arxiv.org/abs/hep-th/9905099}{{\ttfamily hep-th/9905099}}].

\bibitem{Kunduri:2007vf}
H.~K. Kunduri, J.~Lucietti and H.~S. Reall, \emph{{Near-horizon symmetries of
  extremal black holes}},
  \href{https://doi.org/10.1088/0264-9381/24/16/012}{\emph{Class. Quant. Grav.}
  {\bfseries 24} (2007) 4169}
  [\href{https://arxiv.org/abs/0705.4214}{{\ttfamily 0705.4214}}].

\bibitem{Guevara:2023wlr}
A.~Guevara and U.~Kol, \emph{{Self Dual Black Holes as the Hydrogen Atom}},
  \href{https://arxiv.org/abs/2311.07933}{{\ttfamily 2311.07933}}.

\bibitem{Hui:2020xxx}
L.~Hui, A.~Joyce, R.~Penco, L.~Santoni and A.~R. Solomon, \emph{{Static
  response and Love numbers of Schwarzschild black holes}},
  \href{https://doi.org/10.1088/1475-7516/2021/04/052}{\emph{JCAP} {\bfseries
  04} (2021) 052} [\href{https://arxiv.org/abs/2010.00593}{{\ttfamily
  2010.00593}}].

\bibitem{Charalambous:2023jgq}
P.~Charalambous and M.~M. Ivanov, \emph{{Scalar Love numbers and Love
  symmetries of 5-dimensional Myers-Perry black holes}},
  \href{https://doi.org/10.1007/JHEP07(2023)222}{\emph{JHEP} {\bfseries 07}
  (2023) 222} [\href{https://arxiv.org/abs/2303.16036}{{\ttfamily
  2303.16036}}].

\bibitem{Rodriguez:2023xjd}
M.~J. Rodriguez, L.~Santoni, A.~R. Solomon and L.~F. Temoche, \emph{{Love
  numbers for rotating black holes in higher dimensions}},
  \href{https://doi.org/10.1103/PhysRevD.108.084011}{\emph{Phys. Rev. D}
  {\bfseries 108} (2023) 084011}
  [\href{https://arxiv.org/abs/2304.03743}{{\ttfamily 2304.03743}}].

\bibitem{Aharony:1999ti}
O.~Aharony, S.~S. Gubser, J.~M. Maldacena, H.~Ooguri and Y.~Oz, \emph{{Large N
  field theories, string theory and gravity}},
  \href{https://doi.org/10.1016/S0370-1573(99)00083-6}{\emph{Phys. Rept.}
  {\bfseries 323} (2000) 183}
  [\href{https://arxiv.org/abs/hep-th/9905111}{{\ttfamily hep-th/9905111}}].

\bibitem{Maldacena:1998uz}
J.~M. Maldacena, J.~Michelson and A.~Strominger, \emph{{Anti-de Sitter
  fragmentation}},
  \href{https://doi.org/10.1088/1126-6708/1999/02/011}{\emph{JHEP} {\bfseries
  02} (1999) 011} [\href{https://arxiv.org/abs/hep-th/9812073}{{\ttfamily
  hep-th/9812073}}].

\bibitem{Maldacena:1997re}
J.~M. Maldacena, \emph{{The Large N limit of superconformal field theories and
  supergravity}}, \href{https://doi.org/10.1023/A:1026654312961}{\emph{Adv.
  Theor. Math. Phys.} {\bfseries 2} (1998) 231}
  [\href{https://arxiv.org/abs/hep-th/9711200}{{\ttfamily hep-th/9711200}}].

\bibitem{Horowitz:1998pq}
G.~T. Horowitz and S.~F. Ross, \emph{{Possible resolution of black hole
  singularities from large N gauge theory}},
  \href{https://doi.org/10.1088/1126-6708/1998/04/015}{\emph{JHEP} {\bfseries
  04} (1998) 015} [\href{https://arxiv.org/abs/hep-th/9803085}{{\ttfamily
  hep-th/9803085}}].

\bibitem{Cadoni:1993rn}
M.~Cadoni and S.~Mignemi, \emph{{Classical and semiclassical properties of
  extremal black holes with dilaton and modulus fields}},
  \href{https://doi.org/10.1016/0550-3213(94)90644-0}{\emph{Nucl. Phys. B}
  {\bfseries 427} (1994) 669}
  [\href{https://arxiv.org/abs/hep-th/9312171}{{\ttfamily hep-th/9312171}}].

\bibitem{Cadoni:1994uf}
M.~Cadoni and S.~Mignemi, \emph{{Nonsingular four-dimensional black holes and
  the Jackiw-Teitelboim theory}},
  \href{https://doi.org/10.1103/PhysRevD.51.4319}{\emph{Phys. Rev. D}
  {\bfseries 51} (1995) 4319}
  [\href{https://arxiv.org/abs/hep-th/9410041}{{\ttfamily hep-th/9410041}}].

\bibitem{Bredberg2009}
I.~Bredberg, T.~Hartman, W.~Song and A.~Strominger, \emph{{Black Hole
  Superradiance From Kerr/CFT}},
  \href{https://doi.org/10.1007/JHEP04(2010)019}{\emph{JHEP} {\bfseries 04}
  (2010) 019} [\href{https://arxiv.org/abs/0907.3477}{{\ttfamily 0907.3477}}].

\bibitem{Hadar:2020kry}
S.~Hadar, A.~Lupsasca and A.~P. Porfyriadis, \emph{{Extreme Black Hole
  Anabasis}}, \href{https://doi.org/10.1007/JHEP03(2021)223}{\emph{JHEP}
  {\bfseries 03} (2021) 223}
  [\href{https://arxiv.org/abs/2012.06562}{{\ttfamily 2012.06562}}].

\bibitem{Porfyriadis:2021psx}
A.~P. Porfyriadis and G.~N. Remmen, \emph{{Large diffeomorphisms and accidental
  symmetry of the extremal horizon}},
  \href{https://doi.org/10.1007/JHEP03(2022)107}{\emph{JHEP} {\bfseries 03}
  (2022) 107} [\href{https://arxiv.org/abs/2112.13853}{{\ttfamily
  2112.13853}}].

\bibitem{deCesare:2024csp}
M.~de~Cesare, R.~Oliveri and A.~P. Porfyriadis, \emph{{Connecting Gravitational
  Perturbations: from Bertotti-Robinson to Extreme Reissner-Nordstrom}},
  \href{https://arxiv.org/abs/2410.23446}{{\ttfamily 2410.23446}}.

\bibitem{Banerjee:2024zix}
A.~Banerjee, A.~P. Porfyriadis and G.~N. Remmen, \emph{{Accidental Symmetry
  Near Extreme Spinning Black Holes}},
  \href{https://arxiv.org/abs/2412.19880}{{\ttfamily 2412.19880}}.

\bibitem{Chodos:1983zi}
A.~Chodos and E.~Myers, \emph{{Gravitational Contribution to the Casimir Energy
  in Kaluza-Klein Theories}},
  \href{https://doi.org/10.1016/0003-4916(84)90039-3}{\emph{Annals Phys.}
  {\bfseries 156} (1984) 412}.

\bibitem{Higuchi:1986wu}
A.~Higuchi, \emph{{Symmetric Tensor Spherical Harmonics on the $N$ Sphere and
  Their Application to the De Sitter Group SO($N$,1)}},
  \href{https://doi.org/10.1063/1.527513}{\emph{J. Math. Phys.} {\bfseries 28}
  (1987) 1553}.

\bibitem{Chen:2010ik}
B.~Chen and J.~Long, \emph{{Hidden Conformal Symmetry and Quasi-normal Modes}},
  \href{https://doi.org/10.1103/PhysRevD.82.126013}{\emph{Phys. Rev. D}
  {\bfseries 82} (2010) 126013}
  [\href{https://arxiv.org/abs/1009.1010}{{\ttfamily 1009.1010}}].

\bibitem{Starobinsky:1973aij}
A.~A. Starobinski{\v{i}}, \emph{{Amplification of waves during reflection from
  a rotating ``black hole''}}, {\emph{Sov. Phys. JETP} {\bfseries 37} (1973)
  28}.

\bibitem{Starobinskil:1974nkd}
A.~A. Starobinski{\v{i}} and S.~M. Churilov, \emph{{Amplification of
  electromagnetic and gravitational waves scattered by a rotating ``black
  hole''}}, {\emph{Sov. Phys. JETP} {\bfseries 65} (1974) 1}.

\bibitem{Maldacena:1997ih}
J.~M. Maldacena and A.~Strominger, \emph{{Universal low-energy dynamics for
  rotating black holes}},
  \href{https://doi.org/10.1103/PhysRevD.56.4975}{\emph{Phys. Rev. D}
  {\bfseries 56} (1997) 4975}
  [\href{https://arxiv.org/abs/hep-th/9702015}{{\ttfamily hep-th/9702015}}].

\bibitem{Castro:2010fd}
A.~Castro, A.~Maloney and A.~Strominger, \emph{{Hidden Conformal Symmetry of
  the Kerr Black Hole}},
  \href{https://doi.org/10.1103/PhysRevD.82.024008}{\emph{Phys. Rev. D}
  {\bfseries 82} (2010) 024008}
  [\href{https://arxiv.org/abs/1004.0996}{{\ttfamily 1004.0996}}].

\bibitem{Creci:2021rkz}
G.~Creci, T.~Hinderer and J.~Steinhoff, \emph{{Tidal response from scattering
  and the role of analytic continuation}},
  \href{https://doi.org/10.1103/PhysRevD.104.124061}{\emph{Phys. Rev. D}
  {\bfseries 104} (2021) 124061}
  [\href{https://arxiv.org/abs/2108.03385}{{\ttfamily 2108.03385}}].

\bibitem{Ivanov:2022hlo}
M.~M. Ivanov and Z.~Zhou, \emph{{Revisiting the matching of black hole tidal
  responses: A systematic study of relativistic and logarithmic corrections}},
  \href{https://doi.org/10.1103/PhysRevD.107.084030}{\emph{Phys. Rev. D}
  {\bfseries 107} (2023) 084030}
  [\href{https://arxiv.org/abs/2208.08459}{{\ttfamily 2208.08459}}].

\bibitem{tHooft:1979rat}
G.~'t~Hooft, \emph{{Naturalness, chiral symmetry, and spontaneous chiral
  symmetry breaking}},
  \href{https://doi.org/10.1007/978-1-4684-7571-5_9}{\emph{NATO Sci. Ser. B}
  {\bfseries 59} (1980) 135}.

\bibitem{Porto:2016zng}
R.~A. Porto, \emph{{The Tune of Love and the Nature(ness) of Spacetime}},
  \href{https://doi.org/10.1002/prop.201600064}{\emph{Fortsch. Phys.}
  {\bfseries 64} (2016) 723}
  [\href{https://arxiv.org/abs/1606.08895}{{\ttfamily 1606.08895}}].

\bibitem{Charalambous:2024tdj}
P.~Charalambous, \emph{{Love numbers and Love symmetries for p-form and
  gravitational perturbations of higher-dimensional spherically symmetric black
  holes}}, \href{https://doi.org/10.1007/JHEP04(2024)122}{\emph{JHEP}
  {\bfseries 04} (2024) 122}
  [\href{https://arxiv.org/abs/2402.07574}{{\ttfamily 2402.07574}}].

\bibitem{Ivanov:2022qqt}
M.~M. Ivanov and Z.~Zhou, \emph{{Vanishing of Black Hole Tidal Love Numbers
  from Scattering Amplitudes}},
  \href{https://doi.org/10.1103/PhysRevLett.130.091403}{\emph{Phys. Rev. Lett.}
  {\bfseries 130} (2023) 091403}
  [\href{https://arxiv.org/abs/2209.14324}{{\ttfamily 2209.14324}}].

\bibitem{Hod:2013fea}
S.~Hod, \emph{{Purely imaginary polar resonances of rapidly-rotating Kerr black
  holes}}, \href{https://doi.org/10.1103/PhysRevD.88.084018}{\emph{Phys. Rev.
  D} {\bfseries 88} (2013) 084018}
  [\href{https://arxiv.org/abs/1311.3007}{{\ttfamily 1311.3007}}].

\bibitem{Cook:2016fge}
G.~B. Cook and M.~Zalutskiy, \emph{{Purely imaginary quasinormal modes of the
  Kerr geometry}},
  \href{https://doi.org/10.1088/0264-9381/33/24/245008}{\emph{Class. Quant.
  Grav.} {\bfseries 33} (2016) 245008}
  [\href{https://arxiv.org/abs/1603.09710}{{\ttfamily 1603.09710}}].

\bibitem{Cook:2016ngj}
G.~B. Cook and M.~Zalutskiy, \emph{{Modes of the Kerr geometry with purely
  imaginary frequencies}},
  \href{https://doi.org/10.1103/PhysRevD.94.104074}{\emph{Phys. Rev. D}
  {\bfseries 94} (2016) 104074}
  [\href{https://arxiv.org/abs/1607.07406}{{\ttfamily 1607.07406}}].

\bibitem{Gibbons:1987ps}
G.~W. Gibbons and K.-i. Maeda, \emph{{Black Holes and Membranes in Higher
  Dimensional Theories with Dilaton Fields}},
  \href{https://doi.org/10.1016/0550-3213(88)90006-5}{\emph{Nucl. Phys. B}
  {\bfseries 298} (1988) 741}.

\bibitem{Duff:1991pe}
M.~J. Duff, R.~R. Khuri and J.~X. Lu, \emph{{String and five-brane solitons:
  Singular or nonsingular?}},
  \href{https://doi.org/10.1016/0550-3213(92)90025-7}{\emph{Nucl. Phys. B}
  {\bfseries 377} (1992) 281}
  [\href{https://arxiv.org/abs/hep-th/9112023}{{\ttfamily hep-th/9112023}}].

\bibitem{Duff:1993ye}
M.~J. Duff and J.~X. Lu, \emph{{Black and super p-branes in diverse
  dimensions}}, \href{https://doi.org/10.1016/0550-3213(94)90586-X}{\emph{Nucl.
  Phys. B} {\bfseries 416} (1994) 301}
  [\href{https://arxiv.org/abs/hep-th/9306052}{{\ttfamily hep-th/9306052}}].

\bibitem{Gibbons:1994vm}
G.~W. Gibbons, G.~T. Horowitz and P.~K. Townsend, \emph{{Higher dimensional
  resolution of dilatonic black hole singularities}},
  \href{https://doi.org/10.1088/0264-9381/12/2/004}{\emph{Class. Quant. Grav.}
  {\bfseries 12} (1995) 297}
  [\href{https://arxiv.org/abs/hep-th/9410073}{{\ttfamily hep-th/9410073}}].

\bibitem{Horowitz:1991cd}
G.~T. Horowitz and A.~Strominger, \emph{{Black strings and P-branes}},
  \href{https://doi.org/10.1016/0550-3213(91)90440-9}{\emph{Nucl. Phys. B}
  {\bfseries 360} (1991) 197}.

\bibitem{Galtsov:2005thm}
D.~Gal'tsov, S.~Klevtsov, D.~Orlov and G.~Clement, \emph{{More on general
  p-brane solutions}},
  \href{https://doi.org/10.1142/S0217751X06029417}{\emph{Int. J. Mod. Phys. A}
  {\bfseries 21} (2006) 3575}
  [\href{https://arxiv.org/abs/hep-th/0508070}{{\ttfamily hep-th/0508070}}].

\bibitem{Lu:1993vt}
J.~X. Lu, \emph{{ADM masses for black strings and p-branes}},
  \href{https://doi.org/10.1016/0370-2693(93)91186-Q}{\emph{Phys. Lett. B}
  {\bfseries 313} (1993) 29}
  [\href{https://arxiv.org/abs/hep-th/9304159}{{\ttfamily hep-th/9304159}}].

\bibitem{Emparan:2009cs}
R.~Emparan, T.~Harmark, V.~Niarchos and N.~A. Obers, \emph{{World-Volume
  Effective Theory for Higher-Dimensional Black Holes}},
  \href{https://doi.org/10.1103/PhysRevLett.102.191301}{\emph{Phys. Rev. Lett.}
  {\bfseries 102} (2009) 191301}
  [\href{https://arxiv.org/abs/0902.0427}{{\ttfamily 0902.0427}}].

\bibitem{Emparan:2009at}
R.~Emparan, T.~Harmark, V.~Niarchos and N.~A. Obers, \emph{{Essentials of
  Blackfold Dynamics}},
  \href{https://doi.org/10.1007/JHEP03(2010)063}{\emph{JHEP} {\bfseries 03}
  (2010) 063} [\href{https://arxiv.org/abs/0910.1601}{{\ttfamily 0910.1601}}].

\bibitem{Kol:2007rx}
B.~Kol and M.~Smolkin, \emph{{Classical Effective Field Theory and Caged Black
  Holes}}, \href{https://doi.org/10.1103/PhysRevD.77.064033}{\emph{Phys. Rev.
  D} {\bfseries 77} (2008) 064033}
  [\href{https://arxiv.org/abs/0712.2822}{{\ttfamily 0712.2822}}].

\bibitem{Goldberger:2020fot}
W.~D. Goldberger, J.~Li and I.~Z. Rothstein, \emph{{Non-conservative effects on
  spinning black holes from world-line effective field theory}},
  \href{https://doi.org/10.1007/JHEP06(2021)053}{\emph{JHEP} {\bfseries 06}
  (2021) 053} [\href{https://arxiv.org/abs/2012.14869}{{\ttfamily
  2012.14869}}].

\bibitem{Hadad:2024lsf}
T.~Hadad, B.~Kol and M.~Smolkin, \emph{{Gravito-magnetic polarization of
  Schwarzschild black hole}},
  \href{https://doi.org/10.1007/JHEP06(2024)169}{\emph{JHEP} {\bfseries 06}
  (2024) 169} [\href{https://arxiv.org/abs/2402.16172}{{\ttfamily
  2402.16172}}].

\bibitem{Cvetic:2020axz}
M.~Cveti\v{c}, P.~J. Porf\'\i{}rio and A.~Satz, \emph{{Gaussian null
  coordinates, near-horizon geometry and conserved charges on the horizon of
  extremal nondilatonic black $p$-branes}},
  \href{https://doi.org/10.1142/S0218271820410047}{\emph{Int. J. Mod. Phys. D}
  {\bfseries 29} (2020) 2041004}
  [\href{https://arxiv.org/abs/2003.09304}{{\ttfamily 2003.09304}}].

\bibitem{Satoh:1998sg}
Y.~Satoh, \emph{{BTZ black holes and the near horizon geometry of higher
  dimensional black holes}},
  \href{https://doi.org/10.1103/PhysRevD.59.084010}{\emph{Phys. Rev. D}
  {\bfseries 59} (1999) 084010}
  [\href{https://arxiv.org/abs/hep-th/9810135}{{\ttfamily hep-th/9810135}}].

\bibitem{Carlip:1995qv}
S.~Carlip, \emph{{The (2+1)-Dimensional black hole}},
  \href{https://doi.org/10.1088/0264-9381/12/12/005}{\emph{Class. Quant. Grav.}
  {\bfseries 12} (1995) 2853}
  [\href{https://arxiv.org/abs/gr-qc/9506079}{{\ttfamily gr-qc/9506079}}].

\bibitem{Berti:2009kk}
E.~Berti, V.~Cardoso and A.~O. Starinets, \emph{{Quasinormal modes of black
  holes and black branes}},
  \href{https://doi.org/10.1088/0264-9381/26/16/163001}{\emph{Class. Quant.
  Grav.} {\bfseries 26} (2009) 163001}
  [\href{https://arxiv.org/abs/0905.2975}{{\ttfamily 0905.2975}}].

\bibitem{Gray:2024qys}
F.~Gray, C.~Keeler, D.~Kubiznak and V.~Martin, \emph{{Love symmetry in
  higher-dimensional rotating black hole spacetimes}},
  \href{https://arxiv.org/abs/2409.05964}{{\ttfamily 2409.05964}}.

\bibitem{Kapec:2019hro}
D.~Kapec and A.~Lupsasca, \emph{{Particle motion near high-spin black holes}},
  \href{https://doi.org/10.1088/1361-6382/ab519e}{\emph{Class. Quant. Grav.}
  {\bfseries 37} (2020) 015006}
  [\href{https://arxiv.org/abs/1905.11406}{{\ttfamily 1905.11406}}].

\bibitem{Guica:2008mu}
M.~Guica, T.~Hartman, W.~Song and A.~Strominger, \emph{{The Kerr/CFT
  Correspondence}},
  \href{https://doi.org/10.1103/PhysRevD.80.124008}{\emph{Phys. Rev. D}
  {\bfseries 80} (2009) 124008}
  [\href{https://arxiv.org/abs/0809.4266}{{\ttfamily 0809.4266}}].

\bibitem{Lu:2008jk}
H.~Lu, J.~Mei and C.~N. Pope, \emph{{Kerr/CFT Correspondence in Diverse
  Dimensions}},
  \href{https://doi.org/10.1088/1126-6708/2009/04/054}{\emph{JHEP} {\bfseries
  04} (2009) 054} [\href{https://arxiv.org/abs/0811.2225}{{\ttfamily
  0811.2225}}].

\bibitem{Krishnan:2010pv}
C.~Krishnan, \emph{{Hidden Conformal Symmetries of Five-Dimensional Black
  Holes}}, \href{https://doi.org/10.1007/JHEP07(2010)039}{\emph{JHEP}
  {\bfseries 07} (2010) 039} [\href{https://arxiv.org/abs/1004.3537}{{\ttfamily
  1004.3537}}].

\bibitem{Chen:2010zwa}
D.~Chen, P.~Wang and H.~Wu, \emph{{Hidden conformal symmetry of rotating
  charged black holes}},
  \href{https://doi.org/10.1007/s10714-010-1080-7}{\emph{Gen. Rel. Grav.}
  {\bfseries 43} (2011) 181} [\href{https://arxiv.org/abs/1005.1404}{{\ttfamily
  1005.1404}}].

\bibitem{Lowe:2011aa}
D.~A. Lowe and A.~Skanata, \emph{{Generalized Hidden Kerr/CFT}},
  \href{https://doi.org/10.1088/1751-8113/45/47/475401}{\emph{J. Phys. A}
  {\bfseries 45} (2012) 475401}
  [\href{https://arxiv.org/abs/1112.1431}{{\ttfamily 1112.1431}}].

\bibitem{Aminov:2020yma}
G.~Aminov, A.~Grassi and Y.~Hatsuda, \emph{{Black Hole Quasinormal Modes and
  Seiberg\textendash{}Witten Theory}},
  \href{https://doi.org/10.1007/s00023-021-01137-x}{\emph{Annales Henri
  Poincare} {\bfseries 23} (2022) 1951}
  [\href{https://arxiv.org/abs/2006.06111}{{\ttfamily 2006.06111}}].

\bibitem{Bonelli:2021uvf}
G.~Bonelli, C.~Iossa, D.~P. Lichtig and A.~Tanzini, \emph{{Exact solution of
  Kerr black hole perturbations via CFT$_2$ and instanton counting: Greybody
  factor, quasinormal modes, and Love numbers}},
  \href{https://doi.org/10.1103/PhysRevD.105.044047}{\emph{Phys. Rev. D}
  {\bfseries 105} (2022) 044047}
  [\href{https://arxiv.org/abs/2105.04483}{{\ttfamily 2105.04483}}].

\bibitem{Consoli:2022eey}
D.~Consoli, F.~Fucito, J.~F. Morales and R.~Poghossian, \emph{{CFT description
  of BH\textquoteright{}s and ECO\textquoteright{}s: QNMs, superradiance,
  echoes and tidal responses}},
  \href{https://doi.org/10.1007/JHEP12(2022)115}{\emph{JHEP} {\bfseries 12}
  (2022) 115} [\href{https://arxiv.org/abs/2206.09437}{{\ttfamily
  2206.09437}}].

\bibitem{Bautista:2023sdf}
Y.~F. Bautista, G.~Bonelli, C.~Iossa, A.~Tanzini and Z.~Zhou, \emph{{Black hole
  perturbation theory meets CFT2: Kerr-Compton amplitudes from
  Nekrasov-Shatashvili functions}},
  \href{https://doi.org/10.1103/PhysRevD.109.084071}{\emph{Phys. Rev. D}
  {\bfseries 109} (2024) 084071}
  [\href{https://arxiv.org/abs/2312.05965}{{\ttfamily 2312.05965}}].

\bibitem{Arnaudo:2024rhv}
P.~Arnaudo, G.~Bonelli and A.~Tanzini, \emph{{One loop effective actions in
  Kerr-(A)dS black holes}},
  \href{https://doi.org/10.1103/PhysRevD.110.106006}{\emph{Phys. Rev. D}
  {\bfseries 110} (2024) 106006}
  [\href{https://arxiv.org/abs/2405.13830}{{\ttfamily 2405.13830}}].

\bibitem{Arnaudo:2024bbd}
P.~Arnaudo, G.~Bonelli and A.~Tanzini, \emph{{One-loop corrections to near
  extremal Kerr thermodynamics from semiclassical Virasoro blocks}},
  \href{https://arxiv.org/abs/2412.16057}{{\ttfamily 2412.16057}}.

\bibitem{Horowitz:2022mly}
G.~T. Horowitz, M.~Kolanowski and J.~E. Santos, \emph{{Almost all extremal
  black holes in AdS are singular}},
  \href{https://doi.org/10.1007/JHEP01(2023)162}{\emph{JHEP} {\bfseries 01}
  (2023) 162} [\href{https://arxiv.org/abs/2210.02473}{{\ttfamily
  2210.02473}}].

\bibitem{Horowitz:2023xyl}
G.~T. Horowitz, M.~Kolanowski, G.~N. Remmen and J.~E. Santos, \emph{{Extremal
  Kerr Black Holes as Amplifiers of New Physics}},
  \href{https://doi.org/10.1103/PhysRevLett.131.091402}{\emph{Phys. Rev. Lett.}
  {\bfseries 131} (2023) 091402}
  [\href{https://arxiv.org/abs/2303.07358}{{\ttfamily 2303.07358}}].

\bibitem{Horowitz:2024dch}
G.~T. Horowitz, M.~Kolanowski, G.~N. Remmen and J.~E. Santos, \emph{{Sudden
  breakdown of effective field theory near cool Kerr-Newman black holes}},
  \href{https://doi.org/10.1007/JHEP05(2024)122}{\emph{JHEP} {\bfseries 05}
  (2024) 122} [\href{https://arxiv.org/abs/2403.00051}{{\ttfamily
  2403.00051}}].

\bibitem{Horowitz:2024kcx}
G.~T. Horowitz and J.~E. Santos, \emph{{Smooth extremal horizons are the
  exception, not the rule}},
  \href{https://arxiv.org/abs/2411.07295}{{\ttfamily 2411.07295}}.

\bibitem{King:1975}
A.~R. King, J.~P. Lasota and W.~Kundt, \emph{Black holes and magnetic fields},
  \href{https://doi.org/10.1103/PhysRevD.12.3037}{\emph{Phys. Rev. D}
  {\bfseries 12} (1975) 3037}.

\bibitem{Bicak:1985}
J.~Bičák and V.~Janiš, \emph{{Magnetic fluxes across black holes}},
  \href{https://doi.org/10.1093/mnras/212.4.899}{\emph{Monthly Notices of the
  Royal Astronomical Society} {\bfseries 212} (1985) 899}
  [\href{https://arxiv.org/abs/https://academic.oup.com/mnras/article-pdf/212/4/899/3793483/mnras212-0899.pdf}{{\ttfamily
  https://academic.oup.com/mnras/article-pdf/212/4/899/3793483/mnras212-0899.pdf}}].

\bibitem{Bicak:2006hs}
J.~Bi\v{c}\'ak, V.~Karas and T.~Ledvinka, \emph{{Black holes and magnetic
  fields}}, \href{https://doi.org/10.1017/S1743921307004851}{\emph{IAU Symp.}
  {\bfseries 238} (2007) 139}
  [\href{https://arxiv.org/abs/astro-ph/0610841}{{\ttfamily
  astro-ph/0610841}}].

\bibitem{Giribet:2023qhz}
G.~Giribet, J.~La~Madrid, L.~Montecchio, E.~R. de~Celis and P.~Schmied,
  \emph{{Zooming in on the horizon when in its Meissner state}},
  \href{https://doi.org/10.1007/JHEP05(2023)207}{\emph{JHEP} {\bfseries 05}
  (2023) 207} [\href{https://arxiv.org/abs/2302.14140}{{\ttfamily
  2302.14140}}].

\bibitem{Kehagias:2024yzn}
A.~Kehagias, D.~Perrone and A.~Riotto, \emph{{A short note on the Love number
  of extremal Reissner-Nordstr\o{}m and Kerr-Newman black holes}},
  \href{https://doi.org/10.1016/j.physletb.2024.139109}{\emph{Phys. Lett. B}
  {\bfseries 859} (2024) 139109}
  [\href{https://arxiv.org/abs/2406.19262}{{\ttfamily 2406.19262}}].

\bibitem{Couch:1984}
W.~E. Couch and R.~J. Torrence, \emph{{Conformal invariance under spatial
  inversion of extreme Reissner-Nordström black holes}},
  \href{https://doi.org/10.1007/BF00762916}{\emph{General Relativity and
  Gravitation} {\bfseries 16} (1984) 789}.

\end{thebibliography}\endgroup
\end{document}